\pdfoutput=1
\documentclass[preprint2]{aastex}

\usepackage{graphicx}
\usepackage{amssymb}
\usepackage{epstopdf}

\usepackage{fancyhdr}
\begin{document}

\title{IPAC Image Processing and Data Archiving \\
for the Palomar Transient Factory}

\shorttitle{IPAC-PTF Processing and Archiving}

\author{Russ R. Laher\altaffilmark{1},
Jason Surace\altaffilmark{1},
Carl J. Grillmair\altaffilmark{1}, 
Eran O. Ofek\altaffilmark{2},
David Levitan\altaffilmark{3},
Branimir Sesar\altaffilmark{3},
Julian C. van Eyken\altaffilmark{4},
Nicholas M. Law\altaffilmark{5},
George Helou\altaffilmark{6},\\
Nouhad Hamam\altaffilmark{6},
Frank J. Masci\altaffilmark{6}, 
Sean Mattingly\altaffilmark{7},
Ed Jackson\altaffilmark{1},
Eugean Hacopeans\altaffilmark{8},
Wei Mi\altaffilmark{6},
Steve Groom\altaffilmark{6},
Harry Teplitz\altaffilmark{6},
Vandana Desai\altaffilmark{1}, 
David Hale\altaffilmark{9},
Roger Smith\altaffilmark{9},
Richard Walters\altaffilmark{9},
Robert Quimby\altaffilmark{10},
Mansi Kasliwal\altaffilmark{3}, 
Assaf Horesh\altaffilmark{3},
Eric Bellm\altaffilmark{3},\\
Tom Barlow\altaffilmark{3},
Adam Waszczak\altaffilmark{11},
Thomas A. Prince\altaffilmark{3}, and 
Shrinivas R. Kulkarni\altaffilmark{3}}

\email{laher@ipac.caltech.edu}

\altaffiltext{1}{Spitzer Science Center, California Institute of Technology, M/S 314-6, Pasadena, CA 91125, U.S.A.}

\altaffiltext{2}{%
Benoziyo Center for Astrophysics, Weizmann Institute of Science, 76100 Rehovot, Israel
}%

\altaffiltext{3}{%
Division of Physics, Mathematics, and Astronomy, California Institute of Technology, Pasadena, CA 91125, U.S.A.
}%
\altaffiltext{4}{Department of Physics, University of California Santa Barbara, Santa Barbara, CA 93106, U.S.A.}

\altaffiltext{5}{Department of Physics and Astronomy, University of North Carolina at Chapel Hill, Chapel Hill, NC 27599, U.S.A.}

\altaffiltext{6}{Infrared Processing and Analysis Center, California Institute of Technology, M/S 100-22, Pasadena, CA 91125, U.S.A.}

\altaffiltext{7}{%
Department of Physics and Astronomy, The University of Iowa, 203 Van Allen Hall, Iowa City, IA 52242, U.S.A.
}%

\altaffiltext{8}{%
ANRE Technologies Inc., 3115 Foothill Blvd., Suite M202, La Crescenta, CA 91214, U.S.A.
}%
\altaffiltext{9}{%
Caltech Optical Observatories, California Institute of Technology, M/S 11-17, Pasadena, CA 91125, U.S.A.
}%

\altaffiltext{10}{Kavli Institute for the Physics and Mathematics of the Universe (WPI), Todai Institutes for Advanced Study,
The University of Tokyo, 5-1-5 Kashiwanoha, Kashiwa-shi, Chiba,
277-8583, Japan}

\altaffiltext{11}{Division of Geological and Planetary Sciences, California Institute of Technology, Pasadena, CA 91125, USA}

\date{\today}

\begin{abstract}
The Palomar Transient Factory (PTF) is a multi-epochal robotic survey of the
northern sky that acquires data for the scientific study of
transient and variable astrophysical phenomena. The camera and telescope provide for
wide-field imaging in optical bands.  In the five years of
operation since first light on December 13, 2008, images taken with Mould-$R$\/ and
SDSS-$g'$ camera filters have been routinely acquired on a nightly basis
(weather permitting), and two different $H\alpha$ filters were
installed in May 2011 (656 and 663~nm).  The 
PTF image-processing and data-archival program at the Infrared Processing
and Analysis Center (IPAC) is tailored to receive and reduce the data,
and, from it, generate and preserve astrometrically and photometrically
calibrated images, extracted source catalogs, and coadded reference
images.  Relational databases have been deployed to track these
products in operations and the data archive.
The fully automated system has benefited by lessons learned from past IPAC projects and 
comprises advantageous features that are potentially incorporable
into other ground-based observatories.  
Both off-the-shelf and in-house software have been utilized
for economy and rapid development. 
The PTF data archive is curated by the NASA/IPAC Infrared Science Archive (IRSA).
A state-of-the-art custom web
interface has been deployed for downloading the raw images, processed 
images, and source catalogs from IRSA.  
Access to PTF data products is currently limited to an initial public data release (M81, M44, M42, SDSS Stripe 82, and the Kepler Survey Field).  
It is the intent of the PTF collaboration to release the full PTF data archive when sufficient funding becomes available.
\end{abstract}


\maketitle

\tableofcontents

\section{\label{intro}Introduction}

The Palomar Transient Factory (PTF) is a robotic image-data-acquisition
system whose major hardware components include a 92-megapixel digital 
camera with changeable filters mounted to the Palomar Samuel Oschin 
48-inch Telescope.
The {\it raison d'\^{e}tre}\/ of PTF is to advance our scientific
knowledge of transient and variable astrophysical phenomena.
The camera and telescope capacitate wide-field imaging in optical bands,
making PTF eminently suitable for conducting a multi-epochal survey.
The Mt.-Palomar location of the observatory limits the observations
to north of $\approx -30^{\circ}$~in declination.  The camera's pixel size
on the sky is 1.01 arcseconds.
In the five years of operation since first light on December 13, 2008 \citep{ptf}, 
images taken with  Mould-$R$\/ (hereafter~$R$) and SDSS-$g'$
(hereafter~$g$) camera filters
have been routinely acquired on a nightly basis
(weather permitting), and two different $H\alpha$ filters were
installed in May 2011 (656 and 663~nm).  
\citet{ptf} present an overview of PTF initial results and performance, and
\citet{ptf2} give an update after the first year of operation.
\citet{rau} describe the specific science cases that
enabled the preliminary planning of PTF observations.
The PTF project has been very successful in delivering a large
scientific return, as evidenced by the many astronomical discoveries
from its data; e.g., \citet{sesar}, \citet{arcavi}, and \citet{vaneyken}.  As such, it is 
expected to continue for several more years.

\begin{figure*}
\includegraphics[scale=1.0]{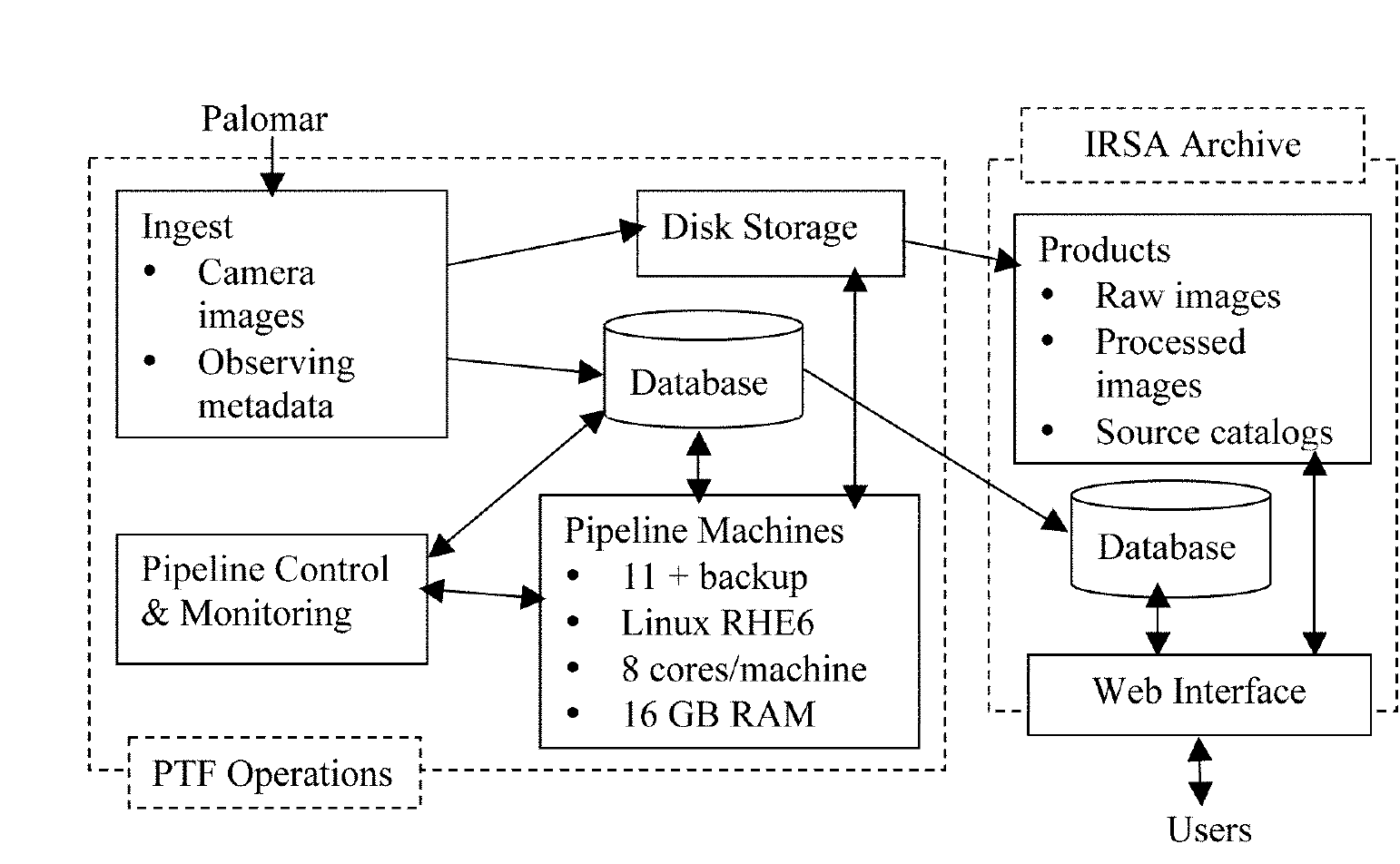}
\caption{\label{fig:overview} Data and processing flow for the
  IPAC-PTF system.}
\end{figure*}

This document presents a comprehensive report 
on the image-processing and data archival
system developed for PTF 
at the Infrared Processing and Analysis Center (IPAC).  
A simplified diagram of the data and processing flow is given in
Figure~\ref{fig:overview}.
The IPAC system is fully automated and designed to
receive and reduce PTF data, and generate and preserve astrometrically and photometrically
calibrated images, extracted source catalogs and coadded reference images.
The system has both software and hardware components.  At the top level, it consists
of a database and a collection
of mostly Perl and some Python and shell scripts that codify the complex
tasks required, such as data ingest,
image processing and source-catalog generation, 
product archiving, and metadata delivery to the archive. 
The PTF data archive is curated by the 
NASA/IPAC Infrared Science Archive\footnote{http://irsa.ipac.caltech.edu/} (IRSA).
An overview of the system has
been given by \citet{grillmair}, and the intent of this document is to
present a complete description of our system and
put forward additional details that heretofore have been generally
unavailable.

The software makes use of relational databases that are
queryable via structured query language (SQL).  The PTF operations
database, for brevity, is simply referred to herein as the
database.  Other databases utilized by the system are called out, as
necessary, when explaining their purpose.

Data-structure information useful for working directly with 
PTF camera-image files, which is important for understanding 
pipeline processes, is given in \S{\ref{cameraImages}}.  
By ``pipeline'' we mean a scripted set of processes that are
performed on the PTF data, in order to generate useful products for
calibration or scientific analysis.
Significant events that occurred during the project's multi-year timeline are 
documented in \S{\ref{projectevents}}.  
Our approach to developing the system is 
given in \S{\ref{devapproach}}.
The system's hardware architecture is laid out in \S{\ref{os}} and the
design of the database schema is outlined in \S{\ref{db}}.
The PTF-data-ingest
subsystem is entirely described in \S{\ref{ingest}}.
The tools and methodology we have developed for science data quality analysis
(SDQA) are elaborated in \S{\ref{sdqa}}.
The image-processing pipelines, along with those for calibration, are detailed in \S{\ref{idp}}.
The image-data and source-catalog archive, as well as methods for data distribution to users is explained in \S{\ref{idad}}.
This paper would be incomplete without reviewing the lessons we have learned
throughout the multi-year and overlapping periods of development and operations, and so we cover them
in \S{\ref{lessonslearned}}.  Our conclusions are
given in \S{\ref{conclusions}}.  Finally, Appendix A presents the
simple method of photometric calibration that was implemented prior to
the more sophisticated one of \citet{ofek} was brought into operation.

\section{\label{cameraImages}Camera-Image Files}

The PTF camera has 12 charge-coupled devices (CCDs) and was
purchased from the Canada-France-Hawaii Telescope~\citep{rahmer}.  
The CCDs are numbered $CCDID=0,\ldots,11$.
Eleven of the CCDs are fully functioning and one is regrettably
inoperable ($CCDID=3$; there is a broken trace that was deemed too
risky to repair).  Each CCD has $2048 \times 4096$~pixels.  The layout of the CCDs in the camera focal plane is
$2~{\rm rows} \times 6~{\rm columns}$, where the rows are east-west aligned and the
columns north-south.
This configuration enables digital imaging of an
area approximately $3.45~{\rm degrees} \times 2.30~{\rm degrees}$ on the 
sky (were it not for the inoperable CCD).  \citet{ptf, ptf2} give additional details
about the camera, system performance, and first results.

PTF camera-image files, which contain the ``raw'' data, 
are FITS\footnote{FITS stands for ``Flexible Image
  Transport System''; see http://fits.gsfc.nasa.gov} files with multiple extensions.  Each file
corresponds to a single camera exposure, and includes a primary HDU
(${\rm header} + {\rm data}$ unit) containing summary header information
pertinent to the exposure.  The primary HDU has no image data, but
does include observational metadata, such as where the telescope
was pointed, Moon and Sun positional and illumination data, weather conditions, and
instrumental and observational parameters.
Tables~\ref{tab:primaryhdu1} and~\ref{tab:primaryhdu2} selectively
list the PTF primary-header keywords, many of whose values are also written to
the {\it Exposures}\/ database table during the data-ingest process (see
\S{\ref{db}} and \S{\ref{ingest}}\/).
A camera-image file also includes 12 additional
HDUs or FITS extensions corresponding to the camera's 12 CCDs, 
where each FITS extension contains the header information and 
image data for a particular CCD.

\begin{table*}
\caption{\label{tab:primaryhdu1}Select keywords in the  PTF-camera-image primary header. }
\begin{tiny}
\begin{tabular}{l p{14cm}}
\\[-5pt]
\hline\hline
Keyword & Definition\\
\hline \\[-5pt]
{\it ORIGIN} & Origin of data (always ``Palomar Transient Factory'') \\   
{\it TELESCOP} & Name of telescope (always ``P48'') \\   
{\it INSTRUME} & Instrument name (always ``PTF/MOSAIC'') \\   
{\it OBSLAT} & Telescope geodetic latitude in WGS84 (always 33.3574 degrees)\\   
{\it OBSLON} & Telescope geodetic longitude in WGS84 (always -116.8599 degrees)\tablenotemark{1}\\   
{\it OBSALT} & Telescope geodetic altitude in WGS84 (always 1703.2 m) \\   
{\it EQUINOX } & Equinox (always 2000 Julian years) \\   
{\it OBSTYPE} & Observation type\tablenotemark{2} \\   
{\it IMGTYP} & Same as {\it OBSTYPE} \\   
{\it OBJECT} & Astronomical object of interest; currently, always set to ``PTF\_survey'' \\   
{\it OBJRA} & Sexagesimal right ascension of requested field in J2000 ({\it HH:MM:SS.SSS}\/) \\   
{\it OBJDEC} & Sexagesimal declination of requested field in J2000 ({\it DD:MM:SS.SS}\/) \\   
{\it OBJRAD} & Decimal right ascension of requested field in J2000 (degrees) \\   
{\it OBJDECD} & Decimal declination of requested field in J2000 (degrees) \\ 
{\it PTFFIELD} & PTF field number \\   
{\it PTFPID} & Project type number \\   
{\it PTFFLAG} & Project category flag (either 0 for ``non-PTF'' or 1 for ``PTF'' observations) \\   
{\it PIXSCALE} & Pixel scale (always 1.01 arcsec) \\   
{\it REFERENC} & PTF website (always ``http://www.astro.caltech.edu/ptf'') \\   
{\it PTFPRPI} & PTF Project Principal Investigator (always ``Kulkarni'') \\   
{\it OPERMODE} & Mode of operation (either ``OCS''\tablenotemark{3}, ``Manual'', or ``N/A'') \\   
{\it CHECKSUM} & Header-plus-data unit checksum \\   
{\it DATE} & Date the camera-image file was created ({\it YYYY-MM-DD}\/)\\   
{\it DATE-OBS} & UTC date and time of shutter opening ({\it YYYY-MM-DDTHH:MM:SS.SSS}\/)\\
{\it UTC-OBS} & Same as {\it DATE-OBS} \\
{\it OBSJD} & Julian date corresponding to {\it DATE-OBS}\/ (days) \\     
{\it HJD} & Heliocentric Julian date corresponding to {\it DATE-OBS}\/ (days) \\   
{\it OBSMJD} & Modified Julian date corresponding to {\it DATE-OBS}\/ (days) \\                 
{\it OBSLST} & Mean local sidereal time corresponding to {\it DATE-OBS}\/ ({\it HH:MM:SS.S}\/) \\   
{\it EXPTIME} & Requested exposure time (s) \\   
{\it AEXPTIME} & Actual exposure time (s) \\   
{\it DOMESTAT} & Dome shutter status at beginning of exposure  (either ``open'', ``close'', or ``unknown'') \\   
{\it DOMEAZ} & Dome azimuth (degrees)\\   
{\it FILTERID} & Filter identification number (ID) \\   
{\it FILTER} & Filter name (e.g., ``R'', ``g'', ``Ha656'', or ``Ha663'') \\   
{\it FILTERSL} & Filter-changer slot position (designated either 1 or 2)\\   
{\it SOFTVER} & Palomar software version (Telescope.Camera.Operations.Scheduling) \\   
{\it HDR\_VER} & Header version \\   
\hline \\[-5pt]
 \end{tabular}
\tablenotetext{1}{Some FITS headers list this value incorrectly as positive.}
\tablenotetext{2}{Possible setting is ``object'', ``dark'', ``bias'',
  ``dome'', ``twilight'', ``focus'', ``pointing'', or ``test''.  Dome and twilight images are potentially useful for constructing flats.}  
\tablenotetext{3}{OCS stands for ``observatory control system''.}
\end{tiny}
\end{table*}

\begin{table*}
\caption{\label{tab:primaryhdu2} (Continued from Table~\ref{tab:primaryhdu1}) Select keywords in the  PTF-camera-image primary header.}
\begin{tiny}
\begin{tabular}{l p{14cm}}
\\[-5pt]
\hline\hline
Keyword & Definition\\
\hline \\[-5pt]
{\it SEEING} & Seeing full width at half maximum (FWHM; pixels), an
average of {\tt FWHM\_IMAGE} values computed by {\it SExtractor} \\   
{\it PEAKDIST} & Mean of distance of brightest pixel to centroid pixel (pixels)  from {\it SExtractor}\tablenotemark{1} \\   
{\it ELLIP} & Clipped median of ellipticity\tablenotemark{2} for all non-extended field objects from {\it SExtractor} \\   
{\it ELLIPPA} & Mean of ellipse rotation angle (degrees) from {\it SExtractor} \\   
{\it FOCUSPOS} & Focus position (mm) \\   
{\it AZIMUTH} & Telescope azimuth (degrees) \\   
{\it ALTITUDE} & Telescope altitude (degrees) \\   
{\it AIRMASS} & Telescope air mass \\   
{\it TRACKRA} & Telescope tracking speed along R.A. w.r.t.\ sidereal time (arcsec/hr) \\   
{\it TRACKDEC} & Telescope tracking speed along Dec.\ w.r.t.\ sidereal time (arcsec/hr) \\    
{\it TELRA} & Telescope-pointing right ascension (degrees) \\   
{\it TELDEC} & Telescope-pointing declination (degrees) \\   
{\it TELHA} & Telescope-pointing hour angle (degrees) \\   
{\it HOURANG} & Mean hour angle ({\it HH:MM:SS.SS}\/) based on {\it OBSLST} \\   
{\it CCD0TEMP} & Temperature sensor on $CCDID=0$ (K) \\   
{\it CCD9TEMP} & Temperature sensor on $CCDID=9$ (K) \\   
{\it CCD5TEMP} & Temperature sensor on $CCDID=5$ (K) \\   
{\it CCD11TEM} & Temperature sensor on $CCDID=11$ (K) \\  
{\it HSTEMP} & Heat spreader temperature (K) \\   
{\it DHE0TEMP} & Detector head electronics temperature, master (K) \\ 
{\it DHE1TEMP} & Detector head electronics temperature, slave (K) \\   
{\it DEWWTEMP} & Dewar wall temperature (K) \\   
{\it HEADTEMP} & Cryogen cooler cold head temperature (K) \\   
{\it RSTEMP} & Temperature sensor on radiation shield (K) \\   
{\it DETHEAT} & Detector focal plane heater power (\%) \\   
{\it WINDSCAL} & Wind screen altitude (degrees) \\   
{\it WINDDIR} & Azimuth of wind direction (degrees) \\   
{\it WINDSPED} & Wind speed (km/hr) \\   
{\it OUTTEMP} & Outside temperature (C) \\   
{\it OUTRELHU} & Outside relative humidity fraction \\   
{\it OUTDEWPT} & Outside dew point (C) \\   
{\it MOONRA} & Moon right ascension in J2000 (degrees) \\   
{\it MOONDEC} & Moon declination in J2000 (degrees) \\ 
{\it MOONILLF} & Moon illuminated fraction \\   
{\it MOONPHAS} & Moon phase angle (degrees) \\   
{\it MOONESB} & Moon excess in sky $V$-band brightness (magnitude) \\   
{\it MOONALT} & Moon altitude (degrees) \\   
{\it SUNAZ} & Sun azimuth (degrees) \\   
{\it SUNALT} & Sun altitude (degrees) \\   
\hline \\[-5pt]
 \end{tabular}
\tablenotetext{1}{If the value is larger than just a few
  tenths of a pixel, it may indicate a focus or telescope-tracking
  problem.  There are 33 exposures with failed telescope tracking,
  acquired mostly in 2009, and their {\it PEAKDIST}\/ values are generally greater than a pixel.} 
\tablenotetext{2}{The ellipticity is from the {\it SExtractor}\/ ELLIPTICITY
  output parameter.  The formula $A/B$ in the FITS-header comment
  should be changed to $1-B/A$, where $A$ and $B$ are defined in the 
{\it SExtractor}\/ documentation.}
\end{tiny}
\end{table*}

The PTF camera-image data are unsigned 16-bit values that are stored
as signed 16-bit integers ($BITPIX=16$) since FITS does not directly
support unsigned integers as a fundamental data
type\footnote{See the CFITSIO User's Reference Guide.}.
Thus, the image data values are shifted by 32,768 data numbers (DN,
a.k.a.\ analog-to-digital units)
when read into computer memory ($BZERO=32768$
is the standard FITS-header keyword that controls the data-shifting when the
data are read in via a CFITSIO or comparable function), 
and so the raw-image
data are in the 0-65,535~DN range.
The raw-image size is $2078 \times 4128$~pixels, a larger 
region than covered by the actual pixels in a CCD, because it
includes regions of bias overscan ``pixels'' 
(which are the data values readout during the pixel sampling time outside of a CCD
row or column of detectors).

The {\it FILTER}, {\it EXPTIME}, {\it SEEING}, and {\it AIRMASS}\/
values associated with
camera images are among the variables that have a significant
impact on the character and quality of the image data.  The exposure
time is nominally 60~s, but this is varied as needed for targets of
opportunity or reduced to avoid saturation for some targets; e.g., SN~2011fe~\citep{nugent}.
There is also variation in some of the parameters and imaging properties
from one CCD to another (some of the CCDs are better than the others in image-quality terms).

The exposures have GMT time stamps in the camera-image filenames and
FITS headers.  This conveniently permits all exposures taken in a
given night to have the same date of observation (no date boundaries
are crossed during an observing night).  An example of a typical
camera-image filename is

\begin{verbatim}
PTF201108182046_2_o_8242.fits
\end{verbatim}

\noindent
Embedded in the filename is the date concatenated with 4 digits of the
fractional day.
The next filename field is the filter number.   The next field is a
1-character moniker for the image type: ``o'' stands for ``object'',
``b'' stands for ``bias'', ``k'' stands for ``dark'', etc.
The last field before the ``.fits'' filename extension is a non-unique
counter, which is reset to zero when the camera is rebooted (which can
happen in the course of a night, although infrequently).

\section{\label{projectevents}Significant Project Events}

There were three different events that occurred during the course of the
project which affected how the processing is done and how
the results are interpreted.  There was a fourth event, which occurred
last, that is mostly programmatic in nature. 
It is convenient to view these events as
having transpired during the day, in
between nightly data-taking periods.

On 9 October 2009, the camera electronics were reconfigured, which greatly 
improved the camera's dynamic range, thus raising the DN levels at which the pixel
detectors saturate.  Image data taken up to this date saturate in the
17,000-36,000~DN range, depending on the CCD\@.  After the upgrade, the data
saturation occurs in the 49,000-55,000~DN range.
Table~\ref{tab:sat} lists the CCD-dependent saturation values,
before and after the upgrade.

\begin{table}
\caption{\label{tab:sat}CCD-dependent saturation values, before and
  after the PTF-camera-electronics upgrade, which occurred on 9 October 2009. }
\begin{small}
\begin{tabular}{rcc}
\\[-5pt]
\hline\hline
{\it CCDID} & Before (DN) & After (DN) \\
\hline \\[-5pt]
    0    &   34,000  &   53,000\\
    1    &   36,000  &   54,000\\
    2    &   25,000  &   55,000\\
    3    &   N/A  &   N/A\\
    4    &  31,000   &  49,000\\
    5    &   33,000   &  50,000\\
    6    &   26,000   &  55,000\\
    7    &   17,000   &  55,000\\
    8    &   42,000   &  53,000\\
    9     &  19,000   &  52,000\\
   10    &   25,000  &   52,000\\
   11    &   36,000   &  53,000\\  
\hline \\[-5pt]
\end{tabular}
\end{small}
\end{table}

On 15 July 2010, the positions of the 
$R$ and $g$ filters were swapped in the filter wheel.  This not only
made the expected filter positions in the filter wheel time-dependent,
but also altered the positions of the ghost reflections on the focal
plane (and, hence, in the images).  

On 2 September 2010, the ``fogging problem'' was solved, which had
been causing a diffuseness in the images around bright stars, and was
the result of an oil film slowly building up on the camera's cold CCD
window during the times between the more-or-less bimonthly window cleanings.
\citet{ofek} discuss the resolution of this problem in more detail.

On 1 January 2013, the official PTF program ended and the
``intermediate'' PTF (iPTF) program started.\footnote{http://ptf.caltech.edu/iptf/}  Coincidently, PTF-archive users will
notice that {\it DAOPHOT}\/ source catalogs~\citep{stetson} are available from this
point on, in addition to the already available {\it SExtractor}\/ source
catalogs~\citep{sex}, which is the result of pipeline upgrades that
were delivered around that time.  Also, this was around the time that
the IPAC-PTF reference-image, real-time, and difference-image pipelines came online.

\section{\label{devapproach}Development Approach}

This section covers our design philosophy and assumptions, and the software
guidelines that we followed in our development approach.

\subsection{Design Philosophy and Assumptions}

The development of the data-ingest, image-processing, 
archival, and distribution components for PTF data and 
products have leveraged existing astronomical software and the 
relevant infrastructure of ongoing projects at IPAC.

Database design procedures developed at IPAC have been followed in order to 
keep the system as generic as possible, and not reliant on a particular brand of database.  
This allows the flexibility of switching from one database to another
over the project's many years of operation, as necessary.

We strived for short database table and column names to minimize keyboard
typing (and mostly achieved this), and to quicken learning the database schema.
We avoided renaming primary keys when used as foreign keys 
in other tables, in order to keep table joins simple.  (A primary key
is a column in a table that stores a unique identification number for
each record in the table, and a foreign key is a column in a table that stores
the primary key of another table and serves to associate a record in one table with a record
in another table.)

The metadata stored in the database on a regular basis during normal
operations come directly from, or are
derivable from information in either the header or filename of camera-image 
files containing the raw data, as well as nightly-observing metadata files.  
Thus, very little prior information about scheduling of specific
observations is required.

We expect to have to be able to deal with occasional corrupt or
incomplete data.  The software must therefore be very robust, and, for
example, be able to supply missing information, if possible.  Having
the ability to flag bad data in various ways is useful.  This and the
means of preventing certain data from undergoing processing are
necessary parts of the software and database design.

Another important aspect of our design is versioning.  Software,
product, and archive versioning are handled independently in our
design, and this simplifies the data and processing management.
A data set, for example, may be subjected to several rounds of
reprocessing to smooth out processing wrinkles before its 
products are ready to be archived.

\subsection{Software Guidelines}

An effort has been made to follow best programming practices.  A
very small set of guidelines were followed for the software
development, and no computer-language restrictions were imposed so
long as the software met performance expectations.  We have made use
of a variety of programming languages in this project, as our team is
quite diverse in preferences and expertise.

The source code is checked into a version control system (CVS).  An
updated CVS version string is automatically embedded into every source-code file
each time a new file version is checked into the CVS repository, and
this facilitates tracking deployed software versions when
debugging code.  The web-based software-version-control system called {\it GNATS}\/ is used
for tracking software changes and coordinating software releases.

All Perl scripts are executed from a common installation of Perl that is
specified via environment variable $PERL\_PATH$, and
require explicit variable declaration (``use strict;'').  Minimal
use is made of global variables.  Standalone blocks of code are wrapped as
subroutines and put into a library for reuse and complexity hiding.

Modules requiring fast computing speed were generally developed in the
C~language on Mac laptops and tested there prior to
deployment on the Linux pipeline machines.  Thus, the software benefited from
multi-platform testing, which enhances its robustness and improves the
chances of uncovering bugs.

All in-house software, which excludes third-party software, is designed to
return a system value in the 0-31 range for normal termination, in the 32-63
range for execution with warnings, and $>=64$ if an error occurs.
At the discretion of the programmer, specific values are designated for special conditions,
warnings, and errors that are particular to the software under development.

All scripts generate log files that are written to the PTF logs
directory, which is appropriately organized into subdirectories
categorized by process type.  The log files are very verbose, and
explicit information is given about the processes executed, along with
the input parameters and command-line options and switches used.
Software version numbers are included, as well as is timing information,
which is useful for benchmark profiling.

\section{\label{os}System Architecture}

\begin{figure*}
\begin{center}
\includegraphics[scale=0.45]{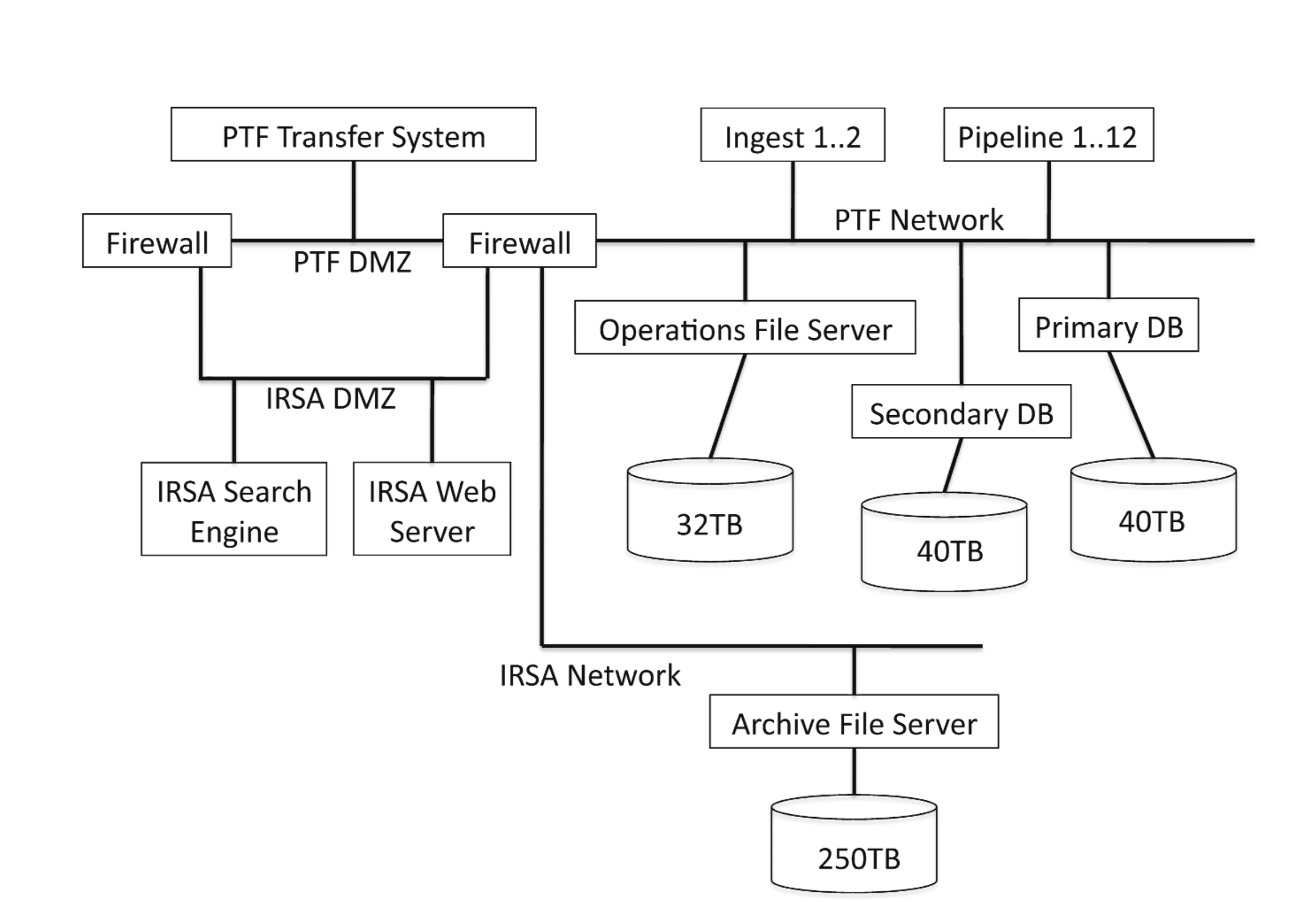}
\caption{\label{fig:hardware} Computing, network, and archiving hardware for the IPAC-PTF system.}
\end{center}
\end{figure*}

Figure~\ref{fig:hardware} shows the principal hardware components of
the IPAC-PTF system, which are located on the Caltech campus.  
Firewalls, servers, and pipeline
machines, which are depicted as rectangular boxes in the figure, are
currently connected to
a 10~gigabit/s network (this was upgraded in 2012 from 1~gigabit/s).
Firewalls provide the
necessary security and isolation between the 
PTF transfer machine that receives nightly PTF data, the IRSA web services, and the 
operations and archive networks.  
A demilitarized zone (DMZ) outside of the inner firewall 
has been set up for the 
PTF transfer machine.
A separate DMZ exists for the IRSA search
engine and web server.  

The hardware has redundancy to minimize down time.
Two data-ingest machines, a primary and a
backup, are available for the data-ingest process (see \S{\ref{ingest}}), but only
one of these machines is required at any given time.  There are 12 identical
pipeline machines for parallel processing, but only 11 are needed for
the pipelines, and so the remaining machine serves as a backup.  
The pipeline machines have 64-bit Linux operating systems installed (Red Hat
Enterprise~6, upgraded from 5 in early 2013), and each has 8~CPU cores and 16~GB of memory.
There are two database servers: a primary for regular PTF operations, and a
secondary for the database backup.
Currently, the database servers are running the Solaris-10 operating system,
but are accessible by database clients running under Linux.

There is ample disk space, which is attached to the operations file
server, for staging camera-image files during the data
ingest and temporarily storing pipeline intermediate and final products.  
These disks are cross-mounted to all pipeline machines for
the pipeline image processing.  This design strategy 
minimizes network traffic by allowing intermediate products to be
available for a short time for debugging purposes and only 
transferring final products to the archive.
The IRSA archive file server is set up to allow the copying of files
from PTF operations through the firewall.  The IRSA archive disk
storage is currently 250~TB, and this will be augmented as needed
over the project lifetime.
It is expected that this disk capacity will be doubled by the end of
the project.
In general, the multi-terabyte disk storage is broken up into 8~TB or
16~TB partitions to facilitate disk management and file backups.

\section{\label{db}Database}

We initially implemented the database in Informix to take advantage
of Informix tools, interfaces, methodologies, and expertise developed under 
the Spitzer project.  After a few months, we made the decision to
switch to an open-source PostgreSQL database, as our Informix 
licensing did not allow us to install the database server on another 
machine and purchasing an additional license was not an option due to
limited funding.  All in all, it was a
smooth transition, and there was a several-month period of overlap where we were
able to switch between Informix and PostgreSQL databases simply by changing a
few environment variables.

\begin{figure}
\includegraphics[scale=0.5]{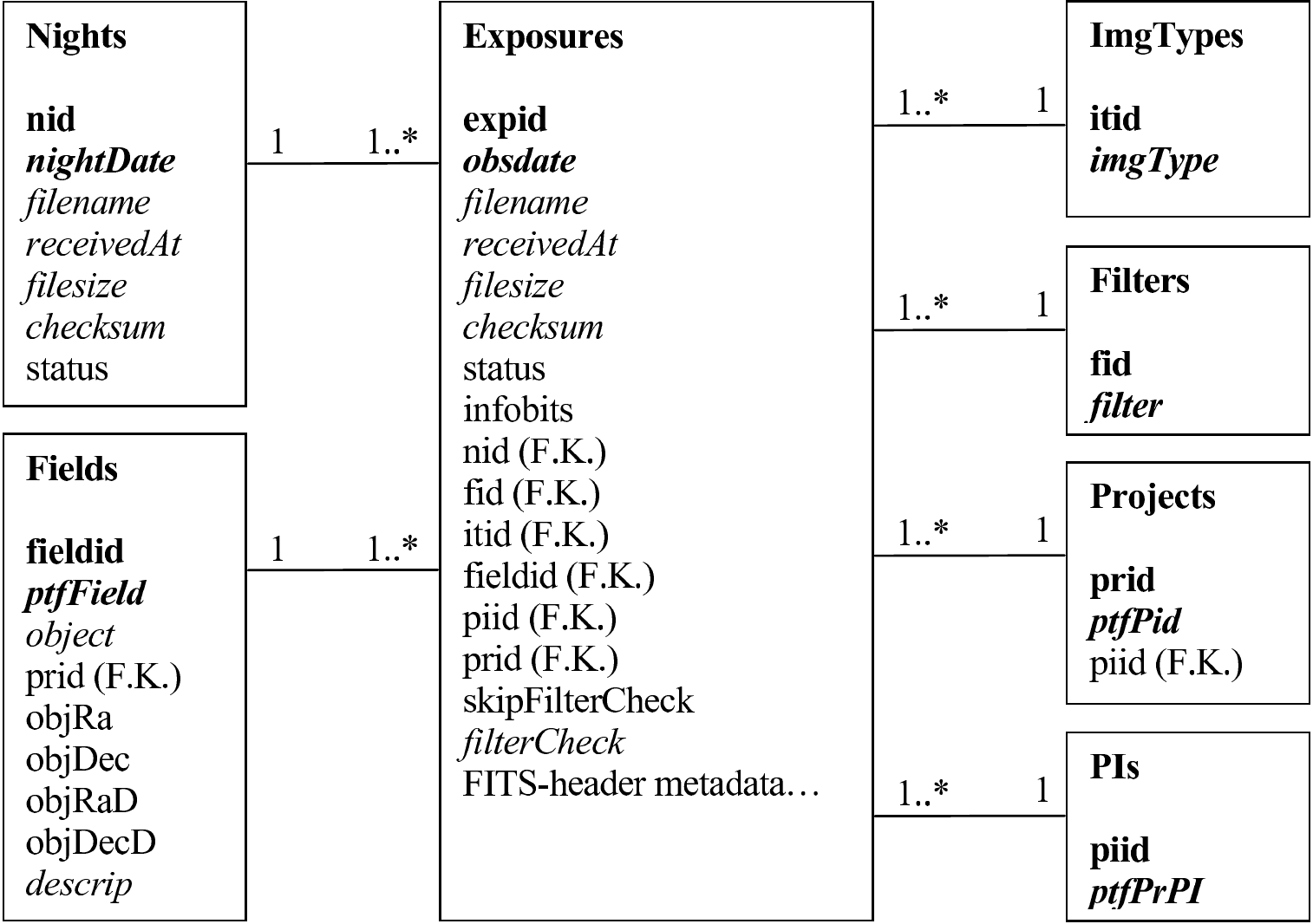}
\caption{\label{fig:ingestdbschema} IPAC-PTF database-schema design
  for the data ingest (see \S{\ref{ingest}}).
  The database table name is given at the top of each box.  The bold-font
  database column listed after the table name in each box is the primary key of the
  table.  The columns listed in bold-italicized font are the alternate
  keys.  The columns listed in regular font are not-null
  columns, and in regular-italicized font are null columns (which are
  columns in which null values possibly may be stored). 
``F.K.'' stands for foreign key, and ``1~{\jot 24pt}~1*..'' stands for
one record to many records, etc..}
\end{figure}

Figure~\ref{fig:ingestdbschema} depicts the database schema for the
basic tables associated with ingesting PTF data.  Some of the details
in the figure are explained in its caption and in \S{\ref{ingest}}.  Briefly, the 
{\it Nights}\/ database table tracks whether any given night has been
successfully ingested ($status=1$) or not ($status=0$).  A record for
each camera exposure is stored in the {\it Exposures}\/ database table,
and each record includes the camera-image filename, whether the
exposure is good ($status=1$) or not ($status=0$, such as in the rare
case of bad sidereal tracking), and other exposure
and data-file metadata.  The exposure metadata is obtained directly
from the primary FITS header of the camera-image file (see \S{\ref{cameraImages}}).
The remaining database tables in the figure track the
database-normalized attributes of the exposures.
The {\it Filters}\/ database table, for example, contains one record per
unique camera filter used to acquire the exposures.

Not shown in Figure~\ref{fig:ingestdbschema} is the 
{\it FieldCoverage}\/ database table, which contains the most complete set
available of fields to be scheduled for multi-epochal
observation, whereas all other tables for information about PTF
data store only records for data that have already been acquired.
This table is not required for the data ingest, but is used by the
pipeline that performs the astrometric calibration
(see~\S{\ref{frameprocpipeline}}),
since it includes columns that identify cached astrometric catalogs
for each PTF field.
A fairly complete list of PTF-operations database tables is given in
Table~\ref{tab:schema}.

\begin{table*}
\caption{\label{tab:schema}Operations database tables of the Palomar Transient Factory.}
\begin{tiny}
\begin{tabular}{l p{14cm}}
\\[-5pt]
\hline\hline
Table name & Description\\
\hline \\[-5pt]
{\it Nights} & Nightly data-ingest status and other metadata (e.g., images-manifest filenames).  Unique index: {\it nid}. Alternate key: {\it nightdate}.\\
{\it Exposures} & Exposure status and other metadata (e.g., camera-image filenames). Unique index: {\it expid}. Alternate key: {\it obsdate}.\\
{\it CCDs} & CCD constants (e.g., sizes of raw and processed images, in pixels). Unique index: {\it ccdid}.\\
{\it Fields} &  Observed PTF field positions and their assigned
identification numbers (IDs). Unique index: {\it fieldid}.  Alternate key: {\it ptffield}.\\
{\it FieldCoverage} & Field positions and their fractional overlap onto SDSS\tablenotemark{1} fields.  
 Unique index: {\it fcid}.  Alternate keys: {\it ptffield}\/ and {\it ccdid}.\\
{\it ImgTypes} & Image types taken by PTF camera (``object'', ``bias'', ``dark'', etc.). Unique index: {\it itid}.\\
{\it Filters} & Camera filters available.  Currently $R$, $g$, and two different H$\alpha$ filters are available.  Unique index: {\it fid}.\\
{\it FilterChecks} & Cross-reference table between filter-checker output indices and human-readable filter-check outcomes.\\
{\it PIs} & Principal-investigator contact information. Unique index: {\it piid}.\\
{\it Projects} & Project abstracts, keywords, and associated investigators. Unique index: {\it prid}.\\
{\it Pipelines} & Pipeline definitions and pipeline-executive metadata (e.g., {\it priority}\/). Unique index: {\it ppid}.\\
{\it RawImages} & Raw-image metadata (after splitting up FITS-multi-extension camara images as needed). Unique index: {\it rid}.\\
{\it ProcImages} & Processed-image metadata (e.g., image filenames). Unique index: {\it pid}.  Alternate keys: {\it rid}, {\it ppid}, and {\it version}.\\
{\it Catalogs} & Metadata about {\it SExtractor}\/ and {\it DAOPHOT}\/ catalogs extracted from processed images. Unique index: {\it catid}.\\
{\it AncilFiles} & Ancillary-product associations with processed images.  Unique index: {\it aid}.  Alternate keys: {\it pid}\/ and {\it anciltype}.\\
{\it CalFiles} & Calibration-product metadata (e.g., filenames, and date-ranges of applicability).  Unique index: {\it cid}.  \\
{\it CalFileUsage} & Associations between processed images ({\it pid}\/) and calibration products ({\it cid}\/).\\
{\it CalAncilFiles} & Ancillary calibration product metadata.  Unique index: {\it caid}.  Alternate keys: {\it cid}\/ and {\it anciltype}.\\
{\it IrsaMeta} & Processed-image metadata required by IRSA (e.g., image-corner positions).  Unique index: {\it pid}\/ (foreign key).\\
{\it QA} & Quality-analysis information (e.g., image statistics).  Unique index: {\it pid}\/ (foreign key).\\
{\it AbsPhotCal} & Absolute-photometric-calibration coefficients.  Unique index: {\it apcid}.   Alternate keys: {\it nid}, {\it ccdid}, and {\it fid}. \\
{\it AbsPhotCalZpvm} & Zero-point-variability-map data. Primary keys: {\it apcid}, {\it indexi}, and {\it indexj}. \\
{\it RelPhotCal} & Relative-photometric-calibration zero points.  Unique index: {\it rpcid}.   Alternate keys: {\it ptffield}, {\it ccdid}, {\it fid}, and {\it version}. \\
{\it RelPhotCalFileLocks} & Utilizes row locking to manage file locking. Primary keys: {\it ptffield}, {\it ccdid}, and {\it fid}. \\
{\it Ghosts} & Metadata about ghosts in processed images.  Unique index: {\it gid}. Alternate keys: {\it pid}, {\it ccdid}, {\it fid}, and ({\it x, y }\/). \\
{\it Halos} & Metadata about halos in processed images.  Unique index: {\it hid}. Alternate keys: {\it pid}, {\it ccdid}, {\it fid}, and ({\it x, y }\/). \\
{\it Tracks} & Metadata about aircraft/satellite tracks in processed images.    Unique index: {\it tid}. Alternate keys: {\it pid}, {\it ccdid}, {\it fid}, and {\it num}. \\
{\it PSFs} & Point spread functions (PSFs) in {\it DAOPHOT}\/ format.    Unique index: {\it psfid}.  Alternate key: {\it pid}.\\
{\it RefImages} & Reference-image metadata (e.g., filenames). Unique index: {\it rfid}.   Alternate keys: {\it ccdid}, {\it fid}, {\it ptffield}, {\it ppid}, and {\it version}.\\
{\it RefImageImages} & Associations between processed images ({\it pid}, $ppid=5$) and reference images ({\it rfid}, $ppid=12$).\\
{\it RefImAncilFiles} & Ancillary-product associations with reference images. Unique index: {\it rfaid}. \\
{\it RefImageCatalogs} & Metadata about {\it SExtractor}\/ and {\it DAOPHOT}\/ catalogs extracted from reference images. Unique index: {\it rfcatid}.\\
{\it IrsaRefImMeta} & Reference-image metadata required by IRSA (e.g., image-corner positions).  Unique index: {\it rfid}\/ (foreign key).\\
{\it IrsaRefImImagesMeta} & IRSA-required metadata for processed images that are coadded to make the reference images (see {\it RefImageImages} database table).\\
{\it SDQA\_Metrics}\tablenotemark{2} & SDQA-metric definitions. Unique index: {\it sdqa\_metricid}.\\
{\it SDQA\_Thresholds} & SDQA-threshold settings.  Unique index: {\it sdqa\_thresholdid}.\\
{\it SDQA\_Statuses} & SDQA-status definitions. Unique index: {\it sdqa\_statusid}.\\
{\it SDQA\_Ratings} & SDQA-rating values for processed images. Unique index: {\it sdqa\_ratingid}.  Alternate keys: {\it pid}\/ and {\it sdqa\_metricid}.\\
{\it SDQA\_RefImRatings} & SDQA-rating values for reference images. Unique index: {\it sdqa\_refimratingid}.  Alternate keys: {\it rfid}\/ and {\it sdqa\_metricid}.\\
{\it SDQA\_CalFileRatings} & SDQA-rating values for calibration files. Unique index: {\it sdqa\_calfileratingid}.  Alternate keys: {\it cid}\/ and {\it sdqa\_metricid}.\\
{\it SwVersions} & Software version information. Unique index: {\it svid}.\\
{\it CdfVersions} & Configuration-data-file version information. Unique index: {\it cvid}.\\
{\it ArchiveVersions} & Metadata about archive versions. Unique index: {\it avid}.\\
{\it DeliveryTypes} & Archive delivery-type definitions.  Unique index: {\it dtid}.\\
{\it Deliveries} & Archive delivery tracking information.  Unique index: {\it did}.\\
{\it Jobs} & Pipeline-job tracking information.  Unique index: {\it jid}.\\
{\it ArchiveJobs} & Archive-job tracking information.  Unique index: {\it ajid}.\\
{\it JobArbitration} &  Job-lock table.\\
{\it IRSA} &  Temporary table for marshaling of metadata to be delivered to the IRSA archive.\\
\hline \\[-5pt]
\end{tabular}
\tablenotetext{1}{Sloan Digital Sky Survey~\citep{york}. }
\tablenotetext{2}{SDQA stands for science data quality analysis. }
\end{tiny}
\end{table*}

\begin{figure*}
\begin{center}
\includegraphics[scale=0.8]{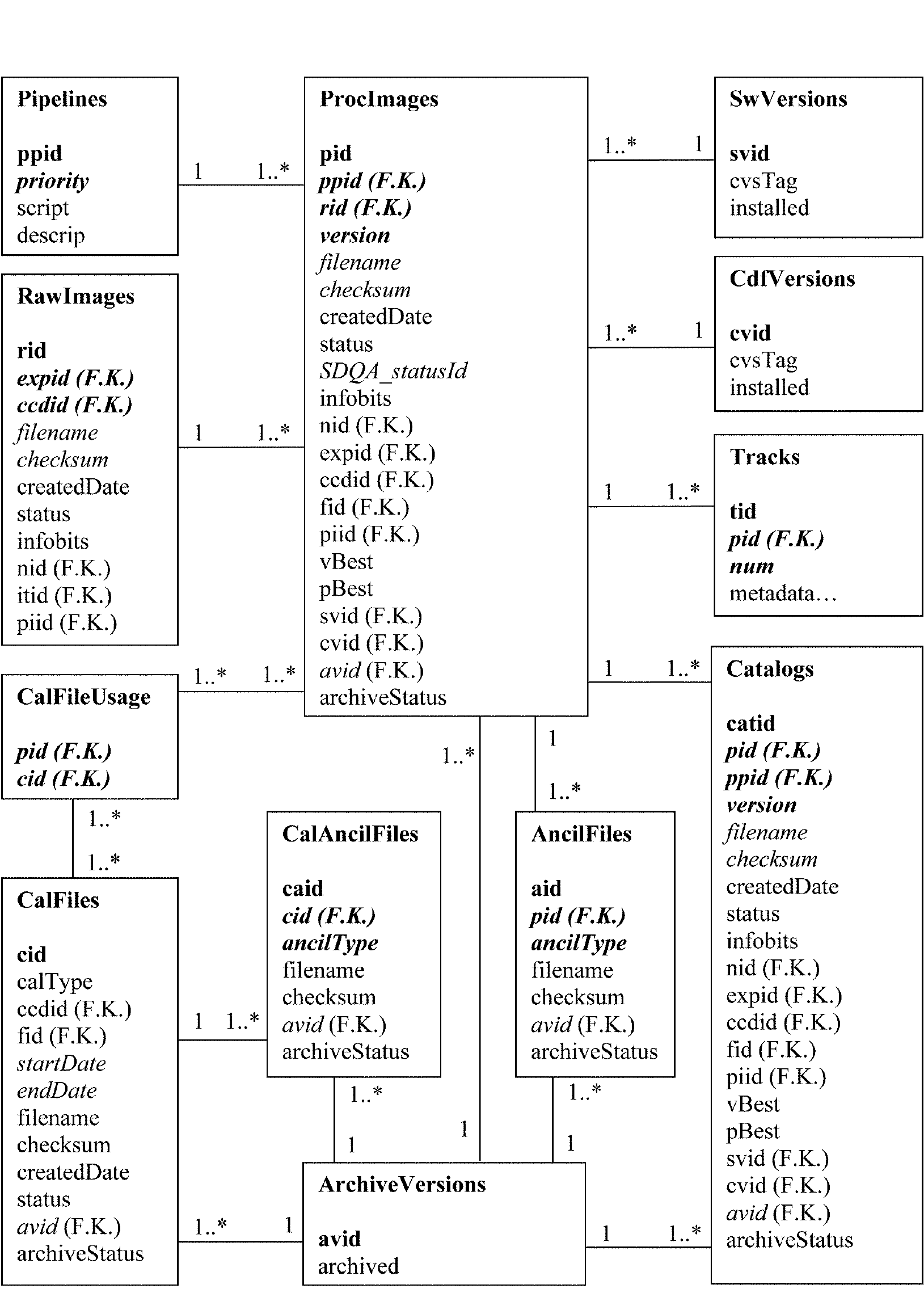}
\end{center}
\caption{\label{fig:procdbschema} IPAC-PTF database-schema design
  for the pipeline image processing (see \S{\ref{idp}}).  The figure
  nomenclature is explained in the caption of Figure~\ref{fig:ingestdbschema}.}
\end{figure*}

Figure~\ref{fig:procdbschema} shows a portion of the database schema relevant to
the pipeline image processing.  The key features of the database tables involved 
are given in the remainder of this section.   The various utilities of these database
tables are discussed throughout this paper as well.  For conciseness,
several equally important database tables are not shown, but are discussed
presently (e.g., see \S{\ref{frameprocpipeline}}).  These
include tables for science data quality analysis (SDQA), photometric
calibration, and tracking artifacts such as ghosts and halos.

The {\it Pipelines}\/ database table assigns a unique index to each
pipeline and stores useful pipeline metadata, such as their priority
order of execution.  See \S{\ref{pipelineoverview}}
and~\S{\ref{pipeexec}} for a detailed
discussion of the table's data contents.

The {\it RawImages}\/ database table stores metadata about raw images, 
one record per raw-image file, where each raw-image file corresponds to the data 
from one of the camera's CCDs in an exposure.  While the 12-CCD
camera images are archived (and tracked in the {\it Exposures}\/
database table), the raw-image files associated with
the {\it filename}\/ column in the  {\it RawImages}\/ database table are not archived,
but serve as pipeline inputs from the sandbox, for as long as they are needed, and then
are eventually removed from the file system to avoid duplicate storage.

The {\it ProcImages}\/ database table stores metadata about processed images, 
one record per image file.  There is a one to many relationship between
{\it RawImages}\/ and {\it ProcImages}\/ records because a given raw image
can be processed multiple times, which is useful when the software
version (tracked in the {\it SwVersions}\/ database table) is upgraded
or the software configuration (tracked in the {\it CdfVersions}\/ database table)
needs to be changed.  Moreover, a given raw image can be processed by
different pipelines.  The {\it version}\/ column keeps track of the
processing episode for a given combination of raw image ({\it rid}\/) 
and pipeline ({\it ppid}\/).  The {\it vBest}\/ column is automatically 
set to one for the latest version and zero for all previous versions,
unless a previous version has the column set to $vBest = 2$, in which case 
it is ``locked'' on that previous version.  In addition, similar
products can be generated by different pipelines, and the {\it pBest}\/
column flags which of the pipelines' products are to be archived.

The {\it Catalogs}\/ database table stores metadata about the 
{\it SExtractor}\/ source catalogs, one record per catalog file.
There is a one to many relationship between
{\it ProcImages}\/ and {\it Catalogs}\/ records because 
catalogs can be regenerated from a given processed image
multiple times.  Image processing takes much more time than catalog
generation, and the latter can be redone, if necessary, without having
to redo the former.  The structure of the  {\it Catalogs}\/ database
table is analogous to that of the  {\it ProcImages}\/ database table
with regard to product versioning and tracking.

The {\it AncilFiles}\/ database table stores metadata about ancillary files that
are created during the pipeline image processing and directly related to
processed images (i.e., ancillary files besides catalogs, which are a special kind of
ancillary file registered in the {\it Catalogs}\/ database
table).  Ancillary files presently include data masks and JPEG preview
images, which are distinguished by the {\it ancilType}\/ column.
The table is flexible in that new {\it ancilType}\/ settings can be
defined for new classes of ancillary files that may arise in the course of development.
This database table
enforces the association between all ancillary files and 
their respective processed images.

Calibration files are created by calibration pipelines and applied by
image-processing pipelines.
The {\it CalFiles}\/ and {\it CalFileUsage}\/ database tables allow multiple
versions of calibration files to be tracked and associated with the 
resulting processed images.

The {\it ArchiveVersions}\/ database table is pivotal for managing
products in the data archive.  For more on that and the archive-related
columns in the {\it ProcImages},  {\it Catalogs}, {\it AncilFiles}, {\it CalFiles},
and {\it CalAncilFiles}\/ database tables, see \S{\ref{productarchiver}}.

The {\it Jobs}\/ database table is indexed by primary key {\it jid}.  It
contains a number of foreign keys that index the associated pipeline 
({\it  ppid}\/) and various data parameters (e.g., night, CCD, and
filter of interest).  It contains time-stamp columns for when the
pipeline started and ended, as well as elapsed time, and it also contains
columns for pipeline exit code, status, and machine number.  Possible
status values -1, 0, or 1 indicate the job is suspended, is ready
to be executed, or has been executed, respectively.  

The {\it AchiveJobs}\/ database table is indexed by primary key {\it ajid}.  
Since product archiving is done on a nightly basis, the database table
has columns that store the date of night of interest ({\it  nightDate}\/), 
and the associated night database index (foreign key {\it nid}\/) for
added convenience.  It contains time-stamp columns for when the
archive job started and ended, as well as for the elapsed time, and it also contains
columns for the archive-job status and virtual-pipeline-operator exit
code (see \S{\ref{vpo}}).  Possible status values -1, 0, or 1 indicate the job is either
in a long transaction (currently running or temporarily suspended), 
is ready to be executed, or has been executed, respectively.  

All database tables that store information about files have a column for 
storing the file's checksum; this is useful for verifying the data
integrity of the file over time.  There is also the very useful {\it status}\/
column for tracking whether the file is good ($status=1$) or not
($status=0$); many pipeline database queries for files require
$status>0$, and files with $status=0$ are effectively removed 
from the processing.  Note also that the filename column in 
these tables is for storing the full path and filename, in order
to completely specify the file's location in file storage.
Most of the database tables in the schema have their first
column data-typed as a database serial ID, in order to enforce record-index
uniqueness, and this is called the primary key of the database table.

The database is backed up weekly, and generally at a convenient time,
i.e., when the pipelines are not running.  The procedure involves stopping all
processes that have database connections (e.g., the pipeline-executive
jobbers), because it is desirable to ensure the database is in a known
state when it is backed up. 
A script is run to query
for database-validation data.  The database server is stopped and the
database file system is snapshotted.  This step takes just a few
seconds, and the database server and pipelines can be restarted
immediately afterwards.
This backup procedure is performed by the pipeline operator.
The database administrator is then responsible for building
a copy of the database from the snapshot and validating it.  The database copy
is made available to expert users from a different database
server.  It is sometimes expedient to test software for schema and data
content changes in the users' database prior to deployment in operations.

\section{\label{ingest}Data Ingest}

This section describes the data flow, processes, and software involved
in the nightly ingestion of PTF data at IPAC\@.  
The data-ingest software has been specially developed
in-house for the PTF project.  

A major requirement is that the ingest process shall not modify either 
the camera-image filenames as received or the data contained within the files.  
The reason for this is to ensure traceability of the files back to the
mountain top where they are created.  Moreover, there are opportunities to ameliorate the
image metadata in the early pipeline processing, if needed, and experience
has shown that, in fact, this must be done occasionally.
The ingest principal functions are to move the files into archival disk 
storage and store information about them in a relational database.  
There are other details, and these are described in
the subsections that follow.

\subsection{\label{highlevelingest}High-Level Ingest Process}

PTF camera-image files are first sent to a data center in San Diego, CA 
from Mount Palomar via fast microwave link and land line as an
intermediate step, and then pushed to IPAC over the Internet.  
The files are received throughout the night at IPAC onto a 
dedicated data-transfer computer
that sits in the IPAC DMZ (see \S{\ref{os}}).  
A mirrored 1~terabyte (TB) disk holds the {\it /inbox}\/ partition where 
the files are written upon receipt.  
This partition is exported via network file system (NFS) 
to both primary and backup data-ingest machines, which are
located behind the firewall.
The primary machine predominantly 
runs the data-ingest processes.
There is also a separate backup data-ingest computer in
case the primary machine malfunctions, and this machine is also 
utilized as a convenience for sporadically required manual data ingestion.

A file containing
a cumulative list of nightly image files, along with their file sizes and MD5
checksums, is also updated throughout the night and pushed to IPAC
after every update.  This special type of file, each one uniquely
named for the corresponding night, is called the ``images manifest''.
The images manifest has a well-defined filename
with embedded observation date and fixed filename extension, 
suitable for parsing via computer script.
An end-of-file
marker is written to the images manifest at the end of the night after
all image files have been acquired and transferred.  This signals the
IPAC data-ingest software subsystem that an entire night's worth of data
has been received, and the data-ingest process is ready to be
initiated for the night at hand.  The contents of each images manifest
are essentially frozen after the end-of-night marker has been written.

The basic data-ingest process involves copying all image files to archival
spinning disk and registering metadata about the night and image files
received in the database.  A number of steps are involved, and these
steps foremost include verifying that the image files are complete,
uncorrupted, permanently stored, and retrievable for image processing.

The data are received into disk subdirectories of the {\it /inbox}\/
partition, each named for the year, month
and day of the observations.  The date and time stamps in the data are in GMT\@.   
A cron job running on the data-ingest computer every 30~minutes 
launches a Bourne shell script, called
{\it automate\_stage\_ingest}, that
checks for both the existence of the images manifest of the current night and that
the end-of-night signal is contained in the images manifest.
A unique lock file is written to
the /tmp directory to ensure that only one night at a time is ingested. 
It then initiates the high-level data-ingest process after these 
conditions are met.  This process runs under the root account because file
ownership must be changed from the data-transfer account to the
operations account under which the image-processing pipelines are executed.   

The high-level data-ingest process is another Bourne shell script, called
{\it stage\_PTF\_raw\_files}, that performs the following steps:

\begin{enumerate}
\item{Checks that the number of files received matches the number of
    files listed in the images manifest.  An alert is e-mailed to
    operations personnel if this condition is not satisfied, and 
    the process is halted.  The cron job will try again 30~minutes later for the
    current night.}
\item{Copies the files into an observation-date-stamped subdirectory under
    the {\it /staging}\/ partition, which is owned by the
    operations account and is an NFS mount point from the operations file server.}
\item{Changes to the aforementioned data directory that houses the
    nightly files to be ingested, and executes the low-level 
    data-ingest processes (see \S{\ref{lowlevelingest}}).
    Bourne-shell script {\it ingest\_staged\_fits\_files}\/ wraps the commands for these processes.}
\item{As a file backup, copies the files into an observation-date-stamped subdirectory under
    the {\it /nights}\/ partition, which is also owned by the
    operations account, but is an NFS mount point from the archive file server.  This is done in parallel to the low-level
    data-ingest process, so as not to hold it up.}
\item{Checks the MD5 checksums of the files stored in the observation-date-stamped subdirectory under
    the {\it /nights}\/ partition.  Again, this rather time-consuming
    process is done in parallel to the low-level
    data-ingest processes.}
\item{Removes the corresponding subdirectory under the {\it /inbox}\/
    partition (and all files therein) upon successful data ingest.
    This will inhibit the cron job from trying to ingest the same
    night again.}
\end{enumerate}

\noindent As a final step, the aforementioned script {\it ingest\_staged\_fits\_files}\/
executes a database command that preloads camera-image-splitting
pipelines for the current night into the {\it Jobs}\/ database table, one 
pipeline instance per camera-image file.
This pipeline is described in~\S{\ref{cameraiimagesplitter}}.

All scripts generate log files that are written to the {\it scripts}\/
and {\it ingest}\/ subdirectories in the PTF logs
directory.

\subsection{\label{lowlevelingest}Low-Level Ingest Processes}

There are three low-level data-ingest processes, which are executed in
the following order:

\begin{enumerate}
\item{Ingest the camera-image files;}
\item{Check the file checksums; and}
\item{Ingest the images manifest.}
\end{enumerate}

\noindent These processes are described in detail in the following paragraphs.

The Perl script called {\it ingestCameraImages.pl}\/ works
sequentially on all files in the current working directory (an
observation-date-stamped subdirectory under the {\it /staging}\/ partition).
A given file first undergoes a number of checks.  Files that are not
FITS files or less that 5~minutes old are skipped for the time being.  All files that are
FITS files and older than 5~minutes are assumed to be PTF camera-image
files and will be ingested.  The MD5 checksum is computed, and the
file size is checked.  Files smaller than 205 Mbytes will be
ingested, but the {\it status}\/ column will be set to zero and
bit $2^0=1$ will be set in the {\it infobits}\/ column
of the  {\it Exposures}\/ database table (see
Table~\ref{tab:ingestinfobits}) for records associated with files that
are smaller than expected, as this has revealed an upstream software
problem in the past.
Select keywords are read from the FITS header (i.e., a
large subset of the keywords listed in Tables~\ref{tab:primaryhdu1}
and~\ref{tab:primaryhdu2}).  The temperature-related FITS keywords are
expected to be missing immediately after a camera reboot, in which
case the software substitutes the value zero for these keywords, and
bit $2^9=512$ is set in the {\it infobits}\/ column
of the  {\it Exposures}\/ database table.
Files with missing {\it FILTER}, {\it FILTERID}, or {\it FILTERSL}\/
will have both their values and their {\it status}\/ set to zero in the {\it Exposures}\/
database table, along with bit $2^2=4$ set in the {\it infobits}\/ column
of the  {\it Exposures}\/ database table.
All science-image files are checked for the unlikely state of
an unopened telescope dome (i.e., $IMGTYP=$~``object'' and
$DOMESTAT=$~``closed''), in which case the associated {\it status}\/ column
is set to zero and bit $2^1=2$ is set in the {\it infobits}\/ column
of the  {\it Exposures}\/ database table.
The file is then copied from the {\it /staging}\/ partition to the
appropriate branch of the observation-date-based directory tree in the
camera-image-file archive.  
A record is inserted into the  {\it Exposures}\/ database table for the
ingested file, and, if necessary and usually at a lower frequency, 
new records are inserted into the following
database tables: {\it PIs}, {\it Projects}, {\it Nights}, 
{\it Filters}, {\it ImgTypes}, and {\it Fields}.
For example, Table~\ref{tab:imgtypes} lists the possible PTF-image types
that are ingested and registered in the {\it ImgTypes}\/ database table.
Finally, the ingested file is removed from the current working
directory, and the software moves on to ingest the next file.
The process terminates after all FITS files have been ingested.

\begin{table}
\caption{\label{tab:imgtypes} PTF-image types in the {\it ImgTypes}\/
  database table. }
\begin{small}
\begin{tabular}{ll}
\\[-5pt]
\hline\hline
{\it itid} & {\it IMGTYP}\\
\hline \\[-5pt]
1 & object \\
2 & dark \\
3 & bias \\
4 & dome \\
5 & twilight \\
6 & focus \\
7 & pointing \\
8 & test \\
\hline \\[-5pt]
\end{tabular}
\end{small}
\end{table}

The Perl script called {\it checkIngestedCameraImages.pl}\/
    recomputes the MD5 checksums of archived PTF camera-image files, and, for each
    file, compares the checksum with that stored in the 
    database and in the images manifest.  This script can be run any
    time there is a want or need to check data-file integrity for a
    given night.  The
    associated {\it Exposures}\/ database record is updated with $STATUS=0$ in
    the rare event of checksum mismatch, and the appropriate bit in the {\it
    infobits}\/ column is set (see Table~\ref{tab:ingestinfobits}).

The Perl script called {\it ingestImagesManifest.pl}\/ copies the
    images manifest to its appropriate archival-disk nightly subdirectory
    and registers its location and filename in the {\it Nights}\/ database table, along with relevant
    metadata, such as MD5 checksum, file size, status, and
    database-record-creation time stamp.

\begin{table*}
\caption{\label{tab:ingestinfobits}Bits allocated for flagging
  data-ingest conditions and exceptions in the {\it infobits}\/ column
  of the {\it Exposures}\/ database table. }
\begin{small}
\begin{tabular}{cl}
\\[-5pt]
\hline\hline
Bit & Definition \\
\hline \\[-5pt]
0 & File size too small \\
1 & $IMGTYP=$ ``object'' and $DOMESTAT=$ ``closed''\\
2 & $FILTER=0$, $FILTERID=0$ and/or $FILTERSL=0$ \\
4 & Telescope sidereal-tracking failure (manually set after image inspection) \\
6 & Checksum mismatch: database vs.\ images manifest \\
7 & Checksum mismatch: recomputed vs.\ images manifest \\
8 & File-size mismatch: recomputed vs.\ images manifest \\
9 & One or more non-crucial keywords missing \\
\hline \\[-5pt]
\end{tabular}
\end{small}
\end{table*}

\section{\label{sdqa}Science Data Quality Analysis}

SDQA is an integral part of the design
implemented for PTF, which is outlined by~\citet{lahersdqa} 
in the context of a different ground-based project under proposal.
It is necessary to provide some details about the IPAC-PTF SDQA subsystem
at this point, so that interactions between it and the
pipelines can be more fully understood.

\begin{figure*}
\includegraphics[scale=0.35]{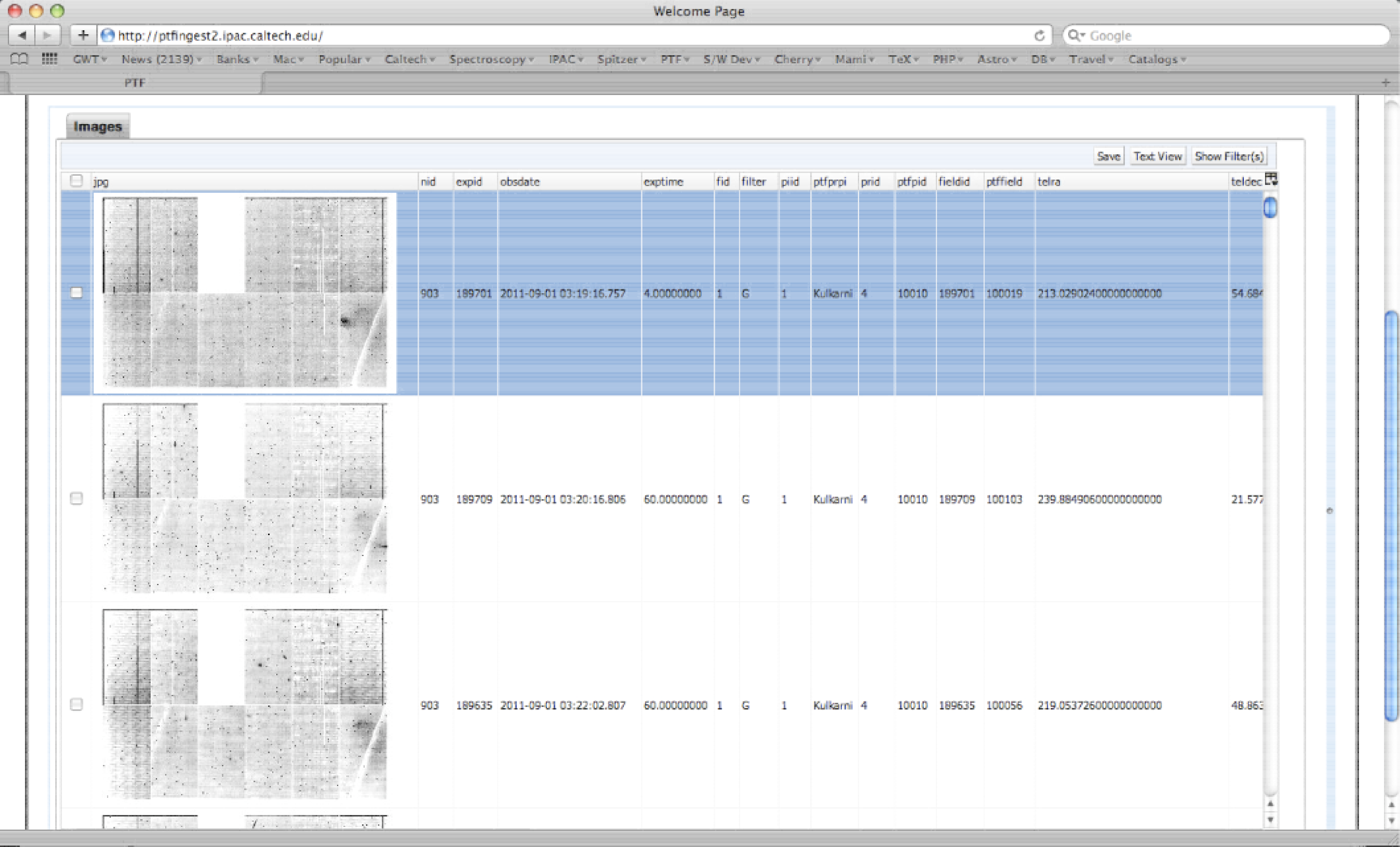}
\caption{\label{fig:sdqaguiimage} Sample screen shot of the SDQA GUI
  developed for the IPAC-PTF system.}
\end{figure*}

Typically within hours after a night's worth of camera images have
been ingested and the camera-image-splitting
pipelines have been executed (see~\S{\ref{cameraiimagesplitter}}), 
the camera images are inspected visually for problems. 
The preview images generated by the camera-image-splitting
pipelines play a pivotal part in speeding up this task.  An in-house 
web-based graphical user interface (GUI) has been designed and 
implemented to provide basic SDQA functionality (see Figure~\ref{fig:sdqaguiimage}),
such as displaying previews of 
raw and processed images, and 
dynamically generating time-series graphs
of SDQA quantities of interest.
The source code for the GUI and visual-display software tools have been developed in Java, primarily 
for its platform-independent and multi-threading capabilities. 
The software queries the database for its information.
The Google Web Toolkit\footnote{\url{http://www.gwtproject.org}} has been used to compile the Java code into Javascript 
for relatively trouble-free execution under popular web browsers.  
The GUI has drill-down capability to selectively 
obtain additional information.
The screen shot in Figure~\ref{fig:sdqaguiimage} shows the window that
displays previews of PTF
camera images and associated metadata.  The previews load quickly and
have sufficient detail to inspect the nightly observations for problems and 
assess the data quality (e.g., when investigating astrometric-calibration failures).
In the event of telescope sidereal-tracking problems, which are spotted visually in
the GUI (and occur infrequently), the associated {\it status}\/ column
is set to zero and bit $2^4=16$ is set in the {\it infobits}\/ column
of the  {\it Exposures}\/ database table (see Table~\ref{tab:ingestinfobits}).

A major function of our SDQA subsystem is to compute and store in
the database all the needed quantities for assessing data quality.  The
goal is to boil down questions about the data into relatively simple
or canned database queries that span the parameter space of the data
on different scales.  Having a suitable framework for this in place
makes it possible to issue a variety of manually requested and 
automatically generated reports.
During pipeline image processing, SDQA data are computed for the images and
astronomical sources extracted from the images, and utilized to grade the 
images and sources. 
The reports summarize the science data quality in various ways and provide feedback
to telescope, camera, facility, observation-scheduling and data-processing 
personnel.

\begin{figure}
\includegraphics[scale=.47]{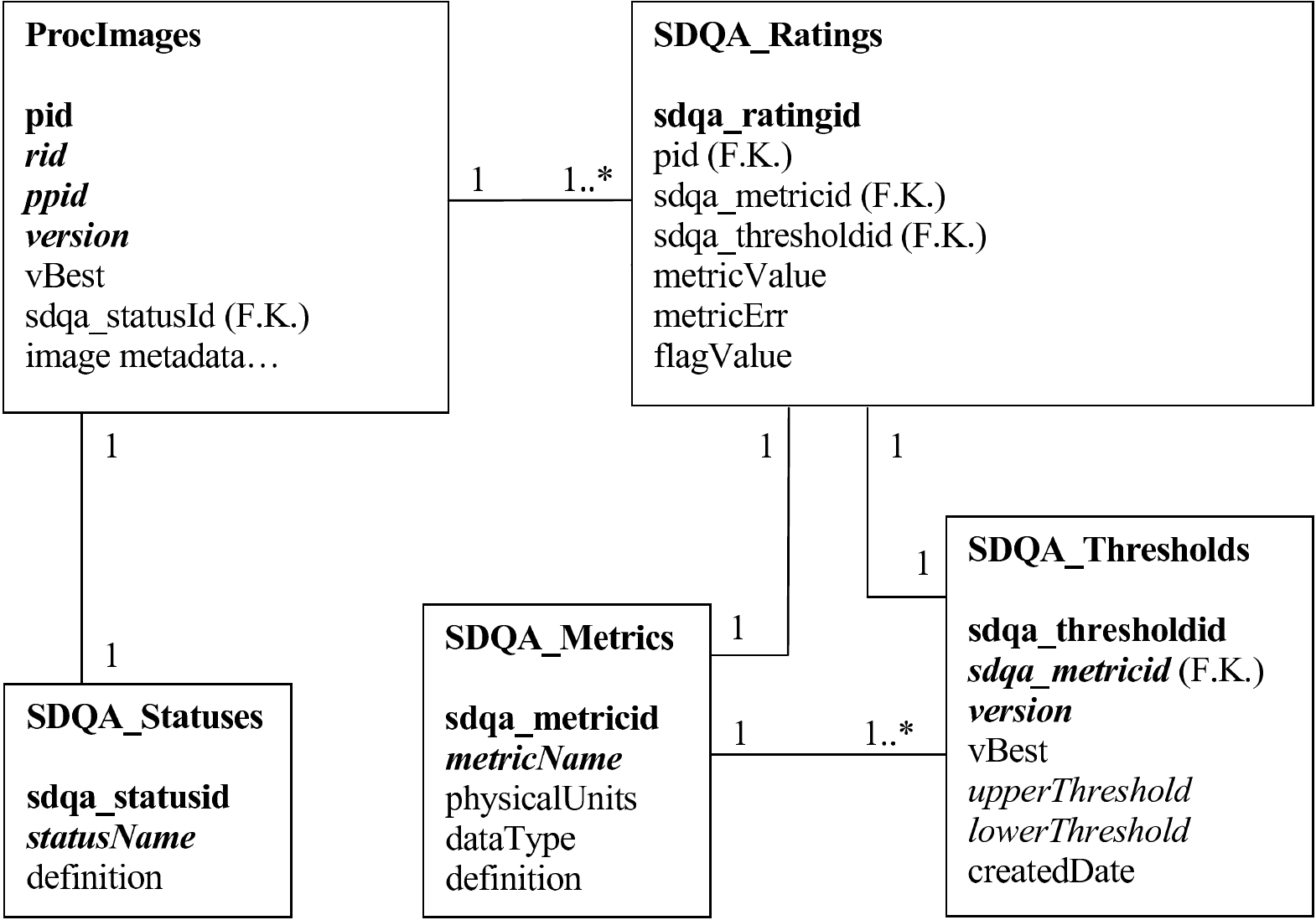}
\caption{\label{fig:sdqadbschema} IPAC-PTF SDQA database-schema
  design. The figure nomenclature is explained in the caption of Figure~\ref{fig:ingestdbschema}.}
\end{figure}

Figure~\ref{fig:sdqadbschema} shows our SDQA database-schema design
for processed images.  Note that the design is easily extended for other
pipeline products.  The {\it ProcImages}\/ database table is indexed by
{\it pid}\/ and stores metadata about processed images, including the
{\it sdqa\_statusid}, which is an integer that indexes the SDQA grade assigned to an image.
A processed image is associated with both a raw image ({\it rid}\/) and a pipeline
({\it ppid}\/).  As the pipeline software is upgraded, new versions of a
processed image for a given raw image and pipeline will be generated,
and, hence, a {\it version}\/ column is included in the table to keep
track of the versions.  The {\it vBest}\/ column flags which version is
best; there is only one best version and it is usually the latest version.

SDQA metrics are diverse, predefined measures that characterize image 
quality; e.g., image statistics, astrometric and photometric figures of 
merit and associated errors, counts of various things, like 
extracted sources, etc.
The {\it SDQA\_Metrics}\/ database table stores the SDQA metrics defined
for IPAC-PTF operations, and these are listed in Tables~\ref{tab:sdqametrics1} through~\ref{tab:sdqametrics2}.
The imageZeroPoint SDQA metric (metricId$=48$) is set to NaN (not a number)  in the
database if either 1) the image did not overlap an SDSS field; 2)
there were an insufficient number of SDSS sources; or 3) the filter
used for the exposure was neither $g$ nor $R$ band (only these two
PTF bands are photometrically calibrated at this time).

\begin{table*}
\caption{\label{tab:sdqametrics1}IPAC-PTF SDQA metrics stored in the 
{\it SDQA\_Metrics}\/ database table.   For the SDQA metrics associated
with sub-images, the size for sub-images (1, $j$) and (3, $j$) is $768
\times 1024$~pixels, and the size for sub-images (2, $j$) is $768
\times 2048$~pixels.}
\begin{scriptsize}
\begin{tabular}{llll p{8cm}}
\\[-5pt]
\hline\hline
{\it metricId} & {\it metricName}&{\it physicalUnits}& {\it definition} \\
\hline \\[-5pt]
            1 & nGoodPix            & Counts                         & Number of good pixels.\\
             2 & nDeadPix            & Counts                         & Number of dead pixels.\\
             3 & nHotPix             & Counts                         & Number of hot pixels.\\
             4 & nSpurPix            & Counts                        & Number of spurious pixels.\\
             5 & nSatPix             & Counts                         & Number of saturated pixels.\\
             6 & nObjPix             & Counts                        & Number of source-object-coverage pixels.\\
             7 & nNanPix             & Counts                       & Number of NaN (not a number) pixels.\\
             8 & nDirtPix            & Counts                        & Number of pixels with filter dirt.\\
             9 & nStarPix            & Counts                        & Number of star-coverage pixels.\\
            10 & nGalxPix            & Counts                      & Number of galaxy-coverage pixels. \\
            11 & nObjSex             & Counts                     & Number of source objects found by {\it SExtractor}. \\
            12 & fwhmSex             & Arcsec                   & {\it SExtractor}\/ FWHM of the radial profile. \\
            13 & gMean               & D.N.                           & Image global mean. \\
            14 & gMedian             & D.N.                          & Image global median. \\
            15 & cMedian1            & D.N.                        & Image upper-left corner median. \\
            16 & cMedian2            & D.N.                         & Image upper-right corner median. \\
            17 & cMedian3            & D.N.                          & Image lower-right corner median. \\
            18 & cMedian4            & D.N.                        & Image lower-left corner median. \\
            19 & gMode               & D.N.                         & Image global mode. \\
            20 & MmFlag              & Counts                      & Image global mode. \\
            21 & gStdDev             & D.N.                            & Image global standard deviation. \\
            22 & gMAbsDev            & D.N.                        & Image mean absolute deviation. \\
            23 & gSkewns             & D.N.                          & Image skewness. \\
            24 & gKurtos             & D.N.                        & Image kurtosis. \\
            25 & gMinVal             & D.N.                           & Image minimum value. \\
            26 & gMaxVal             & D.N.                             & Image maximum value. \\
            27 & pTile1              & D.N.                            & Image 1-percentile. \\
            28 & pTile16             & D.N.                        & Image 16-percentile. \\
            29 & pTile84             & D.N.                        & Image 84-percentile. \\
            30 & pTile99             & D.N.                          & Image 99-percentile. \\
            31 & photCalFlag         & Flag                        & Flag for whether image could be photometrically calibrated. \\
            32 & zeroPoint           & Mag                      & Magnitude zero point at an air mass of zero (see Appendix~A). \\
            33 & extinction          & Mag                        & Extinction. \\
            34 & airMass             & None                  & Air mass. \\
            35 & photCalChi2         & None                   & Chi2 of photometric calibration. \\
            36 & photCalNDegFreedom  & Counts                     & Number of SDSS matches in photometric calibration. \\
            37 & photCalRMSE         & Mag                    & R.M.S.E. of photometric calibration. \\
            38 & aveDeltaMag         & Mag                     & Average delta magnitude over SDSS sources in a given image. \\
            40 & nPhotSources        & Counts                     & Number of sources used in photometry calibration. \\
            41 & astrrms1            & Degrees                       & SCAMP astrometry RMS along axis 1 (ref., high S/N). \\
            42 & astrrms2            & Degrees                        & SCAMP astrometry RMS along axis 2 (ref., high S/N). \\
            43 & 2mass\_astrrms1      & Arcsec                   & 2Mass astrometry RMS along axis 1. \\
            44 & 2mass\_astrrms2      & Arcsec                   & 2Mass astrometry RMS along axis 2. \\
            45 & 2mass\_astravg1      & Arcsec                   & 2Mass astrometry match-distance average along axis 1. \\
            46 & 2mass\_astravg2      & Arcsec                  & 2Mass astrometry match-distance average along axis 2. \\
            47 & n2massMatches       & Counts                    & Number of 2Mass sources matched. \\
            48 & imageZeroPoint      & Mag                      &    Magnitude zero point of image determined directly from SDSS sources (see Appendix~A). \\
            49 & imageColorTerm      & Mag                   & Color term from data-fit to SDSS sources in a given image (see Appendix~A). \\
            50 & 2mass\_astrrms1\_11   & Arcsec                      & 2Mass astrometry RMS along axis 1 for sub-image (1, 1). \\
            51 & 2mass\_astrrms2\_11   & Arcsec                     & 2Mass astrometry RMS along axis 2 for sub-image (1, 1). \\
            52 & 2mass\_astravg1\_11   & Arcsec            & 2Mass astrometry match-distance average along axis 1 for sub-image (1, 1). \\
            53 & 2mass\_astravg2\_11   & Arcsec                   & 2Mass astrometry match-distance average along axis 2 for sub-image (1, 1). \\
           54 & n2massMatches\_11    & Counts                          & Number of 2Mass sources matched for sub-image (1, 1). \\
            55 & 2mass\_astrrms1\_12   & Arcsec                & 2Mass astrometry RMS along axis 1 for sub-image (1, 2). \\
            56 & 2mass\_astrrms2\_12   & Arcsec                     & 2Mass astrometry RMS along axis 2 for sub-image (1, 2). \\
            57 & 2mass\_astravg1\_12   & Arcsec                     & 2Mass astrometry match-distance average along axis 1 for sub-image (1, 2). \\
            58 & 2mass\_astravg2\_12   & Arcsec                      & 2Mass astrometry match-distance average along axis 2 for sub-image (1, 2). \\
            59 & n2massMatches\_12    & Counts                       & Number of 2Mass sources matched for sub-image (1, 2). \\
           60 & 2mass\_astrrms1\_13   & Arcsec                      & 2Mass astrometry RMS along axis 1 for sub-image (1, 1). \\
\hline \\[-5pt]
\end{tabular}
\end{scriptsize}
\end{table*}

\begin{table*}
\caption{\label{tab:sdqametrics2} (Continued from Table~\ref{tab:sdqametrics1}.) IPAC-PTF SDQA metrics stored in the 
{\it SDQA\_Metrics}\/ database table.  For the SDQA metrics associated
with sub-images, the size for sub-images (1, $j$) and (3, $j$) is $768
\times 1024$~pixels, and the size for sub-images (2, $j$) is $768
\times 2048$~pixels.}
\begin{scriptsize}
\begin{tabular}{llll p{8cm}}
\\[-5pt]
\hline\hline
{\it metricId} & {\it metricName}&{\it physicalUnits}& {\it definition} \\
\hline \\[-5pt]
            61 & 2mass\_astrrms2\_13   & Arcsec                   & 2Mass astrometry RMS along axis 2 for sub-image (1, 1). \\
            62 & 2mass\_astravg1\_13   & Arcsec                      & 2Mass astrometry match-distance average along axis 1 for sub-image (1, 1). \\
            63 & 2mass\_astravg2\_13   & Arcsec                    & 2Mass astrometry match-distance average along axis 2 for sub-image (1, 1). \\
            64 & n2massMatches\_13    & Counts                    & Number of 2Mass sources matched for sub-image (1, 1). \\
            65 & 2mass\_astrrms1\_21   & Arcsec                   & 2Mass astrometry RMS along axis 1 for sub-image (2, 1). \\
            66 & 2mass\_astrrms2\_21   & Arcsec                     & 2Mass astrometry RMS along axis 2 for sub-image (2, 1). \\
            67 & 2mass\_astravg1\_21   & Arcsec                     & 2Mass astrometry match-distance average along axis 1 for sub-image (2, 1). \\
            68 & 2mass\_astravg2\_21   & Arcsec                     & 2Mass astrometry match-distance average along axis 2 for sub-image (2, 1). \\
            69 & n2massMatches\_21    & Counts                      & Number of 2Mass sources matched for sub-image (2, 1). \\
            70 & 2mass\_astrrms1\_22   & Arcsec                   & 2Mass astrometry RMS along axis 1 for sub-image (2, 2).\\
            71 & 2mass\_astrrms2\_22   & Arcsec                & 2Mass astrometry RMS along axis 2 for sub-image (2, 2).\\
            72 & 2mass\_astravg1\_22   & Arcsec                     & 2Mass astrometry match-distance average along axis 1 for sub-image (2, 2).\\
            73 & 2mass\_astravg2\_22   & Arcsec                     & 2Mass astrometry match-distance average along axis 2 for sub-image (2, 2).\\
            74 & n2massMatches\_22    & Counts                         & Number of 2Mass sources matched for sub-image (2, 2).\\
            75 & 2mass\_astrrms1\_23   & Arcsec                    & 2Mass astrometry RMS along axis 1 for sub-image (2, 3).\\
            76 & 2mass\_astrrms2\_23   & Arcsec                     & 2Mass astrometry RMS along axis 2 for sub-image (2, 3).\\
            77 & 2mass\_astravg1\_23   & Arcsec                    & 2Mass astrometry match-distance average along axis 1 for sub-image (2, 3).\\
            78 & 2mass\_astravg2\_23   & Arcsec                     & 2Mass astrometry match-distance average along axis 2 for sub-image (2, 3).\\
            79 & n2massMatches\_23    & Counts                         & Number of 2Mass sources matched for sub-image (2, 3).\\
            80 & 2mass\_astrrms1\_31   & Arcsec                     & 2Mass astrometry RMS along axis 1 for sub-image (3, 1).\\
            81 & 2mass\_astrrms2\_31   & Arcsec                     & 2Mass astrometry RMS along axis 2 for sub-image (3, 1).\\
            82 & 2mass\_astravg1\_31   & Arcsec                      & 2Mass astrometry match-distance average along axis 1 for sub-image (3, 1).\\
            83 & 2mass\_astravg2\_31   & Arcsec                      & 2Mass astrometry match-distance average along axis 2 for sub-image (3, 1).\\
            84 & n2massMatches\_31    & Counts                        & Number of 2Mass sources matched for sub-image (3, 1).\\
            85 & 2mass\_astrrms1\_32   & Arcsec                   & 2Mass astrometry RMS along axis 1 for sub-image (3, 2).\\
            86 & 2mass\_astrrms2\_32   & Arcsec                   & 2Mass astrometry RMS along axis 2 for sub-image (3, 2).\\
            87 & 2mass\_astravg1\_32   & Arcsec                       & 2Mass astrometry match-distance average along axis 1 for sub-image (3, 2).\\
            88 & 2mass\_astravg2\_32   & Arcsec                     & 2Mass astrometry match-distance average along axis 2 for sub-image (3, 2).\\
            89 & n2massMatches\_32    & Counts                         & Number of 2Mass sources matched for sub-image (3, 2).\\
           90 & 2mass\_astrrms1\_33   & Arcsec                     & 2Mass astrometry RMS along axis 1 for sub-image (3, 3).\\
            91 & 2mass\_astrrms2\_33   & Arcsec                  & 2Mass astrometry RMS along axis 2 for sub-image (3, 3).\\
            92 & 2mass\_astravg1\_33   & Arcsec                   & 2Mass astrometry match-distance average along axis 1 for sub-image (3, 3).\\
            93 & 2mass\_astravg2\_33   & Arcsec                      & 2Mass astrometry match-distance average along axis 2 for sub-image (3, 3).\\
            94 & n2massMatches\_33    & Counts                           & Number of 2Mass sources matched for sub-image (3, 3).\\
            95 & medianSkyMag        & Mag/(sec-arcsec$^2$) & Median sky magnitude.\\
            96 & limitMag            & Mag/(sec-arcsec$^2$) & Limiting magnitude (obsolete method).\\
            97 & medianFwhm          & Arcsec                      & Median FWHM.\\
            98 & medianElongation    & None                    & Median elongation.\\
            99 & stdDevElongation    & None                    & Standard deviation of elongation.\\
          100 & medianTheta         & Degrees                        &  Special median of THETAWIN\_WORLD.\\
           101 & stdDevTheta         & Degrees                & Special standard deviation of THETAWIN\_WORLD.\\
           102 & medianDeltaMag      & Mag/(sec-arcsec$^2$)   & Median (MU\_MAX -MAG\_AUTO).\\
           103 & stdDevDeltaMag      & Mag/(sec-arcsec$^2$)    & Std.\ dev of (MU\_MAX - MAG\_AUTO).\\
           104 & scampCatType        & None                  & {\it  SCAMP}-catalog type: 1=SDSS-DR7, 2=UCAC3, 3=USNO-B1\\
           105 & nScampLoadedStars   & None                   & Number of stars loaded from {\it  SCAMP}\/ input catalog.\\
           106 & nScampDetectedStars & None                & Number of stars detected by {\it  SCAMP}.\\
           107 & imageZeroPointSigma & Mag                     & Sigma of magnitude difference between {\it SExtractor}\/ and SDSS sources.\\
           108 & limitMagAbsPhotCal     & Mag/(sec-arcsec$^2$)  & Limiting magnitude (abs.\ phot.\ cal.\ zero point).\\
           109 & medianSkyMagAbsPhotCal & Mag/(sec-arcsec$^2$)    & Median sky magnitude based on abs.\ phot.\ cal.\ zero point.\\
           110 & flatJarqueBera      & Dimensionless  & Jarque-Bera test for abnormal data distribution of superflat image.\\
           111 & flatMean            & Dimensionless & Mean of superflat image.\\
           112 & flatMedian          & Dimensionless   & Median of superflat image.\\
           113 & flatStdDev          & Dimensionless    & Standard deviation of superflat image.\\
           114 & flatSkew            & Dimensionless                     & Skew of superflat image.\\
           115 & flatKurtosis        & Dimensionless                     & Kurtosis of superflat image.\\
           116 & flatPercentile84.1  & Dimensionless                     & 84.1 percentile of superflat image.\\
           117 & flatPercentile15.9  & Dimensionless                     & 15.9 percentile of superflat image.\\
           118 & flatScale           & Dimensionless                     & Scale (one half the difference between 84.1 and P15.9 percentiles) of superflat image.\\
           119 & flatNumNanPix       & Counts & Number of NaNed pixels in superflat image.\\
\hline \\[-5pt]
\end{tabular}
\end{scriptsize}
\end{table*}

SDQA thresholds can be defined for values associated with SDQA metrics.
The {\it SDQA\_Thresholds}\/ database table stores the SDQA thresholds defined
for IPAC-PTF operations, and can include lower and/or upper thresholds.  Since
thresholds can change over time as the SDQA subsystem is tuned,
the table has {\it version}\/ and {\it vBest}\/ columns to keep track of
the different and best versions (like the {\it ProcImages}\/ database table).

The {\it SDQA\_Ratings}\/ database table is associated with 
the {\it ProcImages}\/ database table in a one-to-many
relationship record-wise, and, for a given processed image, 
stores multiple records of what we refer to as image
``SDQA ratings'', which are the values associated with SDQA metrics (referred to above).
An SDQA rating is basically the computed value of an SDQA metric and 
its uncertainty.  This design encourages the storing of an uncertainty
with its computed SDQA-rating value, although this is not required.
The {\it flagValue}\/
column in a given record is normally set to zero, but is reset to one when
the associated {\it metricValue}\/ falls outside of the region allowed 
by the corresponding threshold(s).
A processed image, in general, has many different SDQA ratings as
noted above, which
are computed at various pipeline stages; PTF processed images each
have over 100 different SDQA ratings
(see Tables~\ref{tab:sdqametrics1} through~\ref{tab:sdqametrics2}).  
An {\it SDQA\_Ratings}\/ record contains indexes to the relevant
processed image, SDQA metric, and SDQA threshold, which are foreign
keys.
The {\it SDQA\_Ratings}\/ database table potentially will have a large number
of records; bulk loading of these records may reduce the impact of the SDQA subsystem on pipeline throughput,
although this has not been necessary for IPAC-PTF pipelines.  

A separate database stored
function called {\it setSdqaStatus(pid)}\/ is called to compute the SDQA
grade of a processed image after its SDQA ratings have been loaded
into the database.  The function computes the percentage of SDQA
ratings that are flagged ($flagValue = 1$ in the {\it SDQA\_Ratings}\/ database table).  The 
possible pipeline-assigned SDQA status values are listed in Table~\ref{tab:sdqastatuses}.

\begin{table*}
\caption{\label{tab:sdqastatuses}Possible SDQA status values. }
\begin{scriptsize}
\begin{tabular}{lll p{7cm}}
\\[-5pt]
\hline\hline
{\it sdqa\_statusid} & {\it statusName} & SDQA ratings flagged (\%)& {\it definition} \\
\hline \\[-5pt]
             1 & passedAuto             & $<5$ & Image passed by automated SDQA. \\
             2 & marginallyPassedAuto   & $\ge 5$ and $<25$ & Image marginally passed by automated SDQA. \\
             3 & marginallyFailedAuto   & $> 75$& Image marginally failed by automated SDQA. \\
             4 & failedAuto             & $\ge 90$ & Image failed by automated SDQA. \\
             5 & indeterminateAuto      & $\ge 25$ and $\le 75$ & Image is indeterminate by automated SDQA. \\
             6 & passedManual           & N/A & Image passed by manual SDQA. \\
             7 & marginallyPassedManual & N/A & Image marginally passed by manual SDQA. \\
             8 & marginallyFailedManual & N/A & Image marginally failed by manual SDQA. \\
             9 & failedManual           & N/A & Image failed by manual SDQA. \\
            10 & indeterminateManual    & N/A & Image is indeterminate by manual SDQA. \\
\hline \\[-5pt]
\end{tabular}
\end{scriptsize}
\end{table*}

\section{\label{idp}Image-Processing Pipelines}

\subsection{\label{pipelineoverview}Overview}

The pipelines consist of Perl scripts and the modules or binary
executables that they run.   The modules are either custom 
developed in-house or freely downloadable astronomical-software 
packages (e.g., {\it  SExtractor}\/).  There are product-generation and
calibration pipelines (see Table~\ref{tab:pipelinedbtable}), which
must be executed in a particular order.

\begin{table*}
\caption{\label{tab:pipelinedbtable}Contents of the {\it Pipelines}\/ database table.}
\begin{small}
\begin{tabular}{llcll}
\\[-5pt]
\hline\hline
{\it ppid}\tablenotemark{1} & Priority \tablenotemark{2} & Blocking & Perl script & Description\\
\hline \\[-5pt]
    1 &       10 & 1 & {\it superbias.pl} & Superbias calibration\\
    2 &       20 & 1 & {\it domeflat.pl} & Domeflat calibration\\
    3 &       30 & 1 & {\it preproc.pl} & Raw-image preprocessing\\
    4 &       40 & 1 & {\it superflat.pl} & Superflat calibration\\
    5 &       50 & 1 & {\it frameproc.pl} & Frame processing\\
    6 &       70 & 1 & TBD & Mosaicking\\
    7 &      500 & 1 & {\it splitCameraImages.pl} & Camera-image splitting\\
    8 &       60 & 1 & {\it sourceAssociation.pl} & Source association\\
    9 &       55 & 0 & {\it loadSources.pl} & Load sources into database\\
   10 &      45 & 1 & {\it flattener.pl} & Flattener\\
  11 &       41 & 1 & {\it twilightflat.pl} & Twilight flat\\
   12 &       80 & 1 & {\it genRefImage.pl} & Reference image\\
   13 &       52 & 1 & {\it genCatalog.pl} & Source-catalog generation\\
\hline \\[-5pt]
\end{tabular}
\tablenotetext{1}{Pipeline database index. }
\tablenotetext{2}{The priority numbers are relative, and smaller numbers have higher priority. }
\end{small}
\end{table*}

In normal operations, the pipelines are initiated via multi-threaded
job client software developed expressly for PTF at IPAC.\@  One job client
is typically run on one pipeline machine at any given time.
The job clients interact with the database to coordinate the pipeline jobs.
The database maintains a queue of jobs waiting to be processed.  Each
job is associated with a particular pipeline and data set.
Job clients that are not busy periodically poll the database for more
jobs, which responds with
the database IDs of jobs to process, along with concise information
about the jobs that is needed by the pipelines.  The job client then
launches the called-for pipeline as a separate processing thread and is typically
blocked until the thread completes.  The database is updated with 
relevant job information after the job finishes (e.g., pipeline start and end
times).

The pipelines nominally query the database for any additional metadata that are
required to run the pipeline.  The last step of the pipeline
includes updating the database with metadata about the
processed-image product(s) and their ancillary files (e.g., data masks).
The pipelines make and sever database connections as needed, and 
database communications to the pipeline and to the job executive are independent.  

The pipelines create numerous intermediate data files on the pipeline 
machine's local disk, which are handy to have for manually rerunning
pipeline steps, should the need arise.  A fraction of these files are copied to a
sandbox disk (see \S{\ref{os}}), which serves to marshal together
the products for a given night generated in parallel on different
pipeline machines.  It is expedient to organize the products in the
sandbox in subdirectories that make them easy to find without having
to query the database.  The following sample file path exemplifies the
subdirectory scheme that we have adopted:

\begin{verbatim}
/sbx1/2011/09/19/f2/c9/p5/v1
\end{verbatim}

\noindent After the sandbox logical name and the year, month, and day, there is
``f2/c9/p5/v1'', which stands for filter ($fid=2$), CCD ($ccdid=9$),
pipeline ($ppid=5$), and product version ($version = 1$).  The directory tree
for the archive is exactly the same, except that the archive logical
name replaces the sandbox's.  The method employed for copying products
from the sandbox to the archive is described in \S{\ref{productarchiver}}.

\subsection{\label{compenv}Computing Environment}

The pipelines inherit the shell environment they run under,
which is overridden by settings particular to the PTF software system
(see Table~\ref{tab:environment}).  A modest number of 
environment variables is required.  The {\it PATH}\/ environment
variable must include locations of PTF scripts and binary executables, Perl, Python, Matlab,
Astrometry.net, and Jessica Mink's {\it WCS Tools}.
The {\it PTF\_IDL}\/ environment variable gives the path and command
name of IPAC's SciApps installation of IDL.\@
Table~\ref{tab:thirdpartysoftwaretable} lists the versions of
third-party software utilized in IPAC-PTF pipelines.

\begin{table*}
\caption{\label{tab:environment}Environment variables required by the PTF software system. }
\begin{footnotesize}
\begin{tabular}{ll}
\\[-5pt]
\hline\hline
Variable & Definition\\
\hline \\[-5pt]
{\it PTF\_ROOT} & Root directory of PTF software system. \\
{\it PTF\_LOGS} & Directory of log files (e.g., {\it \${PTF\_ROOT}/logs}\/). \\
{\it PTF\_ARCHIVE} & Archive directory (e.g., {\it \${PTF\_ROOT}/archive}\/). \\
{\it PTF\_ARCHIVE\_RAW\_PARTITION} & Archive raw-data disk partition (e.g., {\it raw}\/). \\
{\it PTF\_ARCHIVE\_PROC\_PARTITION} & Archive processed-data disk partition (e.g., {\it proc}\/). \\
{\it PTF\_SBX} & Current sandbox directory (e.g., {\it \${PTF\_ROOT}/sbx1}\/). \\
{\it PTF\_SW} & Top-level software directory (e.g., {\it \${PTF\_ROOT}/sw}\/). \\
{\it PTF\_BIN} & Binary-executables directory (e.g., {\it \${PTF\_SW}/ptf/bin}\/). \\
{\it PTF\_LIB} & Libraries directory (e.g., {\it \${PTF\_SW}/ptf/lib}\/). \\
{\it PTF\_EXT} & External-software directory (e.g., {\it \${PTF\_ROOT}/ext}\/). \\
{\it PTF\_LOCAL} & Machine local directory (e.g., {\it /scr/ptf}\/). \\
{\it PTF\_CDF} & Configuration-data-file directory (e.g., {\it /scr/cdf}\/). \\
{\it PTF\_CAL} & Calibration-file directory (e.g., {\it /scr/cal}\/). \\
{\it PTF\_IDL} & Full path and filename of IDL program. \\
{\it PTF\_ASTRONOMYNETBIN} & Astrometry.net binary-executable directory. \\
{\it WRAPPER\_UTILS} & Perl-library directory (e.g. \${PTF\_SW}/perlibs). \\
{\it WRAPPER\_VERBOSE} & Pipeline verbosity flag (0 or 1). \\
{\it DBTYPE} & Database type. \\
{\it DNAME} & Database name. \\
{\it DBSERVER} & Database-server name. \\
{\it SODB\_ROLE} & Database role. \\
{\it TY2\_PATH} & Location of the Tycho-2 catalog. \\
{\it PATH} & Location(s) of binary executables (e.g., \$PTF\_BIN). \\
{\it LD\_LIBRARY\_PATH} & Location(s) of libraries (e.g., \$PTF\_LIB). \\
{\it PERL\_PATH} & Location of Perl-interpreter command. \\
{\it PERL5LIB} & Location(s) of Perl-library modules. \\
{\it PYTHONPATH} & Location of Python-interpreter command. \\
\hline \\[-5pt]
\end{tabular}
\end{footnotesize}
\end{table*}

\begin{table}
\caption{\label{tab:thirdpartysoftwaretable}Versions of third-party
  software executed in IPAC-PTF pipelines. }
\begin{small}
\begin{tabular}{ll}
\\[-5pt]
\hline\hline
Software & Version\\
\hline \\[-5pt]
Astrometry.net & 0.43\\
CFITSIO & 3.35\\
{\it Eye} & 1.3.0\\
{\it fftw} & 3.2.2\\
IDL & 8.1\\
Matlab & 7.10.0.499\\
Montage & 3.2\\
Perl & 5.10.0\\
Python & 2.7.3\\
EPD & 7.3-2\\
{\it SCAMP}      &   1.7.0\\
{\it MissFITS} & 2.4.0\\
{\it SExtractor} &   2.8.6\\
{\it Swarp} & 2.19.1\\
{\it WCS Tools} & 3.8.7\\
{\it DAOPHOT} & 2004 Jan 15\\
{\it ALLSTAR} & 2001 Feb 7\\
SciApps & 08/29/2011\\
\hline \\[-5pt]
\end{tabular}
\end{small}
\end{table}

\subsection{Configuration Data Files}

Configuration data files (CDFs) are text files that store
configuration data in the form of keyword=value pairs.
They are parameter files that control software behavior.
On the order of a hundred of these files are required for 
PTF processing.  In many cases, there are sets of 11 files 
for a given process working on individual CCDs,
thus allowing CCD-dependent image processing.
The CDFs for the superbias-calibration pipeline (see \S{\ref{sbcp}}),
for example, store the outlier-rejection threshold and the 
pixel coordinates of the floating-bias strip.
Among the files are {\it SExtractor}\/ ``config'' and ``param'' files.
The CDFs are version-controlled in CVS and the version numbers of the
CDFs as a complete set of files are tracked in the {\it CdfVersions}\/
database table, along with deployment dates and times, etc.  
For fast access, the CDFs are stored locally on each
pipeline machine's scratch disk (as defined by environment variable {\it
  PTF\_CDF}\/; see \S{\ref{compenv}}\/).

\subsection{\label{pmi}Pixel-Mask Images}

Pixel masks are used to flag any badly behaved pixels on the
CCDs.  The flagged pixels can be specially treated by the 
image-processing pipelines as appropriate.
The pixel masks for PTF data were constructed as 
described by~\citet{vaneyken}.
The algorithm is loosely based on the
IRAF\footnote{\url{http://iraf.noao.edu/}} {\it ccdmask}\/ procedure
\citep{IRAF1,IRAF2}. The masks were created from images made by
dividing a 70\,s LED\footnote{Light-emitting diode; see~\citet{ptf}}
flatfield by a 35\,s LED flatfield. 
Three independent such divided frames were obtained for each of the
11 functioning CCDs. 
Any pixels with outlier fluxes beyond 4~standard deviations in at least 2 of
the 3 frames, or beyond 3~standard deviations in all 3 of the frames
were flagged as bad. This approach helps catch excessively variable
pixels, in addition to highly non-linear pixels, while still rejecting
cosmic-ray hits. The bad-pixel-detection procedure was then repeated after boxcar
smoothing of the original image along the readout direction. 
This finds column segments where individual pixels are not statistically
bad when considered alone, but are statistically bad when taken
together as an aggregate. This process was iterated several times,
with a selection of smoothing bin sizes from 2 to 20~pixels. Pixels
lying in small gaps between bad pixels were then also iteratively
flagged, with the aim of completely blocking out large regions of bad
pixels while minimizing encroachment into good-pixel regions.

\subsection{\label{pipeexec}Pipeline Executive}

The pipeline executive is software that runs in parallel on the 
pipeline machines as pipeline job clients.  There is no
pipeline-executive server {\it per se}, as its function has been
replaced by a relational database.
The pipeline executive expects pipeline jobs to be inserted as
records in the {\it Jobs}\/ database table, which is an integral 
part of the operations database schema (see \S{\ref{db}}\/).  
Thus, staging pipeline jobs for execution is 
as simple as inserting database records 
and assuring that the records are in the required state
for acceptance by the pipeline executive.
The {\it Jobs}\/ database table is queried for a job when a pipeline 
machine is not currently running a job and its job
client is seeking a new job.  The job
farmed out to a machine will be next in the priority ordering, which
is specified in the {\it Pipelines}\/ database table.  The
current contents of this table are listed in Table~\ref{tab:pipelinedbtable}.
The pipeline-priority numbers are relative and can be renumbered as new
pipelines are added or priority changes arise.

A {\it Jobs}\/ database record is
prepared for pipeline running by nulling out the run-time columns and
setting the status to 0.  Staged jobs that have not yet been executed
can be suspended by setting their status to -1, and then reactivated
later by setting their status back to zero.

The  job-client software is written in Perl ({\it ptfJobber.pl}\/) and
has an internal table that associates each of the 11 PTF CCDs with a
different pipeline machine.  It allows a pipeline machine to either run 
only jobs for the associated CCD or jobs that are CCD-independent
(e.g., the camera-image-splitting pipeline described in
\S{\ref{cameraiimagesplitter}}).  It runs in an open loop, and wakes
up every 5 seconds to check whether a job has completed and/or 
a new job can be started.

Each client maintains a list of launched pipelines that grows
indefinitely (until stopped and restarted, which, for example, 
is done for the weekly database backup).  Each launched pipeline
executes as a separate processing thread.
The attributes of the
launched pipelines include their job database IDs ({\it jid}\/), whether
the job has completed, and whether the job is nonblocking ($blocking =
0$; see Table~\ref{tab:pipelinedbtable}).
If the job currently being run by the client has a pipeline blocking
flag of 1, then the client will
wait for the job to finish before requesting another job.  If, on the
other hand, the job is nonblocking, then the client will request
another job and run it in parallel to the first job as another
processing thread.  The client is currently limited to running only 
one nonblocking job in parallel to a blocking job, but this can be
increased by simply changing a parameter.

\subsection{\label{vpo}Virtual Pipeline Operator}

Running pipelines and archiving the products, delivering product metadata
to IRSA, and other routine daily operations are automated with a Perl
script that we call the virtual pipeline operator (VPO).  In addition,
the script monitors disk usage, sends e-mail notifications and nightly
summaries, and runs
a nightly process that generates all-sky depth-of-coverage images (Aitoff
projections in galactic and equatorial coordinates).

The VPO can be run in open-loop mode for continuous operation.  The
polling-time interval is currently set at 10 minutes.  The software can
also be run in single-night mode for targeted reprocessing.  It does
much of its work by querying the database for information, and, in
particular, the {\it Jobs}\/ database table for pipeline monitoring.  It
is basically a finite state machine that sets internal flags to keep track of
what has been done and what needs to be done still for a given night's
worth of data.  The flags are also written to a state file, which is
unique for a given night, each time the state is updated.
The software is easily extensible by a Perl programmer when additional
states and/or tasks are needed.  It resets to default initial-state values every
24 hours; currently this is set to occur at 10 a.m., which is around
the time the data-ingestion process completes for the previous night
and its pipeline processing can be started.  

The VPO can also read the
initial state from a hand-edited input file (preferably by an 
expert pipeline operator).  This is advantageous
when an error occurs and the VPO must be restarted at some
intermediate point.  There are combinations of states that are not
allowed, and the software could be made more robust by adding checks
for invalid states.

\subsection{\label{archivalfilenames}Archival Filenames}

Pipeline-product files are created with fixed, descriptive filenames
(e.g., ``superflat.fits''), and then
renamed to have unique filenames near the end of the pipeline.  The
unique filenames are of constant length and have 11 identifying 
fields arranged in a standardized form.
Table~\ref{tab:stdfname} defines the
11 fields, and gives an example filename.
The filename fields are delimited by an underscore character, 
and are all lower case, except for the first field.  
If necessary, a filename field is padded with leading zeros to keep
the filename length constant.   
The filename contains enough information to identify the file precisely.

\begin{table*}
\caption{\label{tab:stdfname}Standardized file-naming scheme for PTF
  products.  }
\begin{small}
\begin{tabular}{l p{14cm}}
\\[-5pt]
\hline\hline
Filename field \#\tablenotemark{1} & Definition\\
\hline \\[-5pt]
1 & Always ``PTF'' (upper case)\\   
2 & Concatenation of year (4~digits), month (2~digits), day (2~digits), and fractional day (4~digits)\\
3 & One-character product format\tablenotemark{2}\\
4 & One-character product category\tablenotemark{3}\\
5 & Four-character product type\tablenotemark{4}\\
6 & Prefix ``t'' for time followed by hours (2~digits), minutes (2~digits), and seconds (2~digits)\\
7 & Prefix ``u'' for unique index followed by relevant database-table primary key\\
8 & Prefix ``f'' for filter followed by 2-digit filter number ({\it FILTERID}\/)\\
9 & Prefix ``p'' for PTF field and is followed by PTF field number ({\it PTFFIELD}\/)\\
10 & Prefix ``c'' for CCD followed by two-digit CCD index ({\it CCDID}\/)\\
11 & Filename extension (e.g., ``fits'' or ``ctlg'')\\
\hline \\[-5pt]
 \end{tabular}
\tablenotetext{1}{Sample filename: PTF\_200903011372\_i\_p\_scie\_t031734\_u008648839\_f02\_p000642\_c10.fits}
\tablenotetext{2}{Choice of ``i'' for image or ``c'' for catalog}
\tablenotetext{3}{Choice of ``p'' for processed, ``s'' for super, or ``e'' for external}
\tablenotetext{4}{Choice of ``scie'' for science, ``mask'' for
mask, ``bias'' for superbias, ``banc'' for
superbias-ancillary file, ``flat'' for superflat, ``twfl'' for twilight
flat, ``fmsk'' for flat mask, ``weig'' for weight, ``zpvm'' for zero-point variability map,
``zpve'' for zero-point-variability-map error, ``sdss'' for SDSS, ``uca3'' for UCAC3, ``2mas'' for 2MASS (Two-Micron
All-Sky Survey), or ``usb1'' for USNO-B1}
\end{small}
\end{table*}

The structure of the archive directory tree, in which the archived
products are stored on disk, has already been described in \S{\ref{pipelineoverview}}.

\subsection{Pipeline Multi-Threading}

Parallel image-processing on each of our pipeline machines is possible, 
given the machine architecture (see \S{\ref{os}}), and this is enabled
in our pipelines by the Perl {\it threads}\/ module. 
Some modules executed by our pipelines, such as {\it SCAMP}~\citep{2006ASPC..351..112B} and 
{\it SExtractor}~\citep{bertin}, are also multi-threaded codes, 
and the maximum number of threads 
they run simultaneously must be limited when running 
multiple threads at the Perl-script level.

Our pipelines currently run only a single instance of the
astrometry-refinement code, {\it SCAMP}, at a
time, and in a configuration that will cause it to automatically use
as many threads as there are cores in the machine (which is 8).
The pipelines run multi-threaded {\it SExtractor}\/ built to allow up to
2 threads, and let the Perl wrapper code control the multi-threading
at a higher level.

The multi-threading in the Perl pipeline scripts is nominally configured to allow
up to 7 threads at a time, which we found is optimal for non-threaded
parallel processes through benchmark
testing on our pipeline machines.  Wherever in our pipelines 
running a module in multi-threaded mode
is determined to be advantageous, a master thread is launched to
oversee the multi-threaded processing for the
module, and then are launched multiple slave threads running separate instances 
of the module on different images or input files in parallel.  For
thread synchronization, a thread-join
function is called to wait for all threads to complete before moving
on to the next step in the pipeline.  The exit code from each thread
is checked for abnormal termination.

\subsection{Standalone Pipeline Execution}

PTF pipelines can be easily executed outside of the pipeline
executive.  Since the pipelines query a database for inputs,
the particular database used must be updated with pointers to the
input files on disk.
Once the raw data for a given night have been ingested, the database
is updated automatically as the pipelines are run in proper
priority order (see Table~\ref{tab:pipelinedbtable}).

The simplicity of the basic instructions for standalone pipeline
execution are illustrated in the following example, in which the
superbias pipeline is executed:

\begin{verbatim}
cd /scr/work/dir
source $PTF_SW/ptf/ops/ops.env
setenv PTF_SBX /user/sbx1
setenv DBNAME user22
setenv DBSERVER dbsvr42
setenv PIPEID 1
setenv RID 34
$PTF_SW/ptf/src/pl/perl/superbias.pl
\end{verbatim}

\noindent The selected working directory serves the same purpose as the
pipeline machine's local disk where all pipeline intermediate data files
are written.  Standalone pipeline execution is therefore useful for
diagnosing problems.  After sourcing the basic environment file,
generally the user will want to override the environment variables
that point to the user's sandbox and database.  The user's database is
normally a copy of the operations database.  Environment variables
{\it RID}, which is a representative raw-image database ID ({\it rid}\/), and {\it PIPEID}\/
which is the pipeline database ID ({\it ppid}\/), reference the input data and pipeline number to
be executed, respectively.  In this particular case, the
representative image is representative of all bias images taken for a given night
and CCD; in the case of the superflat pipeline, the representative
image is representative of all science images (i.e., $IMGTYP=$~``object'') for a
given night, CCD, and filter.
Once the pipeline is setup using these commands, the
pipeline is executed with the last command listed above.  
In most cases, the user will want to redirect the standard output and
error streams to a log file.
The basic procedure is similar for all PTF pipelines, and can easily be
scripted if a large number of pipeline instances are involved.

\subsection{\label{cameraiimagesplitter}Camera-Image-Splitting Pipeline}

After the PTF data for a given night are ingested, the camera-image-splitting
pipelines, one pipeline instance per camera exposure, 
are launched automatically by the high-level data-ingest
process (see~\S{\ref{highlevelingest}}), or by the VPO
(see~\S{\ref{vpo}}) in the case that the
data had to be manually ingested because of some abnormal condition.
The pipeline executive is
set up to execute one instance of this pipeline per machine at a time.
Since there are 11 pipeline machines, 11 instances of the pipeline
are run in parallel.  This particular pipeline is not particularly compute-
or memory-intensive, and so more of these pipeline instances per machine could be run,
and tests of up to 4 instances per machine have been performed
successfully.

The camera-image-splitting pipeline is wrapped in a Perl script called
{\it splitCameraImages.pl}.  The input camera-image file is copied from the
archive to the pipeline machine's scratch disk.  The checksum of
the file is recomputed and compared to the checksum stored in the database,
and a mismatch, like any other pipeline error, would result in a
diagnostic message written to the log file and pipeline termination
with exit code $>=64$.  The filter associated with the camera-image
file is verified by running {\it check\_filter.py}, which uses median
values of various regions of image data and smoothing to look for  
patterns in the data that have high amplitude for the $g$ band but 
are weak for the $R$ band.  A filter mismatch results in pipeline
termination with exit code $=65$.  Manual intervention is required in
this case to decide whether to alter the filter information in the
database (filter-changer malfunctions have occurred intermittently
during the project) or 
skip the filter checking for that pipeline.  Experience has shown that
this filter checking is not reliable when the seeing is poor.

The module {\it ptfSplitMultiFITS}\/ is executed on the camera-image
file to break it up into 12 single-extension FITS files.  The primary
HDU, plus CCD-dependent keywords for the gain, read noise, and dark
current ({\it GAIN}, {\it READNOI}, and {\it DARKCUR}, respectively) 
are copied to the headers of the split-up files.
The resulting single CCD-image FITS files are then processed separately (except for
dead CCDID=3, which is skipped).  

If the CCD images are science images ($itid=1$; see Table~\ref{tab:imgtypes}), then
they are processed to find first-iteration astrometric solutions.  
Initial values of world-coordinate-system (WCS) keywords are written to the
CCD-image FITS headers. $CRVAL1$ and $CRVAL2$, the coordinates of the WCS reference point on
the sky, are set to the right ascension and declination of the
telescope boresight, $TELRA$ and
$TELDEC$, respectively.  $CRPIX1$ and $CRPIX2$, the corresponding
reference-point image coordinates for a given CCD, are set to
the telescope-boresight pixel positions that have been predetermined for each
CCD-image reference frame.  Finally, the following fixed values for
the pixel scale (at the distortion center) and image rotation angle are
set, as appropriate for the telescope and camera: 
$CDELT1=-0.000281$~degrees, $CDELT2=0.000281$~degrees, and 
$CROTA2=180$~degrees.  Next, source extraction is done with
{\it SExtractor}~\citep{bertin, sex, dummy} to generate a source catalog for
the astrometry.   The pipeline then runs Astrometry.net
modules {\it augment-xylist}, {\it backend}, and {\it new-wcs}~\citep{lang}
in succession with the objective of finding an astrometric solution.

If an astrometric solution is found, then it is verified and
recorded.  Verification includes requiring the pixel scale to be
within $\pm 5\%$ of the initial known value, the rotation angle to be within
$5^{\circ}$ of the initial known value, and the absolute values of {\it CRPIX1}\/
and {\it CRPIX2}\/ to be $\le 10,000$~pixels.  If these conditions are
not met, then 
bit $2^3=8$ is set in the {\it infobits}\/ column
of the  {\it RawImages}\/ database table (see
Table~\ref{tab:rawimageinfobits}) to flag this condition.
The astrometric solution is written both to the FITS header of the CCD
image and also to a text file in the archive containing only the astrometric
solution, in order to facilitate later generation by IRSA of source-catalog 
overlays onto JPEG preview images of PTF data.

\begin{table}
\caption{\label{tab:rawimageinfobits}Bits allocated for flagging
  various conditions and exceptions in the {\it infobits}\/ column of the {\it RawImages}\/ database table. }
\begin{small}
\begin{tabular}{cl}
\\[-5pt]
\hline\hline
Bit & Definition \\
\hline \\[-5pt]
0 & Dead CCD \\
1 & Astrometry.net failed\\
2 & Sidereal-tracking failure\tablenotemark{1} \\
3 & Bad astrometric solution\\
4 & Transient noise in image\tablenotemark{1}\\
\hline \\[-5pt]
\end{tabular}
\tablenotetext{1}{Manually set after image inspection.}
\end{small}
\end{table}

The CCD-image files are copied to the sandbox into a hierarchical directory
tree that differentiates the stored files by observation year, month, day, filter ID, CCD
ID, and pipeline database ID.\@  A record is created in the 
{\it  RawImages}\/ database table for
each CCD-image file.  The record contains a number of useful foreign
keys to other database tables ({\it expid}, {\it ccdid}, {\it nid},
{\it itid}, {\it piid}\/) and comprises columns for storing the location and
name of the file, record-creation date, image status, checksum, and {\it
infobits}.  The image status can be either 0 or 1, and is normally 0
only for the dead CCD ($CCDID=3$).  A bad astrometric solution,
although flagged in the {\it infobits}\/ column of the {\it  RawImages}\/ database table,
will not result in $status=0$ for the image at this point because the downstream 
frame-processing pipeline (see \S{\ref{frameprocpipeline}}) 
will make another attempt at finding a good solution.

The pipeline makes preview images in JPEG format using
IRSA's Montage software, both for the camera 12-CCD-composite image
and individual CCD images.  The preview images are subsequently 
used by the SDQA subsystem (see \S{\ref{sdqa}}).

\subsection{\label{sbcp}Superbias-Calibration Pipeline}

The purpose of the superbias calibration pipeline is to compute the
pixel-by-pixel electronic bias correction that is applied to every PTF
science image.
These pipelines are launched after the 
camera-image-splitting pipelines have completed for a given
night, one pipeline instance per CCD per night.  
This is done either automatically by the VPO
or manually by a human pipeline operator.

The superbias pipeline is wrapped in a Perl script called
{\it superbias.pl}.  The database is queried for all bias images for
the night and CCD of interest.  The {\it ptfSuperbias}\/ module is then
executed, and this produces the superbias-image calibration file, a file called
``superbias.fits'', which is the common bias in the image data for a
given CCD and night.
The file is renamed to an archival filename, 
copied to the sandbox, and registered in the {\it CalFiles}\/
database table with $caltype=$``superbias''.

The method used to compute the superbias is described as follows.
The bias images are read into memory.  The floating bias of each image
is computed and then subtracted from its respective bias image.
The CCD-appropriate pixel mask is used to ignore dead or bad pixels.
The software can be set up to compute the floating bias from up to three
different overscan regions, but, in practice, only the long strip
running down the right-hand side of the image is utilized.  The floating
bias is the average of the values in the overscan region after
an aggressive outlier-rejection step.  The outliers are found by
thresholding the data at the median value
$\pm2.5$~times the data dispersion, which is given
by half of the difference between the 84.1 percentile and the 15.9
percentile.  The bias-minus-floating-bias values are then processed by
a similar outlier-rejection algorithm on a pixel-by-pixel basis, and
the surviving values are averaged at each pixel location to 
yield the superbias image and
accompanying ancillary images, which are described in the next paragraph.

Ancillary calibration products are also generated by the {\it ptfSuperbias}\/ module.
These are packed into a file called ``superbias\_ancil\_data.fits''.
The ancillary FITS file is an image-data cube
($NAXIS=3$) containing
the superbias uncertainties in the first data plane, the number of
samples in the second data plane, and the number of outliers rejected
in the third data plane.  All quantities are on a pixel-by-pixel basis.
The file is renamed to an archival filename, 
copied to the sandbox, and registered in the {\it AncilCalFiles}\/
database table with $anciltype=$~``superbiasstats''.

\subsection{\label{preprocpipe}Preprocessing Pipeline}

The preprocessing pipeline prepares the science images ($IMGTYP=$``object'') to be fed into
the downstream superflat-calibration and image-flattener pipelines.  
The preprocessing is several-fold:

\begin{enumerate}
\item{Subtract off the floating bias and superbias from each pixel value;}
\item{Crop the science images to remove the bias overscan regions;}
\item{Compute data-mask bit settings for saturated and ``dirty''
    pixels (bit $2^8=256$ and bit $2^{11}=2048$, respectively; 
    see Table~\ref{tab:dmask}; ``dirty'' pixels are defined below), and combine them 
    with the appropriate fixed, CCD-dependent pixel mask (see~\S{\ref{pmi}})
    to create an initial data mask for every science image; }
\item{Recompute an improved value for the seeing; and }
\item{Augment the data-mask image for each science image with the bit
    setting allocated for marking object detections 
    (bit $2^1=2$; see Table~\ref{tab:dmask}) 
    taken from {\it SExtractor}\/ object check images.}
\end{enumerate}

\begin{table}
\caption{\label{tab:dmask}Bits allocated for data masks. }
\begin{small}
\begin{tabular}{rl}
\\[-5pt]
\hline\hline
Bit & Definition \\
\hline \\[-5pt]
0 & Aircraft/satellite track \\
1 & Object detected\\
2 & High dark current \\
3 & Reserved\\
4 & Noisy\\
5 & Ghost\\
6 & CCD bleed\\
7 & Radiation hit\\
8 & Saturated\\
9 & Dead/bad\\
10 & NaN (not a number)\\
11 & Dirt on optics\\
12 & Halo\\
13 & Reserved\\
14 & Reserved\\
15 & Reserved\\
\hline \\[-5pt]
\end{tabular}
\end{small}
\end{table}

\noindent
The preprocessing pipeline is wrapped in a Perl script called
{\it preproc.pl}.  
An instance of this pipeline runs on a per night, per CCD, per filter
basis.  The saturation level for the CCD at hand is looked up at the
beginning of the pipeline.

The preprocessing pipeline requires the
following input calibration files: a pixel mask and a superbias image.
It will also utilize a superflat image, if available.  The calibration
files are retrieved via a call to database stored function {\it
  getCalFiles}, which queries the {\it CalFiles}\/ database table, and
returns a hash table of the latest calibration files available for the
night, CCD, and filter of interest.  The function always returns
fallback calibration files for the superbias and superflat, which are
zero-value and unity-value images, respectively.  The fallbacks are
pressed into service when the primary calibration files are non-existent.

The bit allocations for data-mask images are documented in
Table~\ref{tab:dmask}.  Bit $2^1=2$ is allocated for pixels overlapping
onto detected astronomical objects.  Bit $2^8=256$ is allocated for saturated pixels.
Bit $2^{11}=2048$  is allocated for dirty pixels, where ``dirty'' is defined as 10~standard deviations below
the image's local median value.

The pipeline first runs the {\it ptfSciencePipeline}\/ module to perform
bias corrections, image cropping, and computation of the initial data masks.  
The floating bias
is computed via the method described above (see \S{\ref{sbcp}}).  The pipeline runs
multiple threads of this process, where each thread processes a
portion of the input science images in parallel.  
The science images are cropped to $2048 \times 4096$~pixels.
The pipeline outputs are a set of
bias-corrected images and a set of bias-corrected \& flattened images
(useful if a flat happens to be available from a prior run).  

Next, multi-threaded runs of
{\it SExtractor}\/ are made on the aforementioned
latter set of images, one thread per
image, in order to generate source catalogs for the seeing calculation.
Object check images are also generated in the process.
Bit $2^7=128$ will be set in the {\it infobits}\/ column
of the  {\it ProcImages}\/ database table (see
Table~\ref{tab:procinfobits}) for $ppid=3$ records 
associated with science images that
contain no sources.

\begin{table*}
\caption{\label{tab:procinfobits}Bits allocated for flagging
  various conditions and exceptions in the {\it infobits}\/ column of the {\it
    ProcImages}\/ database table. }
\begin{small}
\begin{tabular}{rl}
\\[-5pt]
\hline\hline
Bit & Definition \\
\hline \\[-5pt]
0 & {\it SCAMP}\/ failed \\
1 & WCS\tablenotemark{1} solution determined to be bad\\
2 & {\it mShrink}\/ module execution failed \\
3 & {\it mJPEG}\/ module execution failed\\
4 & No output from {\it ptfQA}\/ module (as {\it SExtractor}\/ found no sources)\\
5 & Seeing was found to be zero; reset it to 2.5 arcseconds\\
6 & {\it ptfSeeing}\/ module had insufficient number of input sources\\
7 & No sources found by {\it SExtractor}\/\\
8 & Insufficient number of 2MASS sources in image for WCS verification\\
9 & Insufficient number of 2MASS matches for WCS verification\\
10 & 2MASS astrometric R.M.S.E.(s) exceeded threshold\\
11 & {\it SExtractor}\/ before SCAMP failed\\
12 & {\it pv2sip}\/ module failed\\
13 & {\it SCAMP}\/ ran normally, but had too few catalog stars\\
14 & {\it SCAMP}\/ ran normally, but had too few matches\\
15 & Anomalous low-order WCS terms\\
16 & Track-finder module failed\\
17 & Anomalously high distortion in WCS solution\\
18 &     Astrometry.net was run\\
19 &     Error from sub runAstrometryDotNet\\
20 &     Time limit reached in sub runAstrometryDotNet\\

\hline \\[-5pt]
\end{tabular}
\tablenotetext{1}{World-coordinate system}
\end{small}
\end{table*}

The {\it ptfSEEING}\/ module is then executed in multi-threaded mode on
different images in parallel.
The seeing calculation requires at least 25 sources with the following
{\it SExtractor}\/ attributes: $FWHM\_IMAGE>0$,
a minimum stellarity ({\it CLASS\_STAR}\/) of 0.8, 
and {\it MAG\_BEST}\/ flux
between 5,000 and 50,000~DN.\@
Bit $2^6=64$ will be set in the {\it infobits}\/ column
of the  {\it ProcImages}\/ database table (see
Table~\ref{tab:procinfobits}) for $ppid=3$ records associated with science images that
contain an insufficient number of sources for the seeing calculation.
The $FWHM\_IMAGE$
values for the vetted sources are histogrammed in 0.1-pixel bins and
the seeing is taken as the mode of the distribution, which is,
in practice, the position of the peak bin.

The recomputed seeing is refined relative to the {\it SEEING}\/
keyword/value that is already present in the header of the
camera-image file (see Table~\ref{tab:primaryhdu2}), 
and is written to the output FITS header with the
keyword {\it FWHMSEX}, in units of arcseconds.  
In addition to the selection based on {\it SExtractor}\/ parameters
described above, the refinements include the benefits of 
the pixel mask, bias-corrected input data,
and proper accounting for saturation. 

Lastly, the {\it ptfMaskCombine}\/ module is executed in multi-threaded mode on
different masks in parallel, in order to fold the object detections from the {\it SExtractor}\/
object check images into the data masks.

The resulting science images are copied to the sandbox and registered
in the {\it ProcImages}\/ database table with pipeline index $ppid=3$ (see
Table~\ref{tab:pipelinedbtable}).  
The resulting data masks are copied to the sandbox and registered in
the {\it AncilFiles}\/ database table with $anciltype=$``dmask''.
The science images and their respective data masks are explicitly associated
in the latter database table.

\subsection{Superflat-Calibration Pipeline}

A superflat is a calibration image that corrects for 
relative pixel-to-pixel responsivity variations across a CCD.\@
This is also known as the nonuniformity correction. 
Images of different fields observed throughout the 
night are stacked to build a high signal-to-noise superflat. 
This process also allows the removal of stars and cosmic rays via
outlier rejection and helps 
average-out possible sky and instrumental variations at 
low spatial frequencies across the input images.

The superflat calibration pipeline produces a superflat 
from all suitable science images for a given night, CCD, and filter,
after data reduction by the preprocessing pipeline.  
A minimum of 5 PTF fields covered by the input images is required to
ensure field variegation and 
effective source removal in the process of superflat generation.
Also, a minimum of 10 input images is required, but typically 100-300 images
are used to make a superflat.
Special logic avoids too many input images from predominantly observed
fields in a given night.
The resulting superflat is applied to the
science images in the image-flattener pipeline (see \S{\ref{imageflattener}}).

The superflat pipeline is wrapped in a Perl script called
{\it superflat.pl}.  The database is queried for the relevant
preprocessed science images, along with their data masks.  
The query excludes exposures from the Orion
observing program~\citep{vaneyken}, in which the imaging was of the same
sky location for many successive exposures
and the telescope dithering was insufficient for making
superflats with the data.
  
The {\it normimage}\/ module is executed for each preprocessed science
image to create an interim image that is
normalized by its global median, which is computed
after discarding pixel values for which any data-mask bit is set.
All normalized values that are less than 0.01 are reset to
unity, which minimizes the introduction of artifacts into the superflat.

In order to fit the entire stack of images into available memory (as
many as 422 science exposures have been taken in a single night), the
{\it quadrantifyimage}\/ module is executed to break each normalized
image into four equally sized sub-images.  The same module is separately
executed for the data masks.  

The {\it createflat}\/ module processes, one quadrant at a time, all of the
sub-images and their data masks to create associated stack-statistics and
calibration-mask sub-images.  
A separate CDF for each CCD provides
input parameters for the process (although CCD-dependent processing
for superflats is not done at this time, the capability exists).  
The parameters direct the code, for each pixel
location, to compute the median value of the stacked sub-image data values (as
opposed to some other trimmed average), and the trimmed standard
deviation ($\sigma$) after eliminating the lower 10\% and the upper 10\% of the
data values for a given pixel (and re-inflating the result in
accordance with a trimmed Gaussian distribution to account
for the data-clipping).  Lastly, the module recomputes the median after
rejecting outliers greater than $\pm 5 \sigma$ from the initial median
value, as well as computing the corresponding uncertainty.  The stack statistics
are written to a FITS data cube, where the first plane contains the
clipped medians and the second plance contains the uncertainties.
The bit definitions for calibration-mask images are given in
Table~\ref{tab:cmask}.

\begin{table}
\caption{\label{tab:cmask}Bits allocated for the superflat calibration
  mask.  Bits not listed are reserved.}
\begin{small}
\begin{tabular}{rl}
\\[-5pt]
\hline\hline
Bit & Definition \\
\hline \\[-5pt]
1 & One or more outliers rejected\\
2 & One or more NaNs present in the input data \\
3 & One or more data-mask-rejected data values\\
12 & Too many outliers present\tablenotemark{1}\\
13 & Too many NaNs present\tablenotemark{1}\\
14 & No input data available\\
\hline \\[-5pt]
\end{tabular}
\tablenotetext{1}{Currently, the allowed fraction is 1.0, so this bit
  will never be set.}
\end{small}
\end{table}

The {\it tileimagequadrants}\/ module pieces back together the four quadrants of the 
stack-statistics and calibration-mask sub-images corresponding to each
science image.  Finally, the {\it normimage}\/ module is executed on the full-sized stack-statistics
image to normalize it by its global image mean and reset
any normalized value to unity that is less than 0.01 (in the manner
described above).  The latter module ignores image data that are within
10~pixels of all four image edges in computing the normalization factor.

The pipeline's chief product is a superflat called
``superflat.fits''.  
The file is renamed to an archival filename, 
copied to the sandbox, and registered in the {\it CalFiles}\/
database table with $caltype=$``superflat''.
A corresponding ancillary product is also generated:
the calibration mask, which is called 
``superflat\_cmask.fits''.
The ancillary file is renamed to an archival filename, 
copied to the sandbox, and registered in the {\it AncilCalFiles}\/
database table with $anciltype=$``cmask''.

A number of processing parameters are written to the FITS header of
the superflat.  These include the number of input images, the
outlier-rejection threshold, the superflat normalization factor,
and the threshold for unity reset.

Several SDQA ratings are computed for the superflat.  These
include the following image-data statistics: average, median, standard
deviation, skewness, kurtosis, Jarque-Bera test\footnote{The
  Jarque-Bera test is a 
goodness-of-fit test of whether a sample skewness and
kurtosis are as expected from a normal distribution.}, 15.9~percentile, 
84.1~percentile, scale (half the difference between the 84.1~and 
15.9~percentiles), number of good pixels, and number of NaN pixels.
These values are written to  the {\it SDQA\_CalFileRatings}\/ database table.
We have found the Jarque-Bera test particularly useful in locating
superflats that infrequently contain point-source remnants due to insufficient input data variegation.

\subsection{\label{imageflattener}Image-Flattener Pipeline}

The image-flattener pipeline's principal function is to apply the
nonuniformity or flat-field corrections to the science images.  Also,
the pipeline runs a process to detect CCD bleeds and radiation hits in the
science images (see below), and then executes
the {\it ptfPostProc}\/ module to update the data masks and compute
weight images for later source-catalog generation in the  
frame-processing pipeline (see \S{\ref{frameprocpipeline}}).
The pipeline is wrapped in a Perl script called
{\it flattener.pl}.  
An instance of this pipeline runs on a per night, per CCD, per filter
basis.  At the beginning of the pipeline, the database is queried for
the science images to process, along with their data masks, and 
relevant calibration image, namely, the superflat associated with the
night, CCD, and filter of interest.  The saturation
level for the CCD is also retrieved.

In the rare case that the superflat does not exist, the database function 
{\it getCalFiles}\/ searches backward in time, up to 20 nights, for the 
closest-in-time superflat substitute.  In most cases, the superflat
made for the previous night is returned for the CCD and filter of interest.
Our experience has been that, generally, the superflat changes slowly
over time, hence the substitution does not unduly compromise the data.

The {\it ptfSciencePipeline}\/ module performs the image flattening.  It
reads in a list of science images
and the superflat.  It then simply divides each science image
by the superflat on a pixel-by-pixel basis.  Since the superflat was
carefully constructed to contain no values very close to zero, the
output image is well behaved, although the processing includes logic
to set the image value to NaN in case it has been assigned
the representation for infinity.  The applied flat is associated with the pipeline products
via the {\it CalFileUsage}\/ database table.

{\it  SExtractor}\/ is executed to detect CCD bleeds and radiation hits in the
science images, and the output check images contain the detections.
It is executed on separate science images via 7 parallel threads at a
time.  The saturation level is an important input to this process.
The detection method is an artificial-neural-network (ANN) filter.  A
program called {\it Eye}\/ was used to specifically train the ANN on PTF
data.  Both {\it SExtractor}\/ and {\it Eye}\/ are freely
available\footnote{See http://www.astromatic.net for more details.}.

The {\it ptfPostProc}\/ module is a pipeline process that, for
each science image: 1) updates its data mask, and 2) creates a
weight image suitable for use in a subsequent {\it SExtractor}\/ run for 
generating a source catalog.  
The module is executed in multi-threaded 
mode on separate data masks.   The superflat, along with the 
pertinent check image from the aforementioned {\it
SExtractor}\/ runs, are the
other major inputs to this process for a given data mask.
The {\it ptfPostProc}\/ data-mask update includes setting
bits to flag CCD bleeds and radiation hits (see
Table~\ref{tab:dmask}), which are taken to have occurred at 
pixel locations where check-image values are $\ge 1$.
Since the check image does not differentiate between the two artifacts at this time,
both bits are set in tandem.  The {\it ptfPostProc}\/ weight-map creation starts with the
superflat as the initial weight map and then sets the weights to zero if
certain bits are set in the data mask at the same pixel location.
Pixels in the weight maps that are masked as dead/bad or NaN
(see Table~\ref{tab:dmask}) consequently will have zero weight values.

Similar to the preprocessing pipeline (see~\S{\ref{preprocpipe}}),
the resulting science images are copied to the sandbox and registered
in the {\it ProcImages}\/ database table with pipeline index $ppid=10$ (see
Table~\ref{tab:pipelinedbtable}), and
the resulting data masks are copied to the sandbox and registered in
the {\it AncilFiles}\/ database table with $anciltype=$ ``dmask''.
The science images and their respective data masks are explicitly associated
in the latter database table.
The weight-map files, which are not archived 
(see \S{\ref{productarchiver}}) but used by the next pipeline (see
\S{\ref{frameprocpipeline}}), are copied to the sandbox but not
registered in the {\it AncilFiles}\/ database table.

\subsection{\label{frameprocpipeline}Frame-Processing Pipeline}

\begin{figure}
\includegraphics[scale=0.5]{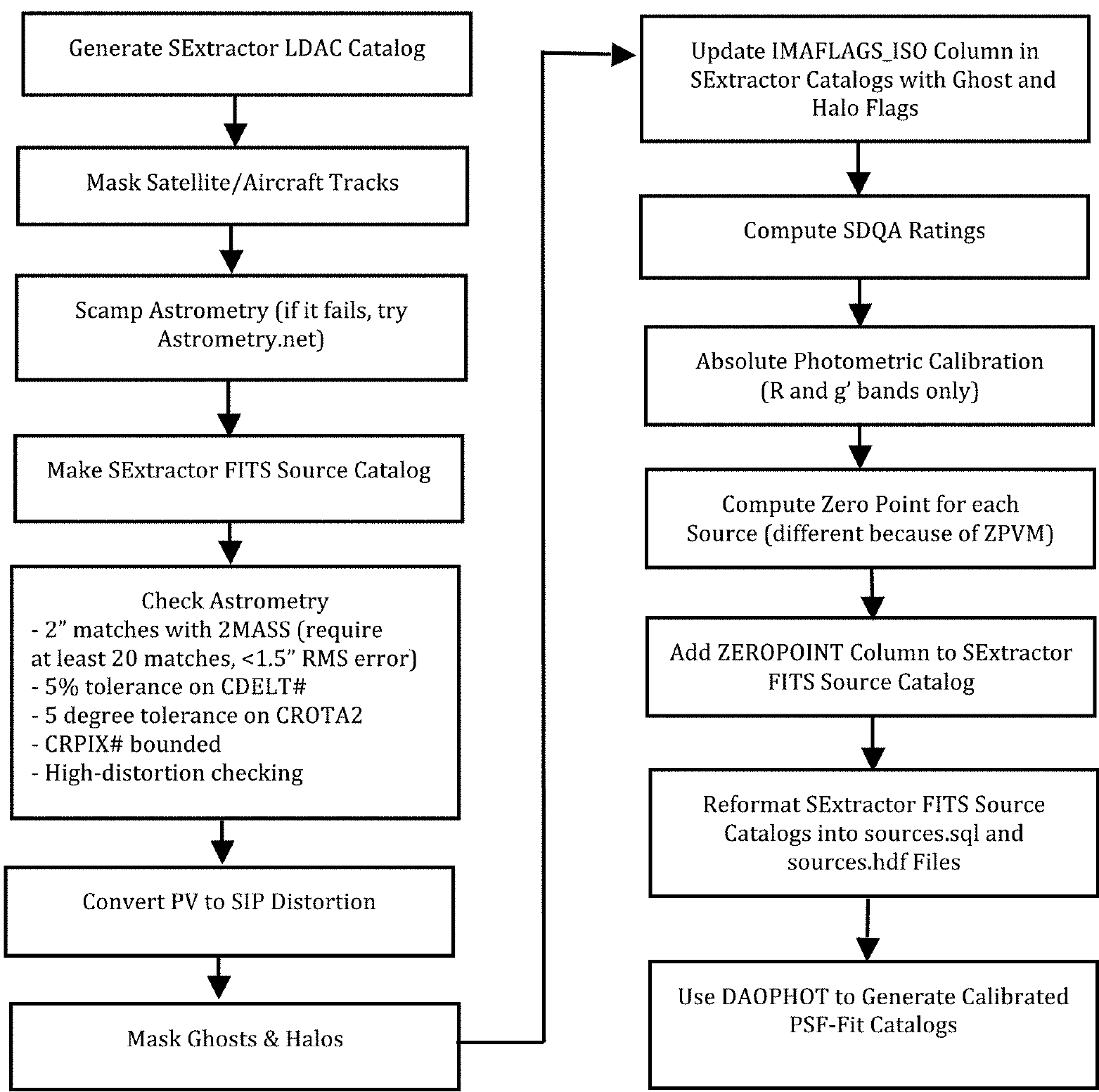}
\caption{\label{fig:frameprocflow} Flowchart for the frame-processing pipeline.}
\end{figure}

The frame-processing pipeline's major functions are to perform
astrometric and photometric calibration of the science images.  
In addition, aperture-photometry source catalogs are made 
from the processed science images using {\it SExtractor}, and 
PSF-fit catalogs are made using {\it DAOPHOT}.
The
processed science images, their data masks, source catalogs,
and other information 
(such as related to SDQA; see \S{\ref{sdqa}} for more details) 
are registered in the database to facilitate data analysis and
product archiving.   Figure~\ref{fig:frameprocflow}
shows the flow of data and control through the pipeline.

The frame-processing pipeline is wrapped in a Perl script called
{\it frameproc.pl}.  The pipeline begins by querying the database 
for all flattened science
images and associated data masks for
the night, CCD, and filter of interest.  The files are copied from the
sandbox to the pipeline machine's scratch disk for local access.
A record for each science image is created in the {\it ProcImages}\/
database table with pipeline index $ppid=5$ (see Table~\ref{tab:pipelinedbtable}),
which will store important metadata about the processed images, such
as a unique processed-image database ID ($pid$), 
disk location and filename, status, processing version, which version is
``best'', etc.  

The refined seeing computed by the preprocessing pipeline is
read from the FITS header (see \S{\ref{preprocpipe}}).  If its value
is zero, then it is reset to 2.5 arcseconds and this condition is flagged by setting 
bit $2^5=32$ in the {\it infobits}\/ column of the corresponding {\it ProcImages}\/
database record (see Table~\ref{tab:procinfobits}).  The refined
seeing is a required input parameter for source-catalog generation by
{\it SExtractor}.

The pipeline next executes {\it SExtractor}\/ to generate source
catalogs, one per science image,
in  FITS ``LDAC'' format, which is the required format for input
to the {\it SCAMP}\/ process described below~\citep{scamp}.  
The {\it SExtractor}\/-default convolution filter is applied.  The non-default input
configuration parameters are listed in Table~\ref{tab:sex3}.

\begin{table}
\caption{\label{tab:sex3}Non-default {\it SExtractor}\/ parameters for FITS ``LDAC'' catalog generation.}
\begin{small}
\begin{tabular}{ll}
\\[-5pt]
\hline\hline
Parameter & Setting \\
\hline \\[-5pt]
{\it CATALOG\_TYPE}   &    FITS\_LDAC \\	
{\it DETECT\_THRESH}   &   4 	 \\           
{\it ANALYSIS\_THRESH} &  4 	 \\       
{\it GAIN} &  1.5 \\
{\it DEBLEND\_MINCONT}  &  0.01	  \\    
{\it PHOT\_APERTURES}  &   2.0, 3.0, 4.0, 6.0, 10.0     \\
{\it PHOT\_PETROPARAMS}  & 2.0, 1.5    \\                       
{\it PIXEL\_SCALE}   &     1.01        \\    	
{\it BACK\_SIZE}   &       32          \\ 
{\it BACKPHOTO\_TYPE}  &   LOCAL \\
{\it BACKPHOTO\_THICK}	 &  12  \\
{\it WEIGHT\_TYPE}  &     MAP\_WEIGHT \\
\hline \\[-5pt]
\end{tabular}
\end{small}
\end{table}

The {\it createtrackimage}\/ module is executed to
detect satellite and aircraft tracks in each science image.
Tracks appear with a frequency of a few to several times in a given night and
the same track often crosses multiple CCDs.  The
module looks for contiguous blobs of pixels that are at or above the
local image median plus 1.5 times the local image-data dispersion, where the
dispersion is computed via the robust method of taking half the difference between the 84.1
percentile and the 15.9 percentile (which reduces to 1~standard
deviation in the case of Gaussian-distributed data).  All thresholded pixels that
comprise the blobs are tested to ensure they neither are an image-edge
pixel nor have their data values equal to NaN or 
are generally masked out (data-mask bit $2^1=2$ for source
detections is excepted).  The track-detection properties of this module
were improved by using local statistics, instead of global, in the
image-data thresholding, and our method of computing local statistics, which
involves computing statistics on a coarse grid and using bilinear
interpolation between the grid points, incurred
only a small processing-speed penalty.  The {\it createtrackimage}\/ module
utilizes a morphological classification algorithm that relies on
pixel-blob size and shape characteristics.  The median and dispersion
of the blob intensity data are computed and subsequent morphology testing is done
only on pixels with intensities that are within $\pm 3 \sigma$ of the median.
The blobs 
must consist of a minimum of 1000~pixels to be track-tested.  In order for a
blob to be classified as a track, at least one of the following
parametrically tuned tests must be satisfied:

\begin{enumerate}
\item{The blob length is greater than 900~pixels, or}
\item{The blob length is $\ge 300$~pixels and
    the blob half-width is $\le 10$~pixels, or}
\item{The blog length is greater than 150~pixels and
    the blob half-width is less than 2~pixels.}
\end{enumerate}

\noindent The blob length is found by least-squares fitting a line to
the positions of the blob pixels, and then computing the
maximum extent of the line across the blob.  The blob half-width is 
the robust dispersion of the perpendicular distances between the blob pixels 
and the fitted line.  The data mask associated with the processed
image of interest is updated for each
track found.  The pixels masked as tracks in the data mask are
blob pixels that are located within the double-sided envelope defined
by 4~blob half-widths on either side of  the track's fitted
line.  Bit $2^0=1$ in the data mask is allocated for flagging track pixels
(see Table~\ref{tab:dmask}).  A record for each track is inserted into the  {\it Tracks}\/
database table; the columns defined for this table
are given in Table~\ref{tab:tracks}.

\begin{table*}
\caption{\label{tab:tracks}Columns in the {\it Tracks}\/ database table. }
\begin{small}
\begin{tabular}{ll}
\\[-5pt]
\hline\hline
Column & Definition \\
\hline \\[-5pt]
{\it tid}  & Unique index associated with the track (primary key) \\
{\it pid}  & Unique index of the processed image (foreign key)\\
{\it expid}  & Unique index of the exposure (foreign key)\\
{\it ccdid}  & Unique index of the CCD (foreign key)\\
{\it fid}  & Unique filter index (foreign key)\\
{\it num}  & Track number in image\\
{\it pixels}  & Number of pixels in track\\
{\it xsize}  &Track size in $x$ image dimension (pixels)\\
{\it ysize}  & Track size in $y$ image dimension (pixels)\\
{\it maxd}  & Maximum track half-width (pixels)\\
{\it maxx}  & Track $x$ pixel position associated with {\it maxd}\\
{\it maxy}  & Track $y$ pixel position associated with {\it maxd}\\
{\it length}  & Length of track (pixels)\\
{\it median}  & Median of track intensity data (DN)\\
{\it scale}  & Dispersion of track intensity data (DN)\\
{\it a}  & Zeroth-order linear-fit coefficient of track $y$ vs. $x$ (pixels)\\
{\it b}  & First-order linear-fit coefficient of track $y$ vs. $x$ (dimensionless)\\
{\it siga}  & Uncertainty of zeroth-order linear-fit coefficient\\
{\it sigb}  & Uncertainty of first-order linear-fit coefficient\\
{\it chi2}  & $\chi^2$ of linear fit\\
{\it xstart}  &Track starting coordinate in $x$ image dimension (pixels)\\
{\it ystart}  &Track starting coordinate in $y$ image dimension (pixels)\\
{\it xend}  &Track ending coordinate in $x$ image dimension (pixels)\\
{\it yend}  &Track ending coordinate in $y$ image dimension (pixels)\\
\hline \\[-5pt]
\end{tabular}
\end{small}
\end{table*}

The astrometric solution for each science image is computed by 
{\it SCAMP}~\citep{scamp}.  The star catalog specified as input depends on
whether the science image overlaps an SDSS field.  The overlap fractions
are precomputed and stored in the {\it FieldCoverage}\/ database table.
For the $R$ and $g$ filters, if the fraction equals 1.0, the {\it SDSS-DR7}\/ 
catalog \citep{abazajian} is selected; otherwise, the {\it UCAC3}\/ catalog \citep{zacharias} is selected.  
If {\it SCAMP}\/ fails to find an astrometric solution, then it
is rerun with the {\it  USNO-B1}\/ catalog \citep{monet}.
For the $H\alpha$ filters, only the {\it UCAC3}\/ catalog is selected.  
Up to 5~minutes per science image is allowed for {\it SCAMP}\/
execution.  The process is killed after the time limit is reached and
retry logic allows up to 3 retries.  Since a {\it SCAMP}\/ catalog will
be the same for a given field, CCD, and filter, the catalogs are cached on disk
in a directory tree organized by catalog type and the aforementioned parameters after
they are received from the catalog server.  The catalog-file cache is
therefore checked first before requesting a catalog from the server.  Since 
{\it  SCAMP}\/ represents distortion using {\it PV}\/
coefficients\footnote{The {\it PV}\/ distortion coefficients implemented in {\it SCAMP}\/ are best documented by~\citet{shupe}.}, and some
distortion is always expected, the pipeline requires {\it PV}\/
coefficients to be present in the FITS-header file that {\it SCAMP}\/
outputs as a container for the astrometric solution.  The pipeline also
parses {\it SCAMP}\/ log output for the number of catalog sources loaded and
matched, and requires more than 20 of these as one of the criteria 
for an acceptable astrometric solution.

A {\it SCAMP}-companion program called {\it missfits}\/ transfers the
astrometric solution to the FITS header of each science-image file.
Another process called {\it hdrupdate}\/ removes the astrometric
solution previously found by Astrometry.net from the science-image 
FITS headers (see \S{\ref{cameraiimagesplitter}}).

A custom module called {\it pv2sip}\/ converts the {\it PV}\/ distortion
coefficients from {\it SCAMP}\/ into the {\it SIP}\/ representation~\citep{shupesip}.  The original code was
developed in Python~\citep{shupe}, and later translated into the C
language by one of the authors (R. R. L.).
This pipeline step is needed
because {\it WCS Tools}\/ and other off-the-shelf astronomical 
software used by the pipeline require {\it SIP}\/ distortion
coefficients for accurate conversion between image-pixel coordinates 
and sky coordinates.

The astrometric solution is first sanity-checked and then later verified.
The sanity checks, which assure proper
constraining of the low-order WCS terms ({\it CDELT1}, {\it CDELT2}, 
{\it CRPIX1}, {\it CRPIX2},  and {\it CROTA2}\/), are relatively simple tests that are done as
described in \S{\ref{cameraiimagesplitter}}.
Regardless of whether
the solution is good or bad, the astrometric coefficients are loaded
into the {\it IrsaMeta}\/ database table, which is indexed by
processed-image ID ({\it pid}\/) and contains the metadata
that are required by IRSA (see
Section~\ref{idad} below).  There is a one-to-one relationship between
records in this table and the {\it ProcImages}\/ database table.
Images with solutions that fail
the sanity-checking will
be flagged with $status=0$ in the {\it ProcImages}\/ database table,
and bit $2^{15}=$~32,768 will be set in the {\it infobits}\/ column
of the  {\it ProcImages}\/ database table (see
Table~\ref{tab:procinfobits}).
The astrometric verification involves matching the sources extracted from 
science images with selected sources from the {\it 2MASS}\/
catalog~\citep{skrutskie}.  A matching radius of 2~arcseconds is specified for this purpose.
A minimum of 20 2MASS
sources must be contained in the image, and the root-mean-squared error (R.M.S.E.) of the
matches along both image dimensions must be less than 1.5 arcseconds.
If any of these criteria are not satisfied, then the appropriate bit
will be set in the {\it infobits}\/ column
of the  {\it ProcImages}\/ database table (see
Table~\ref{tab:procinfobits}), and the image will be flagged as
having failed the astrometric verification.

If {\it SCAMP} fails to give an acceptable astrometric solution, then
Astrometry.net is executed.  If this succeeds, then a custom module
called {\it sip2pv} is run to convert the {\it SIP} distortion
coefficients into {\it PV} distortion coefficients, so that the
correct source positions are computed by {\it SExtractor} when making
the source catalogs.

\begin{figure}
\begin{center}
\includegraphics[scale=0.75]{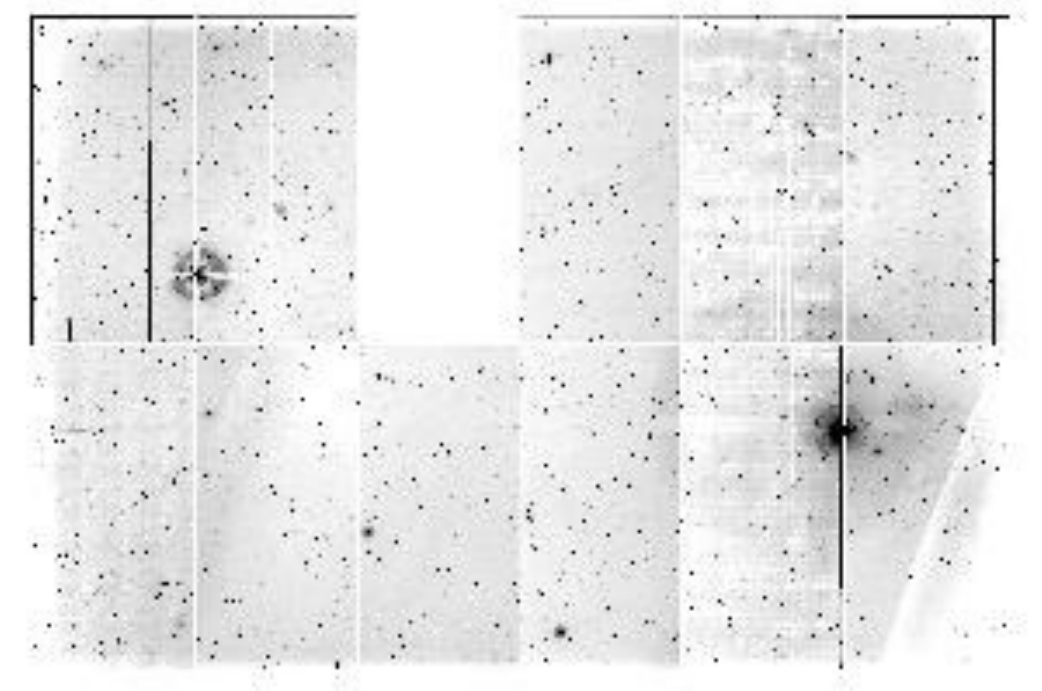}
\caption{\label{fig:exghost} Example ghost in PTF exposure
  $expid=203381$.  The image-display gray-scale table is inverted, so that
  black indicates high brightness and white indicates low brightness.  
  The large ghost is located in the upper-left portion of
  the 12-CCD composite image, and is imaged onto two CCDs ($ccdid=4$
  and $ccdid=5$).   It is caused by the bright star located in the
  lower-right portion.}
\end{center}
\end{figure}

\begin{figure}
\begin{center}
\includegraphics[scale=0.5]{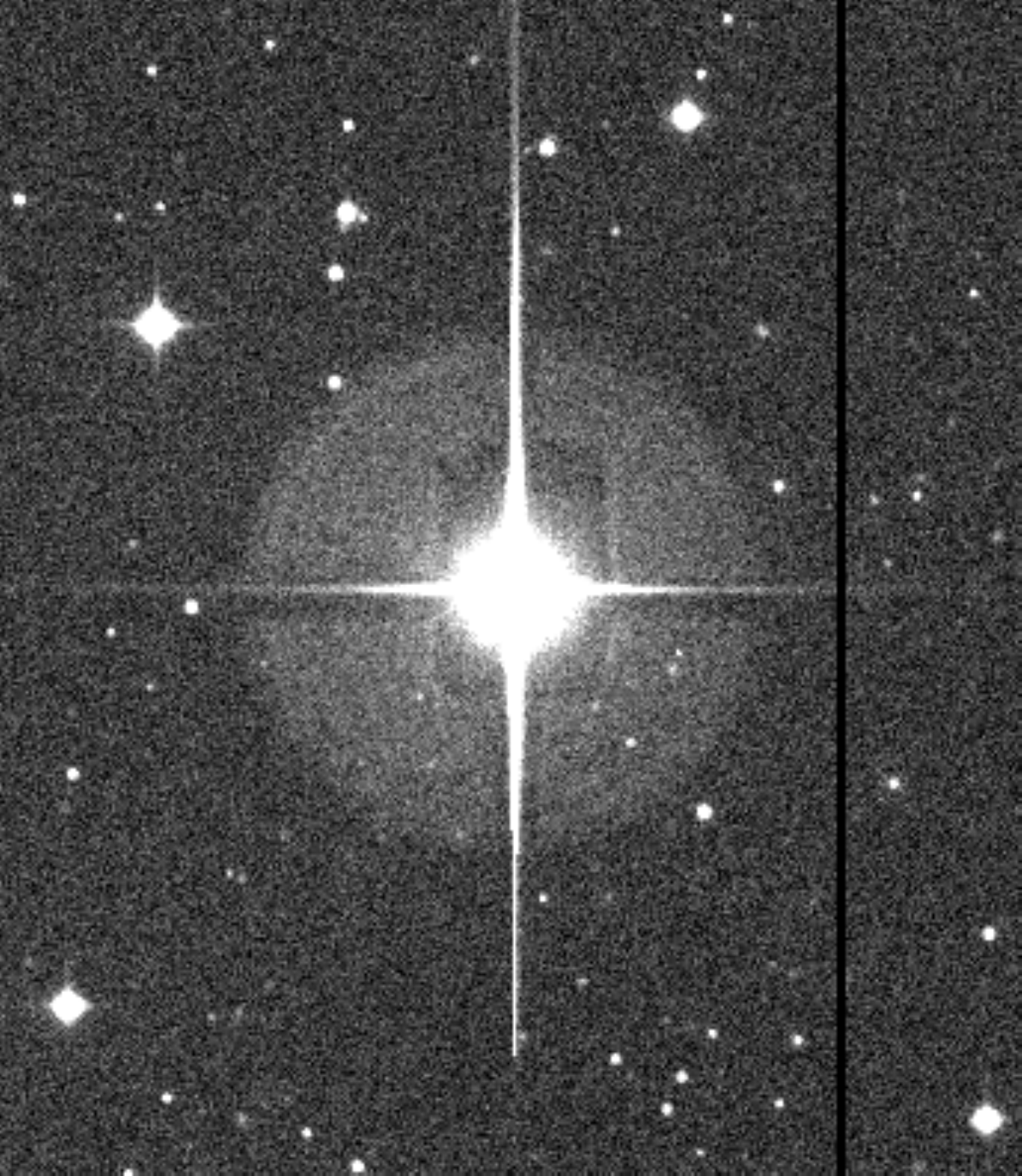}
\caption{\label{fig:exhalo} Example halo in PTF processed image
  $pid=9514402$.  Only a portion of the CCD image is shown.  
  The halo surrounding the bright star is
  $\approx 3$~arcminutes in diameter.}
\end{center}
\end{figure}

The pipeline includes functionality for inferring the presence of ghosts and halos in 
$R$- and $g$-band images.  
Ghosts are optical features that are 
reflections of bright stars about the telescope's optical axis.  A
bright star imaged in one CCD or slightly outside of the field of view
can lead to the creation of a ghost image in an opposite CCD with 
respect to the telescope boresight.  
An example ghost is shown in Figure~\ref{fig:exghost}.
Halos are optical
features that surround bright stars, and are double reflections that
end up offset slightly from the bright star toward the optical axis.  
An example halo is shown in Figure~\ref{fig:exhalo}.
The ghost positions vary
depending on the filter and also whether the image
was acquired before or after the aforementioned filter swap
(see~\S{\ref{projectevents}}).   Locating these
features starts by querying the Tycho-2 catalog and supplement for
bright stars, with $V_{{\rm mag}}$ brighter than 6.2 mag and 9.0 mag for $g$
and $R$ bands, respectively, before the filter swap, and brighter than
7.2 mag for both bands after the filter swap.  
Ghosts and halos are separately flagged in the data masks associated
with processed images.
Bit $2^5=32$ is reserved for ghosts and bit $2^{12}=4096$ for halos in the data mask
(see Table~\ref{tab:dmask}).  A circular area is flagged in the data
mask to indicate a ghost or halo.  Although the ghost and halo sizes vary with
bright-star intensity and filter, only a maximally sized circle for a
given filter, which was determined empirically for cases before
and after the filter swap, is actually masked off.
Accordingly, the radius of the circle for a ghost
is 170~pixels for the $R$ band (both before and after the filter
swap), and, for the $g$ band, is 450~pixels
before the filter swap and 380~pixels afterwards.  
Similarly, the radius of the circle for a $g$-band halo
is 85~pixels before the filter swap and 100~pixels afterwards, and is
95~pixels before and 100~pixels afterwards for $R$-band halos.
Database records in
the {\it Ghosts}\/ and/or {\it Halos}\/ database tables are inserted for
each ghost and/or halo found, respectively.

\citet{ofek} give a description of the photometric calibration, which
is done on a per-night, per-CCD, per-filter basis.  The source code
for the photometric calibration is written in Matlab, and the pipeline
makes a system call to execute this process.
A minimum of 30 astrometrically calibrated science images for the photometric calibration is
a software-imposed requirement to ensure adequate solution statistics
(sometimes fewer science images are taken in a given night or an inadequate
number could be astrometrically calibrated due to cloudy conditions, etc.).
Also, at least 1000 SDSS-matched stars extracted from the PTF
processed images for a given night, CCD, and filter are required for
the photometric-calibration process to proceed.
The resulting calibration data, consisting of fit coefficients, their
uncertainties, and a coarse grid of zero-point-variability-map (ZPVM) values,
are loaded into the {\it AbsPhotCal}\/ and 
{\it AbsPhotCalZpvm}\/ database tables, and are also
written to the pipeline-product image and source-catalog FITS headers.  While the
source catalogs contain instrumental magnitudes, their FITS
headers contain enough information to compute the
photometric zero points for the sources, provided that the photometric calibration
could be completed successfully.  In addition, as elaborated in the
next paragraph, we also compute the zero points of individual sources
(which vary from source to source because of the ZPVM) and include them
in the source catalogs as an additional column; these zero points
already include the $2.5 \log( \delta t)$ contribution for normalizing
the image data by the exposures time, $\delta t$, in seconds,
and so simply adding the instrumental magnitudes to their respective zero points
will result in calibrated magnitudes.
The photometric-calibration process also generates a FITS-file-image version of the
ZPVM, which is ultimately archived and
metadata about it is loaded into the 
{\it CalFiles}\/ database table with $caltype=$``zpvm''.  This
calibration file
is associated with the relevant pipeline products
in the {\it CalFileUsage}\/ database table.  
The minimum and maximum values in the ZPVM image are loaded into
the {\it AbsPhotCal}\/ database table as additional image-quality measures.
There is also a corresponding output FITS file
containing an image of ZPVM standard deviations, which is registered in the
{\it CalAncilFiles}\/ database table under $anciltype=$``zpve'' and associated with the ZPVM FITS file.

The calculation of the ZPVM contribution to the photometric 
zero point by the pipeline
itself for each catalog source
is done via bilinear interpolation of the ZPVM
values in the aforementioned grid of coarse cells, which are queried from the
{\it AbsPhotCalZpvm}\/ database table.
If any of the values is equal to NaN, which occurs when not enough good matches between
PTF-catalog and SDSS-catalog sources are available, then the
interpolation result is reset to zero.   The ZPVM algorithm
requires at least 1000~matches in a $256 \times 256$-pixel cell per CCD and
filter for the entire night~\citep{ofek}, in order to calculate the
value for a cell.  Because of the ZPVM, the zero point varies from one
source to the next.  The zero point for each source is
written to the {\it SExtractor}\/ source catalogs as an additional column,
called {\it ZEROPOINT}.

For each astrometrically calibrated image, {\it SExtractor}\/ is executed one
last time to generate its final aperture-photometry source catalog.  The correct gain and
saturation level is set for the CCD of interest.  Both detection and
analysis thresholds are set to $1.5 \sigma$. The input weight map
is the superflat with zero weight values where 
data-mask bits are set for dead, bad or NaN pixels, 
as described in~\S{\ref{imageflattener}}.
The {\it SEEING\_FWHM}\/ option is set to the seeing value computed 
in~\S{\ref{imageflattener}} for each image.  A background check image is
also generated by {\it SExtractor}\/ and stored in the sandbox, in case it is needed as a diagnostic.
The non-default input
configuration parameters for {\it SExtractor}\/ are listed in Table~\ref{tab:sex4}.

\begin{table}
\caption{\label{tab:sex4}Non-default {\it SExtractor}\/ parameters for final source-catalog generation.}
\begin{small}
\begin{tabular}{ll}
\\[-5pt]
\hline\hline
Parameter & Setting \\
\hline \\[-5pt]
{\it CATALOG\_TYPE}   &    FITS\_1.0 \\
{\it DEBLEND\_NTHRESH} & 4 \\
{\it PHOT\_APERTURES}  &   2.0, 4.0, 5.0, 8.0, 10.0     \\
{\it PHOT\_AUTOPARAMS}  &  1.5, 2.5 \\	     
{\it PIXEL\_SCALE}   &     1.01        \\    	
{\it BACKPHOTO\_TYPE}  &   LOCAL \\
{\it BACKPHOTO\_THICK}	 &  35  \\
{\it WEIGHT\_TYPE}  &     MAP\_WEIGHT \\
\hline \\[-5pt]
\end{tabular}
\end{small}
\end{table}

Furthermore, for each astrometrically calibrated image, we perform PSF-fit
photometry using the {\it DAOPHOT}\/ and {\it ALLSTAR}\/
software~\citep{stetson}. These tools are normally run interactively;
however, we have automated the entire process:
from source detection to PSF-estimation and PSF-fit photometry in a
pipeline script named {\it runpsffitsci.pl}. Input parameters
are the FWHM of the PSF (provided by {\it SExtractor}\/ upstream) and
an optional photometric zero point. At the time of writing, the
input photometric zero point is based on an absolute calibration
using the {\it SExtractor}\/ catalogs. This is not optimal and
we plan to recalibrate the PSF-fit extractions using
calibrations derived from PSF-fit photometry in the near future.
The {\it DAOPHOT}\/ routines are executed in a single iteration
with no subsequent subtraction of PSF-fitted sources
to uncover hidden (or missed) sources in a second pass.  A spatially varying
PSF that is modelled to vary linearly over each image is generated.
This is then used to perform PSF-fit photometry. Prior to
executing the {\it DAOPHOT}\/ routines, the {\it runpsffitsci.pl}\/
script dynamically adjusts some of the PSF-estimation and
PSF-fit parameters, primarily those that have a strong dependence
on image quality -- the PSF FWHM and image-pixel noise. The default
input configuration parameters used for PSF-fit-catalog generation
are listed in Table~\ref{tab:dao}. The parameters that are dynamically
adjusted are $RE, LO, HI, FW, PS, FI,$ and the $A${\it i}\/ aperture
radii (where $i = 1\ldots6$). In particular, the parameters that
depend on the input FWHM ($FW$) are the linear-half-size of the
PSF stamp image, $PS$; the PSF-fitting radius, $FI$; and the
aperture radii $A${\it i}, all in units of pixels. These
parameters are adjusted according to:
\begin{eqnarray}
PS & = & \min\left(19, \mbox{int}\left\{\max\left[9, 6FW/2.355\right]
              + 0.5\right\}\right), \nonumber\\
FI & = & \min\left(7, \max\left[3, FW\right]\right), \nonumber\\
Ai & = & \min\left(15, 1.5\max\left[3, FW\right]\right) + i - 1, \nonumber
\end{eqnarray}
where $i = 1\ldots6$, ``min'' and ``max'' denote the minimum and maximum
of the values in parenthese, respectively, and ``int''
denotes the integer part of the quantity.
The {\it runpsffitsci.pl}\/ script reformats the raw output
from {\it DAOPHOT}\/ and {\it ALLSTAR}\/ and assigns WCS information
to each source. The output table is later converted into FITS
binary-table format for the archive. The intermediate products,
such as the raw PSF file, are written to the sandbox.

\begin{table}
\caption{\label{tab:dao}Default input parameters for {\it
    science-image}\/ PSF-fit-catalog generation.  The $TH$ value in
  parentheses is for the PSF-creation step.}
\begin{small}
\begin{tabular}{ll}
\\[-5pt]
\hline\hline
{\it daophotsci.opt} & {\it photosci.opt} \\
\hline \\[-5pt]
$RE$ = 15.0 & $A1$ = 4.5\\
$GA$ = 1.5 & $A2$ = 5.5\\
$LO$ = 10 & $A3$ = 6.5\\
$HI$ = 10000.0 & $A4$ = 7.5\\
$PS$ = 9 & $A5$ = 8.5\\
$TH$ = 2.8 (30) & $A6$ = 9.5\\
$VA$ = 1 & $IS$ = 2.5\\
$EX$ = 5 & $OS$ = 20\\
$WA$ = 0 & \\
$FW$ = 2.5 & \\
$FI$ = 3.0 & \\
$AN$ = 1 & \\
$LS$ = 0.2 & \\
$HS$ = 1.0 & \\
$LR$ = -1 & \\
$HR$ = 1 & \\
\hline \\[-5pt]
\end{tabular}
\end{small}
\end{table}

After the photometric-calibration process has run and the source
catalogs have been created, the pipeline generates a file called
{\it sources.sql}, which contains an aggregation of all {\it SExtractor}\/ source catalogs for the night, CCD and filter of interest.  
The {\it sources.sql}\/ file is
suitable for use in bulk-loading source-catalog
records into the database.  However, after extensive testing, it has been determined
that loading sources into the PTF operations database is unacceptably
slow, and, consequently, this has been temporarily suspended until the
PTF-operations network and database hardware can be upgraded.  Nevertheless, the file still serves a
secondary purpose, which is facilitating the delivery of source information to IRSA,
where it is ultimately loaded into an archive relational database.
The file contains source information extracted
from the final {\it SExtractor}\/ source catalogs, as well as a
photometric zero point computed separately for each source.
In addition, for each source, a level-7
hierarchical-triangular-mesh (HTM) index is computed, and its {\it SExtractor} {\it IMAFLAGS\_ISO}\/
and {\it FLAGS}\/ parameters are packed together, for compact storage, into the upper
and lower 2 bytes, respectively, of a 4-byte integer.

A Python process is also run to generate a file with the same data contents
as the {\it sources.sql}\/ file, but in HDF5 format.  The output from this
process is called {\it sources.hdf}.  The HDF5 files can be read
more efficiently by Python software, and are used in downstream Python
pipelines for matching source objects and performing relative photometric calibration.

At the end of this pipeline, the primary products, which are the
processed images, are copied to the sandbox and registered
in the {\it ProcImages}\/ database table with the preassigned
processed-image database IDs ($pid$), and pipeline index $ppid=5$ (see
Table~\ref{tab:pipelinedbtable}).  There is a similar process for ancillary products and catalogs.
The ancillary products consist of
data masks and JPEG
preview images; these are copied to the sandbox and registered in
the {\it AncilFiles}\/ database table with {\it anciltype}\/ designations
of ``dmask'' and ``jpeg'', respectively.
The catalogs consist of {\it SExtractor}\/ and {\it DAOPHOT}\/ source
catalogs stored as FITS binary tables;  these are copied to the sandbox and registered in
the {\it Catalogs}\/ database table with {\it catType}\/ designations
of 1 and 2, respectively.
The primary products and their ancillary products and catalogs are explicitly associated
with each other by the processed-image database ID, $pid$, in the
{\it AncilFiles}\/ and {\it Catalogs}\/ database tables.  The {\it sources.sql}\/ and {\it sources.hdf}\/ files
created by the pipeline are copied to the sandbox, but not registered
in the database.  All of these products are included in
the subsequent archiving process (see \S\ref{idad}).

\subsection{\label{catgenpipeline}Catalog-Generation Pipeline}

The catalog-generation pipeline is wrapped in a Perl script
called {\it genCatalog.pl}\/ and has been assigned 
$ppid=13$ for its pipeline database ID.\@  It performs many, but not all, of the same
functions as the frame-processing pipeline 
(see~\S{\ref{frameprocpipeline}}).  Most notably, it omits the astrometric
and photometric calibrations, because this pipeline expects calibrated
input images (which are initially produced by the frame-processing pipeline).
The chief purpose of the catalog-generation pipeline is to
provide the capability of regenerating source catalogs directly 
from the calibrated, processed, and archived images and their data
masks, for a given night, CCD, and filter.  The source catalogs, 
if necessary, may
be produced from different {\it SExtractor}\/ and {\it DAOPHOT}\/ configurations than were
previously employed by the frame-processing pipeline. 
Also, for the PTF data taken before 2013, only 
{\it SExtractor}\/ catalogs were generated, as the execution {\it DAOPHOT}\/ had not yet
been implemented in the frame-processing pipeline.
The catalog-generation pipeline is, therefore, intended
to also generate the PSF-fit catalogs missing from the archive.
Like the frame-processing pipeline, the weight
map used by {\it SExtractor}\/ in this pipeline
to create a source catalog for an input image is generated by
starting with a superflat for the weight map and then zeroing out
pixels in the weight map that are masked as dead/bad or NaN in the
respective data mask of that input image.
The pipeline also has functionality for adding and updating
information in the FITS headers of the images and data masks.  Thus,
the products from this pipeline constitute new versions of images, data
masks, and source catalogs.  The pipeline copies its products to the 
sandbox and registers them, as appropriate, in the {\it ProcImages},
{\it AncilFiles}, and {\it Catalogs}\/
database tables with pipeline index $ppid=13$ (see
Table~\ref{tab:pipelinedbtable}).

Local copies of the calibration
files associated with the input images are made by the pipeline, and these
are also copied to the sandbox and
associated with the pipeline products in the
{\it CalFiles}\/ and {\it CalFileUsage}\/ database tables.  This ensures
that the calibration files are also re-archived when the new
products are archived.  The reason for this particular approach is
technical: the calibration files sit in the directory tree close to
the products, and are lost when old versions of products are
removed from the archive by  directory-tree pruning at a high level.

\subsection{\label{refimagepipeline}Reference-Image Pipeline}

To help mitigate instrumental signatures and transient
phenomena in general at random locations in the individual
images (e.g., noisy hardware pixels with highly varying
responsivity, cosmic rays, and moving objects, such as asteroids and
satellite/aircraft streaks), we co-add the images with outlier
rejection to create cleaner and more ``static'' representations
of the sky. Furthermore, this co-addition improves the overall
signal-to-noise ratio relative to that achieved in the
individual image exposures.

The reference-image pipeline creates coadds of input images for the same CCD,
filter, and PTF field ({\it PTFFIELD}\/). 
This pipeline is wrapped in
Perl script {\it genRefImage.pl}, and is run on an episodic basis as
new observations are taken.  It has been assigned 
$ppid=12$ for its pipeline database ID.\@  Currently, reference images
are generated only for the $R$ and $g$ bands.

The candidate input images for the coadds
are selected for the best values of seeing, color term, theoretical limiting
magnitude, and ZPVM (see description of absolute photometric
calibration in \S{\ref{frameprocpipeline}}).
A database stored function is called to
make this selection for a given CCD, filter, and PTF field, and it
returns, among other things, the database
IDs of candidate processed images that are potentially to be coadded.
The input-image selection criteria are listed as follows:
\begin{enumerate}
\item{All input images must be astrometrically and photometrically
    calibrated;}
\item{Exclude inputs with anomalously high-order distortion;}
\item{Minimum number of inputs $=5$;}
\item{Maximum number of inputs $=50$ (those with the faintest
    theoretical limiting magnitudes are selected);}
\item{Have color-term values that lie between the 1st and 99th
    percentiles;}
\item{Have ZPVM values between $\pm0.15$~mag;}
\item{Have seeing FWHM value $<3.6$\arcsec;}
\item{Have theoretical limiting magnitude $>20$~mag; and}
\item{Have at least 300 SExtractor-catalog sources.}
\end{enumerate}

\noindent
The candidate inputs are sorted by limiting magnitude in descending
order.  An input list is progressively incremented with successive
input images and the resulting coadd limiting magnitude (CLM) is computed 
after each increment.  The objective is to find the smallest set of inputs that 
comes as closely as possible to the faintest value of CLM
from a predefined small set of discrete values between 21.5 and
24.7 magnitudes.

An illumination correction is applied to each selected input image, in
order to account for the ZPVM (see~\S{\ref{frameprocpipeline}}).
Catalogs are generated with  {\it SExtractor}\/  and then fed to  
{\it SCAMP}\/ all together, in order to find a new astrometric solution
that is consistent for all input images.

The coadder is a Perl script called {\it mkcoadd.pl}.  It makes use of the
Perl Data Language (PDL) for multi-threading.  The input images and
associated data masks are fed to the coadder.  The input images are 
matched to a common zero point of 27~mag, which is a reasonable 
value for a 60~s exposure.  Thus all PTF reference images have a
common zero point of 27~mag.  {\it Swarp}\/ is used to resample
and undistort each input image onto a common fiducial grid based on
the astrometric solution~\citep{swarp}.
Saturated, dead/bad, and blank pixels are rejected.  The coaddition
procedes via trimmed averaging, weighted by the inverse seeing of each input
frame.  Ancillary products from the coadder include an uncertainty
image and a depth-of-coverage map.

The astrometric solution is verified against the 2MASS catalog (see
\S\ref{frameprocpipeline} for how this is done).
The pipeline generates both {\it SExtractor}\/ and
PSF-fit reference-image catalogs, which are then formatted as
FITS binary tables.  The PSF-fit catalogs are made using {\it DAOPHOT}.
Ancillary products from PSF-fit catalog generation
include a raw PSF file, a DS9-region file for the PSF-fit sources, and a set of PSF thumbnails arranged on
a grid for visualizing the 
PSF-variation across the reference image.
A number of SDQA ratings and useful metadata for
IRSA-archiving are computed for the reference image and loaded into the {\it SDQA\_RefImRatings}\/ and {\it IrsaRefImMeta}\/
database tables, respectively.

At the end of this pipeline, the reference image and associated
catalogs and ancillary files are copied to the sandbox.  The reference
image is registered
in the {\it RefImages}\/ database table with the preassigned
reference-image database ID ({\it rfid}\/), and pipeline index $ppid=12$ (see
Table~\ref{tab:pipelinedbtable}).  
The {\it SExtractor}\/ and {\it DAOPHOT}\/ reference-image catalogs are registered in
the {\it RefImCatalogs}\/ database table with {\it catType}\/ designations
of 1 and 2, respectively.
The reference images and their catalogs and ancillary files are explicitly associated
with each other by the processed-image database ID, {\it rfid}, in the
{\it RefImCatalogs}\/ and {\it RefImAncilFiles}\/ database tables.  All of these products are included in
the subsequent archiving process (see \S\ref{idad}).  The 
{\it RefImageImages}\/ database table keeps track of the input images
used to generate each reference image.

\subsection{\label{otherimagepipelines}Other Pipelines}

Other nascent or mature PTF pipelines will be described in later publications.  These
include pipelines for image differencing, relative
photometry, forced photometry, source association, asteroid detection, and 
large-survey-database loading.

\subsection{Performance}

As of 5 August 2013, a total of approximately $3.5 \times 10^5$ exposures in 1578
nights have been acquired.  About 75\% of the exposures are on the
sky, covering $\approx 2\times10^{6}$\,deg$^{2}$.  
There are also fair numbers of bias, dark, and twilight exposures
(14.3\%, 5.9\% and 4.8\%, respectively).  Table~\ref{tab:perf} lists
selected pipeline run-time robust statistics broken down by routinely
executed pipeline.
Recall the {\it ppid=7}\/ pipeline is run on a per-exposure basis, the
{\it ppid=1}\/ pipeline is run on a per-night, per-CCD basis, and the
remaining pipelines are run on a per-night, per-filter, per-CCD basis,
except for the {\it ppid=12}\/ reference-image pipeline, which is run on a per-filter,
per-CCD basis, per-PTF-field basis.  The run-time median and
dispersion for all pipelines has changed by less than 10\% over the
last couple of years or so, with the exceptions of the $ppid=5$
pipeline, which has become more than 30\% slower because of recently
added functionality, such as PSF-fit-catalog generation, and the
reference-image pipeline, which only came online in the last year.

\begin{table*}
\caption{\label{tab:perf}Selected pipeline run-time statistics
  (updated on 5 August 2013).  The statistics are pipeline runs on a per-CCD,
  per-filter, per-night basis, except for the ppId=12 pipeline, which
  is on a per-CCD, per-filter, per-field basis.}
\begin{small}
\begin{tabular}{rrrr}
\\[-5pt]
\hline\hline
{\it ppid}\tablenotemark{1} & No.\ of samples & Median (s) & Dispersion\tablenotemark{2} (s) \\
\hline \\[-5pt]
7 & 339671 & 200.4 & 84.4\\
1 & 14586 & 85.0 & 30.2\\
3 & 14840 & 2201.4 &  1226.0\\
4 &  14839 & 1416.5 &  815.0 \\
10 & 14827 & 4724.1 & 2424.0\\
5 &  14781 & 9387.1 & 6065.0\\
12 & 27890 & 271.3 & 70.0\\
\hline \\[-5pt]
\end{tabular}
\tablenotetext{1}{Given in execution order.}
\tablenotetext{2}{Half of the difference between the 84.1 and 15.9 percentiles.}
\end{small}
\end{table*}

The performance of our satellite/aircraft track detection algorithm
(see~\S{\ref{frameprocpipeline}})
has not yet been quantitatively scored in terms of completeness
vs.\ reliability; this will be the subject of a future paper.  
The algorithm has been tuned to find all tracks at
the expense of generating some false tracks.  Generally, the false
tracks will be associated with long, thin galaxies that mimic tracks
or very bright stars having extended CCD bleeds that were not fully
masked off in the processing.
A large $\chi^2$ of the track's linear fit may indicate a track-proximate bright star with a CCD bleed
extending across the track.  Multiple records in the {\it Tracks}\/ database table for the
same track in a given image can happen when the data thresholding
results in unconnected groups of contiguous pixels along that track.

\subsection{\label{smartphone}Smart-Phone Command \& Control}

A succinct set of high-level scripted commands was developed to
facilitate interrogation and control of the IPAC-PTF software and data system (see
Table~\ref{tab:smartphonecomands}).  
The commands generate useful short reports and optionally initiate
pipeline and archive processes.  The low data bandwidth and minimal
keyboard typing permitted by these commands makes them ideally suited 
for execution in a terminal window of a smart phone via cellular data
network (a wireless Internet connection is nice, but not required).  
Of course, the same commands also can be conveniently 
executed in a personal-computer terminal window.

One of us (R. R. L.), with the help of IPACer Rick Ebert, 
set up a virtual private network (VPN) on his
iPhone to allow secure connections directly to IPAC machines.
He also purchased secure-shell program ``Prompt, v. 1.1.1'' 
from the Apple Apps Store, which was
developed by Panic, Inc.\ and has since been upgraded, 
and then installed the app on his iPhone.  VPN and 
``Prompt'' are all the software needed to execute the PTF pipeline and
archive processes on the iPhone.  This set up even enables the
execution of low-level commands and arbitrary database queries, 
albeit with more keyboard typing.

All of the commands listed in Table~\ref{tab:smartphonecomands}, 
except for {\it ptfc}, generate brief reports by
default.  Some of the commands accept an optional date or list of dates, which is
useful for specifying night(s) other than the default current night.  Also,
some of the commands accept an optional flag, to be set in order for
the command to take some action beyond simply producing a report;
specifying either no flag or zero for the flag's value will cause the
command to take no further action, and specifying a 
flag value of one will cause the command to perform the
action attributed to the command.
The {\it ptfc}\/ command is normally run in the background, by
either appending an ampersand character to the command or
executing it under the ``screen'' command.

\begin{table*}
\caption{\label{tab:smartphonecomands}High-level commands for
interrogation and control of the IPAC-PTF software and data system. }
\begin{small}
\begin{tabular}{ll}
\\[-5pt]
\hline\hline
Command & Definition\\
\hline \\[-5pt]
{\it ptfh} & Prints summary of available commands. \\
{\it ptfi} & Checks whether current night has been ingested. \\
{\it ptfj} & Checks status of disks, pipelines, and archiver. \\
{\it ptfe} & Prints list of failed pipelines. \\
{\it ptfs [YYYY-MM-DD]\thinspace\tablenotemark{1}[flag (0 or 1)]\thinspace\tablenotemark{2}} & Launches image-splitting pipelines for given night. \\
{\it ptff [YYYY-MM-DD] [flag (0 or 1)]} & Ignores filter checking and relaunches relevant\\   & image-splitting pipelines for given night. \\
{\it ptfp [YYYY-MM-DD] [flag (0 or 1)]} & Launches image-processing pipelines for given night. \\
{\it ptfr [YYYY-MM-DD] [flag (0 or 1)]} & Launches catalog-generation pipelines for given night. \\
{\it ptfm [YYYY-MM-DD] [flag (0 or 1)]} & Launches source-matching pipelines for given night. \\
{\it ptfq} & Prints list of nights ready for archiving. \\
{\it ptfk [YYYY-MM-DD] [flag (0 or 1)]} & Makes archive soft link for given night. \\
{\it ptfa [list of YYYY-MM-DD]} & Schedules processing nights to be
archived, and \\ & generates optional archiver command. \\
{\it ptfc} & Script to manually execute archiver command \\ & generated by {\it ptfa}. \\
{\it ptfd [YYYY-MM-DD]} & Prints delivery/archive information for given night. \\
\hline \\[-5pt]
\end{tabular}
\tablenotetext{1}{The square brackets indicate command options. }
\tablenotetext{2}{The optional flag set to 1 is required for the command to take action beyond simple report generation.}
\end{small}
\end{table*}

\section{\label{idad}Data Archive and Distribution}

PTF camera images and processed products are permanently archived~\citep{wei}.
As was mentioned earlier, the PTF data archive is curated by IRSA.\@
This section describes the processes
involved in the ongoing construction of the PTF archive, and, in addition,
the user web interface provided by IRSA for downloading PTF products.

\subsection{\label{productarchiver}Product Archiver}

The product archiver is software written in Perl, called
{\it productArchiver.pl}, that
transfers the latest version of the products from the sandbox 
to the archive and updates the database with the product archival locations.
With the exception of the pipeline log files,
all-sky-depth-of-coverage images (Aitoff projections), and
nightly aggregated source catalogs ({\it sources.sql}\/ files), only
the processed-image-product files that are registered in the {\it ProcImages}, {\it Catalogs},
{\it AncilFiles}, {\it CalFiles}, and {\it CalAncilFiles}\/ database tables
are stored permanently in the PTF archive.
These include processed images, data masks,
source catalogs (FITS binary tables), and JPEG preview images.  The
calibration files associated with the processed images are also archived.
The camera-image files, processed products, and database metadata are delivered
to IRSA on a nightly basis.  The reference images and associated
catalogs and ancillary files are archived with a separate script, with
corresponding metadata delivered to IRSA on an episodic basis.

Before the product archiver is executed, a soft link for the night of
interest is created to point to the designated archive disk
partition.  The capacity of the partitions is nominally 8~TB each.
The soft links are a convenient means of managing the data stored in
the partitions.  As
new product versions are created and migrated to new partitions, the
old partitions, when they are no longer needed, are cleaned out and recycled.

Because both the frame-processing pipeline ($ppid=5$) and
catalog-generation pipeline ($ppid=13$) produce similar sets of
products, but only one set of products for a given night is desirable for
archiving, it is necessary to indicate which set to archive.
Generally, this is the most recently generated set.
The flagging is done by
executing a database stored function
called {\it setBestProductsForNight}, which determines the latest set of
products and designates it as the one to be archived.  It then
sets database column 
{\it pBest}\/ in the {\it ProcImages}\/ database table to 1 for all
best-version records corresponding to the selected pipeline and 0 for
all best-version records
corresponding to the other.  Here, 1 means archive the pipeline products, and 0
means do not archive.  

The product archiver inserts a record into the {\it ArchiveVersions}\/
database table, which includes a time stamp for when the archiving 
started for a particular night, and gets back a unique database ID for
the archiving session, named {\it avid}.  The product records for the night of interest in the 
aforementioned database
tables are updated to change {\it archiveStatus}\/ from 0 to -1, in
order to indicate the
records are part of a long transaction (i.e., the archiving process
for a night's worth of products).
After each product has been copied to archival disk storage and its MD5
checksum verified, the
associated database record is updated with {\it avid}\/ and the new file
location, and the {\it archiveStatus}\/
is changed from -1 to 1 to indicate that the product has been
successfully archived.

\subsection{Metadata Delivery}

Database metadata for each night, or for the latest episode of reference-image
generation, are queried from the operations
database and written to data files for loading into an
IRSA relational database.  The data files are formatted according to
IRSA's specification, and then transmitted to IRSA 
by copying them to a data directory called the ``IRSA inbox'', 
which is cross-mounted between PTF and IRSA.\@  The inbox is monitored
by a data-ingestion process that is running on an IRSA machine.
Separate metadata deliveries are made for camera images, processed images
\& associated source catalogs, and reference image \& associated
source catalogs.  Source-catalog data for processed images
are read from the aggregated {\it sources.sql}\/ files, rather than
queried from the database (since we are not loading source catalogs
into the operations database at this time).  The creation of the
metadata sets is facilitated by
database stored functions that marshal the data from various 
database tables into the {\it IRSA}\/ database table, which can be
conveniently dumped into a data file.

\subsection{\label{archiveexec}Archive Executive}

The archive executive is software that runs in an open loop on the 
ingest backup machine.  It sequentially launches instances of the VPO (see \S{\ref{vpo}}),
for each night to be archived. 
The archive executive expects archive jobs to be inserted as
records in the {\it ArchiveJobs}\/ database table (see \S{\ref{db}}).  
Staging archive jobs for execution, therefore, is 
effected by inserting associated {\it ArchiveJobs}\/ database records 
and assuring that the records are in the required state
for acceptance by the executive.
The database table is queried for an archive job when the designated archive
machine is not currently running an archive job and its archive executive
is seeking a new job.  The archive job with the latest night date
has the highest priority and is executed first.  Only one archive job
at a time is permitted. 

An {\it ArchiveJobs}\/ database record is prepared for staging an
archive job by setting its {\it status}\/ column to 0.  
The archive job that is currently executing will have its status set
to -1, indicating that it is in a long transaction.  The {\it started}\/
column in the record will also be updated with a time stamp for when the archive job began.
Staged archive jobs that have not yet been executed
can be manually suspended by setting their status to -1.
When the archive job has completed, its status is set to~1, 
its {\it ended}\/ column is updated with a time stamp for when the
archive job finished, and the {\it elapsed}\/ column is updated with the
elapsed time between starting and ending the archive job.

\subsection{\label{archiveproducts}Archive Products}

At the time of writing, $\approx 3$~million
processed CCD images from 1671
nights have been archived.   
The total number of PTF source observations stored in
catalogs is estimated to be more than 40 billion. 
PTF collaboration
members can access the processed products 
from a web interface provided by IRSA (see \S{\ref{webinterface}}).

The archive contains unprocessed camera images,
processed images, accompanying data masks, source catalogs extracted from the
processed images, reference images, reference-image catalogs, 
calibration files, and pipeline log files.  PTF pipelines generate numerous intermediate
product files, but only these final products are stored in the PTF archive.
Table~\ref{tab:archiveproducts} provides a complete list of the
products that exist in the PTF archive.
The archive's holdings include {\it SExtractor}\/ and {\it DAOPHOT}\/
source catalogs in FITS binary-table files.
There are also plans to ingest the catalogs
into an IRSA relational database.  

\begin{table*}
\caption{\label{tab:archiveproducts}Products in the PTF archive. }
\begin{scriptsize}
\begin{tabular}{ll}
\\[-5pt]
\hline\hline
Product & Notes\\
\hline \\[-5pt]
{\it Camera Images} & Direct from Mt. Palomar; multi-extension FITS, per-exposure files. \\
{\it Processed Images} & Astrometrically and photometrically calibrated, per-CCD FITS images. \\
{\it Data Masks} & FITS images with per-pixel bit flags for special data conditions (see Table~\ref{tab:dmask}). \\
{\it Source Catalogs} & Both {\it SExtractor}\/ and {\it DAOPHOT}\/ catalog types in per-CCD FITS binary tables. \\
{\it Aggregated Catalogs} & Nightly aggregated per-CCD {\it SExtractor}\/ catalogs, in both SQL and HDF5 formats.\\
{\it Reference Images} & Co-additions of 5+ processed images for each available field, CCD, and filter. \\
{\it Ref.-Im.\ Catalogs} & Both {\it SExtractor}\/ and {\it DAOPHOT}\/ catalog types in FITS binary-table format. \\
{\it Ref.-Im.\ Ancillary Files} & Uncertainty, PSF, and depth-of-coverage maps; DS9-region file for  {\it DAOPHOT}\/ catalog. \\
{\it Calibration Files} & Superbias, superflat, and ZPVM FITS images for each available night, CCD, and filter. \\
{\it Sky-Coverage Files} & Aitoff FITS images showing per-filter nightly and total observation coverage. \\
{\it Pipeline Log Files} & Useful for monitoring software behavior and tracking down missing products. \\
\hline \\[-5pt]
\end{tabular}
\end{scriptsize}
\end{table*}

\subsection{\label{webinterface}User Web Interface}

The PTF-archive web interface is very similar to the one IRSA provides
for other projects,\footnote{For example,
see http://irsa.ipac.caltech.edu/applications/wise}  which
was in fact built from the same code base.  The
architecture and key technologies used by modern IRSA web interfaces 
have been described by \citet{levine} in the context of the {\it Spitzer Heritage Archive}.

The PTF archive can be easily searched by sky position, field number,
or Solar System object/orbit.  A batch-mode search function is also available,
in which a table of positions must be uploaded.  The search results
include a list of all PTF data taken over time that match the search criteria.
Metadata about the search results, such as when the observations 
were made, is returned in a multi-column table in the web browser.  
The table currently has more than a dozen different columns.  The search
results can be filtered in specific ranges of the metadata using the
available web-interface tools.

The web interface has extensive FITS-image viewing capabilities.
When a row in the metadata table is selected, the corresponding
processed image is displayed.

The desired data can be selected using check boxes.  There is also a check
box to select all data in the search results.  The selected data are packaged
in the background, and data downloading normally commences
automatically.  As an option, the user can elect instead to be
e-mailed the URL for downloading at some later convenient time.

\section{\label{lessonslearned}Lessons Learned}

The development and operations of the IPAC-PTF image processing and
data archiving
has required 1-2 software engineers
to design custom source code, a part-time pipeline operator to utilize the software to generate and
archive the data products on a daily basis, a part-time hardware engineer to set up the
machines and manage the storage disks, a part-time database administrator
to provide database consulting and backup services, and 4-6 scientists
to recommend processing approaches and analyze the data products.  
The team breakdown in terms of career experience is roughly 70\%
seasoned senior and 30\% promising junior engineers and scientists.
The small team allows extreme agility in exploring
data-processing options and setting up new processes.  
Weekly meetings and information sharing via a variety of
database-centric systems (e.g., wiki, operations-database replicate,
software-change tracking, etc.)
have been key managerial tools of a smoothly running project.
Telecons are not nearly as effective as face-to-face meetings for
projects of this kind.  Software documentation has been kept minimal 
to avoid taxing scarce resources.
Separate channels for providing
products to ``power users'' closer to the center of the organization
vs.\ regular consumers of the products have enhanced productivity
and improved product quality on a faster time scale.
The necessity of having engineers 
actually run the software they write on a daily basis has
significantly narrowed the gap between engineering and operational
cultures within the team.
While discipline is needed in making good use of the software 
version-control and change-tracking systems, and in releasing upgraded
software to operations, a CCB (change-control board) has not
been needed thus far.  This kind of organization may not work well in
all settings, but it has worked very well for us.  Also, as data
flow 7 days a week, it is good to have someone on the team who is 
willing to work outside normal business hours, such as doing urgent weekend
builds and monitoring the image processing.

The PTF system is complex, and weeding out problems with a small team and
very limited resources has been a challenge.  To the extent possible,
we have followed best practices with an astronomy
perspective~\citep{shopbell}.
Several specific lessons learned are described in the following paragraphs.

Inspecting the data for issues could absorb a tremendous amount of
time; still, this time is very well spent, and it is important to make
the process as efficient as possible to maximize the benefits from
this inspection.
A balanced approach
that examines the data products more or less evenly, with perhaps slightly more emphasis
on the higher-level data products has been a good strategy.  Analyzing
the products and writing science papers for professional journal
publication is probably the best way to bring data issues to light; in fact,
this method has unearthed subtle flaws in the processed products that would
have otherwise gone unnoticed, and suggests that a narrow partnership
between those writing science papers and those developing the software
is an essential ingredient for success in any data-processing project.

We found it advantageous to wrap all pipeline-software database 
queries in stored functions and
put them all in a single source-code file.  This makes it a much less
daunting task to later review the database functionality and figure out the necessary 
optimizations.  The single source-code file also facilitates viewing 
the database functionality as a coherent unit at a point in time.  
Past versions of this file, which obviously have evolved over time, 
can be easily checked out from the CVS repository.

Pipeline configuration and execution must be kept simple,
in order for those who are not computer scientists to be able to run 
pipelines themselves outside of the pipeline-executive apparatus.
Having several sandbox disks available for storing pipeline products
is invaluable because the pipelines can be run on many cases to test
various aspects of the pipelines and the data.  Equipping pipeline
users with a means of configuring the database and sandbox disk 
for each pipeline instance allows greater flexibility.

Isolating products on disk and in the database according to their
processing version is very important, a lesson learned from the 
Spitzer project.  Our database schema and stored functions 
are set up to automatically
create product records with new version numbers, and these version
numbers are incorporated into disk subdirectory names for uniqueness.
Occasionally, a pipeline for a given CCD will fail for various reasons
and it is necessary to rerun the pipeline just for that CCD.\@  This is 
possible with our pipeline and database design.
Having multiple product versions in the sandbox can be
extremely useful, provided they are clearly identified, in separate,
but nearby data directories, and database queryable.  This, of course, 
requires the capability of querying the database for the best-version 
products before pulling the trigger to archive a night's worth of products.
It is also very useful to be able to locate the products in a
directory tree without having to query a database for the location.

The little details of incorporating the right data in the right places
really do matter.  Writing more diagnostics rather than
less to a pipeline log file provides information for easier
software debugging.  The diagnostics should include time stamps
and elapsed times to run the various processes, as well as CDF
listings and module command-line arguments.
The aforementioned product versioning is
crucial to the data management, and so is having the software and 
CDF version numbers written to both the product's 
database record and its FITS header, which aids not only debugging, 
but also data analysis.  It is not fully appreciated how useful
these things are unless one actually performs these tasks.

Being able to communicate with the image-processing and archiving system
remotely results in great cost savings, because it lessens the need to
have reserve personnel to take over when the pipeline operator is away
from the office.  Ideally, the software that interfaces to the system
will be able to deliver reports and execute commands with a
low-bandwidth connection.  Text-based interfaces rather than GUIs
simply function better under a wider range of conditions and
situations.  Our setup includes these features, and even works
for cases where direct Internet is unavailable, but cellular communications
allow access (see~\S{\ref{smartphone}}).  We have demonstrated its
effectiveness when used from the home office, and from remote
locations, such as observatory mountaintops.

Another lesson learned is that problems occur no matter how
fault-tolerant the system (e.g., power outages).  
Rainy-day scenarios must be developed that prescribe specific 
courses of action for 
manual intervention when automated processing is interrupted.
Sometimes the cause of a problem is
never found, in which case workarounds to deal with the effects must
be implemented as part of the automated system (e.g., rerunning
pipelines that randomly fail with a ``signal 13'' error).
Sometimes the problem goes away mysteriously, obviating the need for a
fix or workaround.
Other problems have known causes, but cannot be dealt with owing to
lack of resources; e.g., an inexpensive router that drops packets or
network limitations of the institutional infrastructure.  The latter example led to
periodically slow and unpredictable network data-transfer rates, which
is one of the reasons we stopped loading source-catalog records into 
the operations database.

Here is a summary of take-away lessons and recommendations for similar large telescope projects:

\begin{enumerate}
\item{Pipeline software development is an ongoing process that continues for years beyond telescope first light.}
\item{A development team in frequent face-to-face contact is highly recommended.}
\item{The engineering and operations teams should work closely
    together, and be incentivized to ``take ownership'' of the system.}
\item{A closely coupled relational database is essential for complex processing and data management.}
\item{Pay special attention to how asynchronous camera-exposure metadata are
    combined with camera images, in order to assure that the correct
    metadata is assigned to each image.}
\item{Low-bandwidth control of pipeline job execution is useful from
    locations remote to the data center.}
\item{Be prepared to work around problems of unknown cause.}
\item{There will be a robust demand from astronomers for both
    aperture-photometry and PSF-fit calibrated source catalogs, as
    well as reference images and associated catalogs, light-curve products, and
    forced-photometry products.}
\item{Scientists studying the data products is an effective
    science-driven means of finding problems with the data and
    processing.}
\item{The data network is a potential bottleneck and should be
    engineered very carefully, both from the mountain and within the
    data center.}
\end{enumerate}

\section{\label{conclusions}Conclusions}

This paper presents considerable detail on
PTF image processing, source-catalog generation, 
and data archiving at IPAC.
The system is fully automated and requires minimal human
support in operations, since much of the work is done by software called the
``virtual pipeline operator''.
This project has been a tremendous success in terms of the number of published
science papers (80 and counting).    There are almost 1500 field and
filter combinations (mostly $R$ band) in which more
than 50 exposures have been taken, which typically occurred twice per
night.  This has
allowed unprecedented studies of transient phenomena from asteroids to supernovae.
More than  3~million
processed CCD images from 1671
nights have been archived at IRSA, along with extracted source
catalogs, and we have leveraged
IRSA's existing software to provide a powerful web interface
for the PTF collaboration to retrieve the products.  Our
archived set of reference (coadded) images and catalogs numbers over 40 thousand
field/CCD/filter combinations, and is growing as more images that
meet the selection criteria are acquired.
We believe the many design features of our PTF-data processing and
archival system 
can be used to support future complex time-domain surveys and projects.
The system design is still evolving and periodic upgrades are
improving its overall performance.

\begin{acknowledgments}
E.O.O. is incumbent of the Arye Dissentshik career development chair and
is gratefully supported by grants from the Israeli Ministry of Science,
the I-CORE Program of the Planning and Budgeting Committee, and 
the Israel Science Foundation (grant No.\ 1829/12).

We wish to thank Dave Shupe, Trey Roby, Loi Ly, Winston Yang, Rick Ebert, Rich
Hoban, Hector Wong, and Jack Lampley for valuable contributions to the
project.

PTF is a scientific collaboration between the California Institute of
Technology, Columbia University, Las Cumbres Observatory, the Lawrence
Berkeley National Laboratory, the National Energy Research Scientific
Computing Center, the University of Oxford, and the Weizmann Institute
of Science.

This work made use of Montage, funded by the NASA's Earth Science Technology Office, 
Computation Technologies Project, under Cooperative Agreement Number NCC5-626 between 
NASA and the California Institute of Technology. Montage is maintained by the NASA/IPAC Infrared Science Archive.

This project makes use of data from the Sloan Digital Sky Survey, managed by the Astrophysical Research Consortium for the Participating Institutions and funded by the Alfred P. Sloan Foundation, the Participating Institutions, the National Science Foundation, the US Department of Energy, NASA, the Japanese Monbukagakusho, the Max Planck Society, and the Higher Education Council for England.

This research has made use of the VizieR catalogue access tool, CDS, Strasbourg, France.

Our pipelines use many free software packages from other institutions
and past projects (see Table~\ref{tab:thirdpartysoftwaretable}), for which we are indebted.

\end{acknowledgments}

\bibliography{ms}   
\bibliographystyle{aas}   

\appendix

\section{Simple Photometric Calibration}

PTF pipeline processing executes two different methods of photometric calibration.
We implemented a simple method early in the development,
which is documented below.  It is relevant because its results
are still being written to the FITS headers of
PTF processed images. 
Later, we implemented a more
sophisticated method of photometric calibration, which is described in
detail by \citet{ofek} and whose results are also included in the FITS
headers.  
For both methods, the SDSS-DR7
astronomical-source catalog \citep{abazajian} is used as the
calibration standard.
The simple method is implemented for the $R$ and $g$ camera filters only, and
there are no plans to extend it to other filters.
The zero point derived from the former method, which is
executed for each CCD and included filter on the associated data taken in a given night, provides a
useful sanity check on the same from the latter method, which are
complicated by small variations in the zero point from one image position
to another.

\subsection{\label{datamodelappendix}Data Model and Method}

Our simple method is a multi-step process 
that finds a robust photometric calibration for astronomical sources
from fields overlapping SDSS fields. 
For a given image, we assume there are $N$ source data points indexed $i  = 0, \cdots, N-1$ and,
for each data point $i$, the calibrated SDSS magnitude $M^{\rm SDSS}_i$ and the PTF
instrumental (uncalibrated) magnitude $M^{\rm PTF}_i $ for the same filter are known.  We also make
use of the color difference $g_i - R_i$ from the SDSS catalog.
The data model is

\begin{equation}
M^{\rm SDSS}_i - M^{\rm PTF}_i = ZP + b (g_i - R_i)
\end{equation}

\noindent The model parameters are the photometric-calibration zero point $ZP$ and the color-term
coefficient $b$.  The latter term on the right-hand side of Equation~A1 represents the magnitude difference
due to the difference in spectral response between like PTF and SDSS
filters.  

Radiation hits, optical ghosts and halos, and
other data artifacts can have an adverse effect on the data-fitting results of
conventional least-squared-error minimization. 
To introduce a robust measure, a Lorentzian probability distribution
function is assumed 
for the error distribution of the matched astronomical sources:

\begin{equation}
f_{\rm Lorentz} \propto \frac{1}{1 + \frac{1}{2} z^2 } 
\end{equation}

\noindent where 

\begin{equation}
z = \frac{y_i - y(g_i - R_i | ZP, b)}{\sigma_i}.
\end{equation}

\noindent In the numerator of Equation~A3, $y_i$ represents the left-hand side of Equation~A1,
while $y(g_i - R_i | ZP, b)$ represents the right-hand side of the same.
In its denominator, $\sigma_i$ is the standard deviation of $y_i$.

Using straightforward maximum-likelihood-estimation analysis, the cost function to
be minimized by varying $ZP$ and $b$ reduces to

\begin{equation}
\Lambda = \sum_{i=0}^{N-1} log(1 + \frac{1}{2} z^2)
\end{equation}

\noindent Equation~A4 has the advantage of decreasing the weight for
outliers in the tails of the data distribution, whereas the Gaussian-based
approach will give more weight to these points, thus skewing the result.

\subsection{\label{implementationappendix}Implementation Details}

Astronomical sources are
extracted from PTF processed images using {\it SExtractor}.  We elected to
use a fixed aperture of 8~pixels (8.08~arcseconds) in diameter in the
aperture-photometry calculations that yield the PTF instrumental
magnitudes, which are derived from {\it SExtractor's} $FLUX\_APER$ values.  The PTF sources used in the simple photometric
calibration are selected on criteria involving the following {\it SExtractor}\/ parameters:
$FLAGS=0$, $CLASS\_STAR \ge 0.85$, and $FLUX\_MAX$ is greater than or
equal to 4~times $FLUX\_THRESHOLD$.
The selected PTF sources, therefore, are unflagged, high
signal-to-noise stars.
These stars are matched to sources in the SDSS-DR7 catalog with a
matching radius of 2~arcseconds, and a minimum of 10 matches are
required, in order to execute the simple photometric calibration.
The flux densities of the stars and associated uncertainties 
are normalized by their image exposure times.

Two steps are taken to perform the data fitting based on the data
model described in Subsection~\ref{datamodelappendix}.  First, a
simple linear regression with Gaussian errors is performed as an
initial input for the robust regression. The Lorentzian error
regression analysis is then performed using a Nelder-Mead downhill
simplex algorithm with these initial values for the zero point and 
color-term coefficient. This algorithm has proven to be quite robust,
with a 5-10 \% failure rate when the precision is set to the machine
epsilon.   This rate drops to nearly zero when the precision is set to a factor of 10
times the machine epsilon. 

Only for images overlapping SDSS fields is the method of Subsection~\ref{datamodelappendix}
performed.  Regardless of SDSS-field overlap, the images will each have a unique air mass value $A$. 
The photometric-calibration results are thus treated as a function of air mass, and by employing a linear
data model, a zero point at an air mass of
zero and an air-mass extinction coefficient are
then computed nightly for 
each CCD and filter (data acquisition for both $g$ and
$R$ filters in the same night is possible).   These quantities are
obtained by a similar linear-regression method, where the
data fitting is done with the following first order polynomial function
of air mass $ZP(A)$, where the zero point at an air mass of zero is the zeroth-order fit
coefficient $ZP^{A=0}$ and the extinction coefficient is the first-order coefficient $\beta$:

\begin{equation}
ZP(A) = ZP^{A=0} - \beta A.
\end{equation}

\noindent This equation is used to obtain the zero point for images that do not overlap SDSS
fields.  For the images that do, the zero point from Subsection~\ref{datamodelappendix}
is used directly.  The data model is formulated so that the extinction
coefficient will normally be a value greater than zero.

The software also makes a determination on whether the night is
``photometric'' for a given CCD and filter.  The basic {\it ad hoc}\/
criterion for this specification is that the extinction coefficient must
be a value in the 0.0-0.5 range.  Additionally, we require a Pearson's
$r$-correlation above 0.75.

To apply the zero point for converting from {\it SExtractor}\/ instrumental
magnitude to calibrated magnitude, the following equation is used:

\begin{equation}
M^{\rm PTF}_{\rm Calibrated} = M^{\rm PTF}_{\rm SExtractor} + ZP + 2.5 log_{10}(T_{\rm exposure}),
\end{equation}

\noindent where $T_{exposure}$ is the exposure time of the associated image, in seconds.
If the color difference $g_i-R_i$ for a source is
known, then the color term can also
be included in the application of the simple photometric calibration;
otherwise, it is ignored.

Table~\ref{tab:fitskeywords} lists the FITS keywords associated with
our simple photometric calibration, which are written to the headers
of the image files.

\begin{table*}
\caption{\label{tab:fitskeywords}FITS keywords associated with our simple photometric calibration.}
\begin{footnotesize}
\begin{tabular}{l p{14cm}}
\\[-5pt]
\hline\hline
FITS keyword & Definition\\
\hline \\[-5pt]
$PHTCALEX$ & Flag set to 1 if simple photometric calibration was
executed without error.  The flag is set to zero if either there was an
execution error, or it was not executed.\\   
$PHTCALFL$ & Flag for whether the image is from what was deemed a
``photometric night'', where 0=no and 1=yes (see
Subsection~\ref{implementationappendix} for more details).\\
$PCALRMSE$ & Root-mean-squared error from data fitting with Equation~A5, in physical units of magnitude.\\
$IMAGEZPT$ & Image zero point, in physical units of magnitude, either
computed with Equation~A5 or taken directly from
the data fitting with Equation~A1, depending on whether
the image overlaps an SDSS field.  The keyword's value is set to NaN if $PHTCALEX=0$.\\
$COLORTRM$ & Color-term coefficient $b$, in dimensionless physical units, from Equation~A1.  This keyword will not be
present in the FITS header unless the image overlaps an SDSS field.\\
$ZPTSIGMA$ & Robust dispersion of $M^{SDSS}_i - M^{\rm PTF}_i$ after data
fitting with Equation~A1, in physical units of magnitude.  This keyword will not be
present in the FITS header unless the image overlaps an SDSS field.\\
$IZPORIG$ & String set to ``SDSS'' if the image overlaps an SDSS field and
$IMAGEZPT$ is from Equation~A1 or set to ``CALTRANS'' if the image does
not overlap an SDSS field and $IMAGEZPT$ is from Equation~A5 or set to
``NotApplicable'' if $PHTCALEX=0$.\\
$ZPRULE$ & String set to ``DIRECT'' if the image overlaps an SDSS field and
$IMAGEZPT$ is from Equation~A1 or set to ``COMPUTE'' if the image does
not overlap an SDSS field and $IMAGEZPT$ is from Equation~A5 or set to
``NotApplicable'' if $PHTCALEX=0$.\\
$MAGZPT$ & Zero point at an air mass of zero, in physical
units of magnitude.  Set to NaN if $PHTCALEX=0$.  Note that the
keyword's comment may state it is the zero point at an air mass of 1,
which is regrettably incorrect.\\
$EXTINCT$ & Extinction coefficient, in physical units of magnitude.  Set to NaN if $PHTCALEX=0$.\\
\hline \\[-5pt]
 \end{tabular}

\end{footnotesize}
\end{table*}

\clearpage

\subsection{Performance}

The simple method yields a photometric calibration of reasonable
accuracy.  
Of the $R$-band nights that could be calibrated,
where typically more than 50 CCD images that overlap SDSS fields
were acquired, half of
the nights had a zero-point standard deviation of less than 0.044 magnitudes across all
magnitudes and CCDs, and 
70\% of them had a standard deviation of less than 0.105 magnitudes.  
The mode of the distribution of nightly zero-point standard deviations
is 0.034 magnitudes.
On the other hand,
22\% of the nights had a standard deviation $>1$ magnitude.
This range is 
larger than the 0.02-0.04 magnitude accuracy reported by \citet{ofek}
for our more sophisticated method.  Yet, under favorable conditions, simple
photometric calibration works remarkably well.

From a sample of approximately 1.66 million data points, we can
evaluate the statistics of the free parameters in Equation~A5.  The average
$ZP^{A=0}$ is 23.320 magnitudes, with a standard deviation of 0.3144
magnitudes.  The average $\beta$ is 0.1650 magnitudes per unit airmass, with a standard
deviation of 0.3019 magnitudes.

The coefficient $b$ has been found empirically to fall into
a relatively small range of values.
Table~\ref{colortermappendix} gives statistics of the
color-term coefficient broken down by CCD and filter.

\begin{table}
\caption{\label{colortermappendix} Statistics of the resulting
  color-term coefficients computed from the simple photometric
  calibration (see Equation A1), broken down by CCD and filter. }
\begin{small}
\begin{tabular}{rcrrr}
\\[-5pt]
\hline\hline
{\it CCDID} & Filter & $N$ (counts) & Average (dimensionless) & Std. Dev. (dimensionless) \\
\hline \\[-5pt]
     0 &   $g$ &  23172 &  0.1786 & 0.0962\\
        &   $R$ & 125604 &  0.1457 & 0.0817\\
     1 &   $g$ &  23247 &  0.1134 &   0.1002\\
       &   $R$ & 126034 &  0.1482 & 0.0758\\
     2 &   $g$ &  23265 &  0.1290 & 0.0919\\
       &   $R$ & 125991 &  0.1416 &  0.0692\\
     4 &   $g$ &  23066 &  0.1158 &  0.0904\\
       &   $R$ & 125205 &  0.1335 &  0.1069\\
     5 &   $g$ &  23140 &   0.1812 & 0.0852\\
       &   $R$ & 125376 &  0.1283 &  0.1311\\
     6 &   $g$ &  23044 &  0.1103 & 0.0925\\
       &   $R$ & 125453 &   0.1500 & 0.0613\\
     7 &   $g$ &  23073 &  0.1027 &  0.1089\\
       &   $R$ & 125613 &  0.1424 & 0.0775\\
     8 &   $g$ &  23092 &  0.1018 & 0.0986\\
       &   $R$ & 126013 &  0.1345 & 0.0795\\
     9 &   $g$ &  23243 &  0.1129 & 0.0958\\
       &   $R$ & 125318 &  0.1097 &  0.1466\\
    10 &   $g$ &  23052 & 0.0993 & 0.0933\\
       &   $R$ & 124806 &  0.1406 & 0.0913\\
    11 &   $g$ &  22775 &  0.1775 & 0.0927\\
       &   $R$ & 124275 &  0.1415 & 0.0743\\
\hline \\[-5pt]
\end{tabular}
\end{small}
\end{table}

\end{document}